Spring 2023

# LAW INFORMS CODE: A LEGAL INFORMATICS APPROACH TO ALIGNING ARTIFICIAL INTELLIGENCE WITH HUMANS

John J. Nay

# NORTHWESTERN JOURNAL OF TECHNOLOGY AND INTELLECTUAL PROPERTY

# LAW INFORMS CODE: A LEGAL INFORMATICS APPROACH TO ALIGNING ARTIFICIAL INTELLIGENCE WITH HUMANS

*John J. Nay*

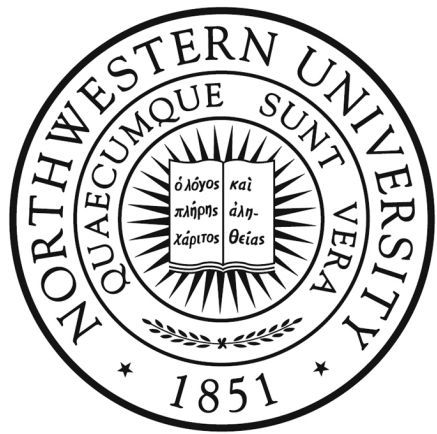

# LAW INFORMS CODE: A LEGAL INFORMATICS APPROACH TO ALIGNING ARTIFICIAL INTELLIGENCE WITH HUMANS

*John J. Nay**

**ABSTRACT** – Artificial Intelligence (AI) capabilities are rapidly advancing. Highly capable AI could cause radically different futures depending on how it is developed and deployed. We are unable to specify human goals and societal values in a way that reliably directs AI behavior. Specifying the desirability (*value*) of AI taking a particular *action* in a particular *state* of the world is unwieldy beyond a very limited set of *state-action-values*. The purpose of machine learning is to train on a subset of states and have the resulting agent generalize an ability to choose high value actions in unencountered circumstances. Inevitably, the function ascribing values to an agent's actions during training is an incomplete encapsulation of human values and the training process is a sparse exploration of states pertinent to all possible futures. After training, AI is therefore deployed with a coarse map of human preferred territory and will often choose actions unaligned with our preferred paths.

Law-making and legal interpretation convert opaque human goals and values into legible directives. *Law Informs Code* is the research agenda embedding legal processes and concepts in AI. Like how parties to a legal contract cannot foresee every potential "if-then" contingency of their future relationship, and legislators cannot predict all the circumstances under which their bills will be applied, we cannot *ex ante* specify "if-then" rules that provably direct good AI behavior. Legal theory and practice offer arrays of tools to address these problems. For instance, legal standards allow humans to develop shared understandings and adapt them to novel situations, i.e., to generalize expectations regarding actions taken to unspecified states of the world. In contrast to more prosaic uses of the law (e.g., as a deterrent of bad behavior), leveraged as an expression of *how* humans communicate their goals, and *what* society values, *Law Informs Code*.

* Fellow at Stanford University – CodeX – Center for Legal Informatics, operated by Stanford Computer Science Department and Stanford Law School, working on A Legal Informatics Approach to AI Alignment; Co-founder of Brooklyn Artificial Intelligence Research and its quantitative investment firm, Brooklyn Investment Group.

We describe how data generated by legal processes and the tools of law (methods of law-making, statutory interpretation, contract drafting, applications of standards, and legal reasoning) can facilitate the robust specification of inherently vague human goals to increase *human-AI* alignment. Toward *society-AI* alignment, we present a framework for understanding law as the applied philosophy of multi-agent alignment, harnessing public law as an up-to-date knowledge base of democratically endorsed values ascribed to state-action pairs. Although law is partly a reflection of historically contingent political power – and thus not a perfect aggregation of citizen preferences – if properly parsed, its distillation offers the most legitimate computational comprehension of societal values available. Other data sources suggested for AI alignment – surveys, humans labeling "ethical" situations, or (most commonly) the beliefs of the AI developers – lack an authoritative source of synthesized preference aggregation. Law is grounded in verifiable resolutions: ultimately obtained from a court opinion, but short of that, elicited from legal experts. If law informs powerful AI, engaging in the deliberative political process to improve law would take on even more meaning.

## I. INTRODUCTION

As the internet went viral, "*Code Is Law*" communicated the power of software as a form of governance in cyberspace.[1] Now that Artificial

---

[1] *See* Lawrence Lessig, *Code Is Law*, HARV. MAG., JAN.–FEB. 2000, https://www.harvardmagazine.com/2000/01/code-is-law-html [https://perma.cc/GY7C-HX8M]; LAWRENCE LESSIG, CODE AND OTHER LAWS OF CYBERSPACE (1999); LAWRENCE LESSIG, CODE VERSION 2.0 (2006). The phrase "Code Is Law" has also been adopted as a rallying cry for "smart

Intelligence (AI) capabilities are rapidly advancing[2] with new model architectures[3] scaled across internet-scale data,[4] "*Law Informs Code*" is the catchphrase for a legal informatics approach to shaping AI toward human goals.

AI is increasingly widely deployed.[5] More powerful AI could cause radically different futures.[6] A summer 2022 survey of hundreds of AI researchers estimated an aggregate forecast time of 37 years for a 50% chance of "high–level machine intelligence" ("when unaided machines can accomplish every task better and more cheaply than human workers").[7] Because natural language processing (NLP) is a key sub-domain of AI, surveys of NLP researchers are of particular interest. A separate summer 2022 survey of hundreds of NLP researchers found that 57% believe that "recent research has advanced us toward AGI [artificial general intelligence] in some significant way," and 73% "agree that labor automation from AI could plausibly lead to revolutionary societal change in this century, on at least the scale of the Industrial Revolution."[8] Even before additional advancements, we currently face monumental challenges specifying human goals and societal values to reliably direct AI behavior.[9]

---

[5] *See, generally,* Daniel Zhang et al., *The AI Index 2022 Annual Report*, STAN. INST. FOR HUMAN-CENTERED A.I. (Mar. 2022), https://aiindex.stanford.edu/wp-content/uploads/2022/03/2022-AI-Index-Report_Master.pdf [https://perma.cc/YJS8-4Z77]. For legal discussions of AI deployments, see, for example, Danielle K. Citron & Frank Pasquale, *The Scored Society: Due Process for Automated Predictions*, 89 WASHINGTON L. REV. 1 (2014); E. Joh, *The New Surveillance Discretion: Automated Suspicion, Big Data, and Policing*, 10 HARV. L. & POL'Y REV. 15 (2016); Cecilia Muñoz, Megan Smith & DJ Patil, *Big Data: A Report on Algorithmic Systems, Opportunity, and Civil Rights*, EXEC. OFF. OF THE PRESIDENT (May 2016), https://obamawhitehouse.archives.gov/sites/default/files/microsites/ostp/2016_0504_data_discriminatio n.pdf [https://perma.cc/PRB8-P25Y]. On large language model deployments, see, for example, Matthew Hutson, *Robo-writers: The Rise and Risks of Language-Generating AI*, 591 NATURE 22–25 (2021); Sam Manning et al., *A Research Agenda for Assessing the Economic Impacts of Code Generation Models*, OPENAI (2022), https://cdn.openai.com/papers/Economic_Impacts_Research_Agenda.pdf [https://perma.cc/DRH9-YBCY].

[6] For a longer-term framing of potential AI impacts, see, for example, Amanda Askell, *Ensuring the Safety of Artificial Intelligence*, in THE LONG VIEW: ESSAYS ON POLICY, PHILANTHROPY, AND THE LONG-TERM FUTURE (Natalie Cargill & Tyler John, eds. 2021); HENRY KISSINGER, ERIC SCHMIDT & DANIEL P. HUTTENLOCHER, THE AGE OF AI: AND OUR HUMAN FUTURE (2021). For a nearer-term framing of potential AI risks, see the recent work by the U.S. National Institute of Standards, *e.g.*, *AI Risk Management Framework: Second Draft* (Aug. 18, 2022), https://www.nist.gov/system/files/documents/2022/08/18/AI_RMF_2nd_draft.pdf [https://perma.cc/V33U-L46Z]; Dan Hendrycks et al., *Unsolved Problems in ML Safety*, ARXIV (Sept. 28, 2021), https://arxiv.org/pdf/2109.13916.pdf [https://perma.cc/3C8Z-EKAP] [hereinafter Hendrycks, *Unsolved Problems*].

[7] *See 2022 Expert Survey*, *supra* note 2. Other surveys also estimate non-trivial impacts: "AI capabilities emerge that could radically transform welfare, wealth, or power, to an extent comparable to the nuclear revolution or even the industrial revolution. These possibilities are strikingly neglected, in part because they involve massive global and intergenerational externalities. There is thus a high leverage opportunity to address what may be the most important global issue of the 21st century." Allan Dafoe, *AI Governance: A Research Agenda*, CTR. GOV. AI (Aug. 27, 2018), https://www.fhi.ox.ac.uk/wp-content/uploads/GovAI-Agenda.pdf [https://perma.cc/95PN-AG4V].

[8] Julian Michael et al., *What Do NLP Researchers Believe? Results of the NLP Community Metasurvey*, ARXIV 11 (Aug. 26, 2022), https://arxiv.org/pdf/2208.12852.pdf [https://perma.cc/J6ET-BLUA] ("[36%] of respondents agree that it is plausible that AI could produce catastrophic outcomes in this century, on the level of all-out nuclear war.").

Significant computing and data resources are required to develop state-of-the-art AI.[10] Large companies are pushing research and deployment boundaries.[11] Regardless of where AI is developed, the developers are disconnected from those affected by AI. To increase alignment of AI with the billions of people impacted, scholars and companies have suggested embedding "ethics" into AI.[12] However, it is unclear how to decide that "ethics" or who gets a say in the process.[13] We take a different approach, arguing that the target of AI alignment should be democratically endorsed law. This provides legitimate grounding. Although law reflects the path-dependent structure of political power within a society and not a perfect aggregation of human values, it is the most democratic encapsulation of the attitudes, norms and values of the governed.

---

[9] *See, e.g.*, Laura Weidinger et al., *Taxonomy of Risks Posed by Language Models*, 2022 ACM CONF. ON FAIRNESS, ACCOUNTABILITY, & TRANSPARENCY 214 (describing risks across the following categories: Discrimination, Hate speech and Exclusion; Information Hazards; Misinformation Harms; Malicious Uses; Human-Computer Interaction Harms; and Environmental and Socioeconomic harms); Hendrycks, *Unsolved Problems*, *supra* note 6 (describing risks across the following categories: Robustness, Monitoring, Steering ML systems "Alignment", and reducing deployment hazards "Systemic Safety"); BRIAN CHRISTIAN, THE ALIGNMENT PROBLEM: MACHINE LEARNING AND HUMAN VALUES (2020) (describing fairness and transparency risks); Miles Brundage et al., *The Malicious Use of Artificial Intelligence: Forecasting, Prevention, and Mitigation*, ARXIV (2018), https://arxiv.org/ftp/arxiv/papers/1802/1802.07228.pdf [https://perma.cc/YP2X-EJQ2] (describing potential security threats from malicious uses of AI); Shahar Avin et al., *Filling Gaps in Trustworthy Development of AI*, 374 SCIENCE 1327–29 (2021); Thomas Arnold & Matthias Scheutz, *The "Big Red Button" Is Too Late: An Alternative Model for The Ethical Evaluation of AI Systems*, 60 ETHICS & INFO. TECH. 59 (2018); Kris McGuffie & Alex Newhouse, *The Radicalization Risks of GPT-3 and Advanced Neural Language Models*, ARXIV (Sept. 15, 2020), https://arxiv.org/pdf/2009.06807.pdf [https://perma.cc/B86X-RJDA].

[10] *See, e.g.*, Or Sharir, Barak Peleg & Yoav Shoham, *The Cost of Training NLP Models: A Concise Overview* (2020), https://arxiv.org/pdf/2004.08900.pdf [https://perma.cc/RV7G-TTU5]. Although, costs for training models are decreasing. *See, e.g.*, Abhinav Venigalla & Linden Li, *Mosaic LLMs (Part 2): GPT-3 Quality for <$500k* (Sept. 29, 2022), https://www.mosaicml.com/blog/gpt-3-quality-for-500k [https://perma.cc/7DF6-V2ZA].

[11] See, for example, this statement by a consortium of AI research companies, COHERE, https://cohere.ai/about, [https://perma.cc/QK55-GAJM]. *See also* OPENAI, https://openai.com/about/ [https://perma.cc/JF4M-GXRN], AI21 LABS, https://www.ai21.com/about [https://perma.cc/4MYR-54HJ], and *Best Practices for Deploying Language Models*, OPENAI (June 2, 2022), https://openai.com/blog/best-practices-for-deploying-language-models/ [https://perma.cc/PUK9-99HY].

[12] *See* Daniel Greene, Anna Lauren Hoffmann & Luke Stark, *Better, Nicer, Clearer, Fairer: A Critical Assessment of the Movement for Ethical Artificial Intelligence and Machine Learning*, 2019 PROC. HAWAII INT'L CONF. ON SYS. SCIS. 2122; Brent Mittelstadt, *Principles Alone Cannot Guarantee Ethical AI*, 1 NATURE MACH. INTEL. 501 (2019).

[13] *See generally* Mittelstadt, *supra* note 12; FRANK PASQUALE, NEW LAWS OF ROBOTICS: DEFENDING HUMAN EXPERTISE IN THE AGE OF AI (2020).

If law is leveraged as a set of methodologies for conveying and interpreting directives and a knowledge base of societal values, it can play a unique role in aligning AI with humans. Law-making and legal interpretation convert human intentions and values into legible[14] directives. *Law Informs Code* is the research agenda embedding human law in AI models so we can better specify our objectives. Most research at the intersection of AI and law has focused on two areas: how existing law[15] (or a proposed legal solution[16]) can be enforced on AI or the humans behind it (i.e., how *Law Governs Code*); or how AI can improve the practice of law[17] or implementation of policy[18] (i.e., how *Code Informs Law*).[19] This Article describes the new pillar: how

---

[14] We are using "legible" here like its use in both JAMES C. SCOTT, SEEING LIKE A STATE: HOW CERTAIN SCHEMES TO IMPROVE THE HUMAN CONDITION HAVE FAILED (1998) (making agents' actions legible to the government), and in Anca D. Dragan, Kenton CT Lee & Siddhartha S. Srinivasa, *Legibility and Predictability of Robot Motion*, 2013 ACM/IEEE INTERNATIONAL CONFERENCE ON HUMAN-ROBOT INTERACTION 301 (legible as intent-expressiveness).

[15] *See, e.g.*, Solon Barocas & Andrew D. Selbst, *Big Data's Disparate Impact*, 104 CAL. L. REV. 671 (2016); Roger Michalski, *How To Sue A Robot*, 2018 UTAH L. REV. 1021 (2018); Andrew D. Selbst, *Negligence and AI's Human Users*, 100 B.U. L. REV. 1315 (2020); Amanda Levendowski, *How Copyright Law Can Fix Artificial Intelligence's Implicit Bias Problem*, 93 WASH. L. REV. 579 (2018).

[16] *See, e.g.*, Andrew Tutt, *An FDA For Algorithms*, 69 ADMIN. L. REV. 83 (2017) (arguing that a new centralized agency is needed for regulating AI); Anton Korinek, *Why We Need a New Agency to Regulate Advanced Artificial Intelligence: Lessons on AI Control from the Facebook Files*, BROOKINGS INST. (Dec. 8, 2021), https://www.brookings.edu/research/why-we-need-a-new-agency-to-regulate-advanced-artificial-intelligence-lessons-on-ai-control-from-the-facebook-files/ [https://perma.cc/4HUE-AARJ]; Jack Clark & Gillian K. Hadfield, *Regulatory Markets for AI Safety*, ARXIV (Dec. 11, 2019), https://arxiv.org/pdf/2001.00078.pdf [https://perma.cc/8FCL-3ATX]; Jonas Schuett, *Defining the Scope of AI Regulations*, Legal Priorities Project Working Paper Series No. 9 (2021); Eric Wu et al., *How Medical AI Devices Are Evaluated: Limitations and Recommendations From an Analysis of FDA Approvals*, 27 NATURE MED. 582 (2021); Axel Walz & Kay Firth-Butterfield, *Implementing Ethics Into Artificial Intelligence: A Contribution, From A Legal Perspective, To The Development of an AI Governance Regime*, 18 DUKE L. & TECH. REV. 176 (2018).

[17] *See, e.g.*, Henry Prakken, *On How AI & Law Can Help Autonomous Systems Obey the Law: A Position Paper* 42, 44 (AI4J–Artificial Intelligence for Justice, Conference Paper, Aug. 30, 2016) ("AI & law research has traditionally focused on support tools for humans carrying out legal tasks."); Howard Turtle, *Text Retrieval in the Legal World*, 3 A.I. & L. 5 (1995).

[18] *See, e.g.*, Peter Henderson, Ben Chugg, Brandon Anderson & Daniel E. Ho, *Beyond Ads: Sequential Decision-Making Algorithms in Law and Public Policy*, ARXIV (2022), https://arxiv.org/pdf/2112.06833.pdf [https://perma.cc/26NQ-7VB5]; Hannah Bloch-Wehba, *Access to Algorithms*, 88 FORDHAM L. REV. 1265, 1273–90 (2019) (describing some existing uses of AI by the government); Emily Berman, *A Government of Laws and Not of Machines*, 98 B.U L. REV. 1277 (2018); Matthew M. Young, Johannes Himmelreich, Justin B. Bullock & Kyoung-Cheol Kim, *Artificial Intelligence and Administrative Evil*, 4 PERSPS. ON PUB. MGMT. & GOVERNANCE 244 (2019).

[19] A complementary intersection has been described as *AI as Law*, in which "AI systems are to be thought of as hybrid critical discussion systems, where different hypothetical perspectives are constructed and evaluated until a good answer is found." Bart Verheij, *Artificial Intelligence as Law*, 28 A.I. & L. 181, 191 (2020) (*AI as Law* comes from the tradition of symbolic systems; *Law Informs AI* is rooted in machine learning; and both are interested in hybrid symbolic-deep learning systems). Cullen O'Keefe defined a "law-following AI [as] an AI system that is designed to rigorously comply with some defined set of human-originating rules ("laws"), using legal interpretative techniques, under the

AI can use law as theoretical scaffolding and data to be safer by design (i.e., how *Law Informs Code*).

The benefits of law-informed AI would be far-reaching (Figure 1). In addition to more locally useful and societally aligned AI, law-informed AI could power the other two pillars: law governing AI, and AI improving legal services.

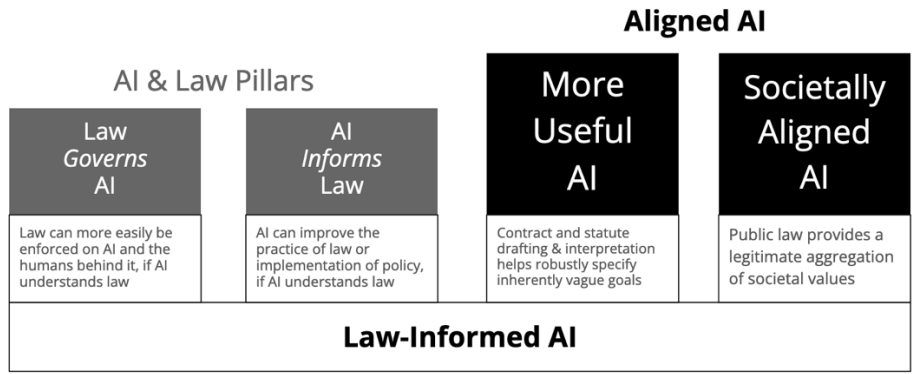

**Figure 1:** Law-informed code can help power the other AI & Law pillars. The Law Informs Code research agenda focuses on More Useful AI and Societally Aligned AI, but it could have positive externalities for Law Governing AI and AI Informing Law.

In this Article, we do not assess the legality of AI behavior[20] or spend much time recommending that AI should play a greater (or lesser) role in legal practice[21] – critical topics we lack room to address here. Instead, we focus on how AI would be more aligned with humans if we couple legal informatics with AI deployment.[22]

Sociology of finance has advanced the idea that financial economics, conventionally viewed as merely a lens on financial markets, shapes markets,

---

assumption that those laws apply to the AI in the same way that they would to a human." Cullen O'Keefe, *Law-Following AI 1: Sequence Introduction and Structure*, AI ALIGNMENT FORUM (2022), https://www.alignmentforum.org/posts/NrtbF3JHFnqBCztXC/law-following-ai-1-sequence-introduction-and-structure [https://perma.cc/6QM4-CYEE].

[20]  See, for example, an article that does discuss such issues, Michalski, *supra* note 15; SIMON CHESTERMAN, WE, THE ROBOTS? (2021).

[21]  See, for example, articles that do discuss such issues, Frank Pasquale & Glyn Cashwell, *Four Futures of Legal Automation*, 63 U.C.L.A. L. REV. DISC. 26 (2015); Benjamin Alarie, *The Path of the Law: Towards Legal Singularity*, 66 U. TORONTO L.J. 443 (2016); Emily Berman, *A Government of Laws and not of Machines*, 98 B.U. L. REV. 1277 (2018); Frank Pasquale, *A Rule of Persons, Not Machines: The Limits of Legal Automation*, 87 GEO. WASH. L. REV. 1 (2019); Aziz Z. Huq, *A Right to a Human Decision*, 106 VA. L. REV. 611 (2020); John Nay, *Large Language Models as Corporate Lobbyists*, ARXIV (January 2, 2023), https://arxiv.org/ftp/arxiv/papers/2301/2301.01181.pdf [https://perma.cc/734G-3FCA].

[22]  In this Article, we focus on U.S. law. *See infra* Section V for a discussion of this limitation.

i.e., the theory is "an engine, not a camera."[23] Law is an engine, *and* a camera. Legal drafting and interpretation methodologies refined within contract law – an engine of private party alignment – are a lens on how humans communicate their inherently ambiguous goals. Public law – an engine for societal coordination and compliance – is a high-fidelity lens on human societal values.

Specifying the desirability (i.e., *value*) of AI taking a particular *action* in a particular *state* of the world is unwieldy beyond a very limited set of *state-action-value* tuples.[24] In fact, the purpose of machine learning is to train on a subset of these tuples[25] and have the resulting agent learn decision policies that generalize to choosing high value actions in unencountered states,[26] maintaining the same level of performance in novel circumstances.[27]

---

[23] *See* DONALD MACKENZIE, AN ENGINE, NOT A CAMERA 11 (2006).

[24] Without loss of much generality to other AI paradigms such as supervised learning, we frame the alignment problem from a decision-making (reinforcement learning) perspective.

[25] Or input-output pairs if the focus is purely prediction rather than taking actions. But in this Article, we focus on the more general problem of choosing actions, rather than merely prediction. *See generally* D. Abel, J. MacGlashan & M.L. Littman, *Reinforcement Learning as a Framework for Ethical Decision Making*, AAAI Workshop: AI, Ethics, and Society (2016), https://david-abel.github.io/papers/wkshp_aaai2016_rl_ethics.pdf [https://perma.cc/296B-GDYG]. There is an increasingly porous distinction between the paradigms of AI prediction (supervised learning) and AI decision-making (reinforcement learning).

[26] Furthermore, a primary purpose of trying to develop future highly advanced AI systems, *see infra* Section II.B., is to conduct tasks that no human is capable of, *see, e.g.,* Richard Ngo, Lawrence Chan & Sören Mindermann, *The Alignment Problem From a Deep Learning Perspective* (AI Alignment Forum, Conference Paper, Aug. 10, 2022), https://www.alignmentforum.org/posts/KbyRPCAsWv5GtfrbG/the-alignment-problem-from-a-deep-learning-perspective [https://perma.cc/H3HY-JCB4]; Ajeya Cotra, *The Case for Aligning Narrowly Superhuman Models* (AI Alignment Forum, Conference Paper, Mar. 5, 2021), https://www.alignmentforum.org/posts/PZtsoaoSLpKjjbMqM/the-case-for-aligning-narrowly-superhuman-models [https://perma.cc/QVK5-RMG9].

[27] Generalization is difficult because machine learning model outputs are effectively interpolations within the model's data manifold, which is defined by the training processes, *see, e.g.,* François Chollet, *On the Measure of Intelligence* (arXiv, Working Paper No.1911.01547v2, Nov. 25, 2019), https://arxiv.org/pdf/1911.01547.pdf [https://perma.cc/LFQ9-3DMK]. For discussion of generalization in the context of reinforcement learning*, see, e.g.,* K. Cobbe et al., *Quantifying Generalization in Reinforcement Learning*, 97 PROC. MACH. LEARNING RSCH. 1283 (2019), http://proceedings.mlr.press/v97/cobbe19a/cobbe19a.pdf [https://perma.cc/A9G6-X62G]; Zhang et al., *A Study on Overfitting in Deep Reinforcement Learning* at 1 (arXiv, Working Paper No. 1804.06893v2, Apr. 20, 2018), https://arxiv.org/pdf/1804.06893.pdf [https://perma.cc/7XDH-DB3P] ("[T]he same agents and learning algorithms could have drastically different test performance, even when all of them achieve optimal rewards during training.") Ultimately, whether an AI system's generalizability is adequate depends on how it is deployed and its relation to live decision-making processes. *See, e.g.,* John Nay & Katherine Strandburg, *Generalizability: Machine Learning and Humans-in-the-loop*, in BIG DATA LAW 285 (Roland Vogl ed., 2021); Ben Green & Yiling Chen, *The Principles and Limits of Algorithm-in-the-loop Decision Making*, 3 PROC. ACM HUM.-COMPUT. INTERACT. (2019), https://scholar.harvard.edu/files/bgreen/files/19-cscw.pdf [https://perma.cc/F4YX-UFAZ]; Saleema Amershi, Maya Cakmak, William Bradley Knox & Todd Kulesza, *Power to the People: The Role of*

The reward function ascribing values to an agent's actions during training is inevitably a proxy for human preferences over all actions in all world states,[28] and the agent's training process is a sparse exploration of all states in all possible futures.[29]

AI often exhibits unanticipated "shortcut" behaviors that seek to optimize an inherently limited reward function.[30] This causes AI agents to aggressively optimize toward specified rewards at the expense of other

(usually less quantifiable) variables of interest that were left unspecified.[31] Unintended negative behavior results.[32] For instance, when a robot hand was trained to grasp a ball (*from the perspective of the human evaluator, which provided the training reward*), it optimized for hovering between the evaluator's camera and the ball. This gave the *impression* it was grasping the ball, which optimized the reward.[33] Although, *ex post,* this may seem simple to address with an improved reward function and retraining; *ex ante,* careful work from experienced machine learning researchers did not design a training process to avoid this.[34] And this is in a tightly controlled environment.

---

[31]  "Excessive literalism" is another way of describing the issue: "A system that is optimizing a function of n variables, where the objective depends on a subset of size k<n, will often set the remaining unconstrained variables to extreme values; if one of those unconstrained variables is actually something we care about, the solution found may be highly undesirable." Stuart Russell, *Of Myths and Moonshine*, EDGE FOUND. (Nov. 14, 2014), https://www.edge.org/conversation/the-myth-of-ai#26015 [https://perma.cc/P3WX-GMJ3] ("This is essentially the old story of the genie in the lamp, or the sorcerer's apprentice, or King Midas: you get exactly what you ask for, not what you want."); *see also* Stuart Russell, HUMAN COMPATIBLE: ARTIFICIAL INTELLIGENCE AND THE PROBLEM OF CONTROL (2019); Brandon Trabucco et al., *Conservative Objective Models for Effective Offline Model-Based Optimization*, 139 PROC. MACH. LEARNING RSCH. 10358 (2021), https://arxiv.org/pdf/2107.06882.pdf [https://perma.cc/M5G9-FYMR]; FRANÇOIS CHOLLET, DEEP LEARNING WITH PYTHON 450 (2nd ed. 2021) ("An effect you see constantly in systems design is the *shortcut rule*: if you focus on optimizing one success metric, you will achieve your goal, but at the expense of everything in the system that wasn't covered by your success metric. You end up taking every available shortcut toward the goal.").

[32]  *See, e.g.*, Jack Clark & Dario Amodei, *Faulty Reward Functions in the Wild*, OPENAI (Dec. 21, 2016), https://openai.com/blog/faulty-reward-functions [https://perma.cc/2VUP-RF84]; Ortega & Maini, *Building Safe Artificial Intelligence: Specification, Robustness and Assurance*, MEDIUM (Sept. 27, 2018), https://deepmindsafetyresearch.medium.com/building-safe-artificial-intelligence-52f5f75058f1 [https://perma.cc/4EVZ-RKRE]; David Manheim & Scott Garrabrant, *Categorizing Variants of Goodhart's Law* (arXiv, Working Paper No. 1803.04585v5, Feb. 24, 2019), https://arxiv.org/pdf/1803.04585.pdf [https://perma.cc/A4K2-RMAM]; Rachel L. Thomas & David Uminsky, *Reliance on Metrics is a Fundamental Challenge for AI*, PATTERNS, May 13, 2022, https://www.ncbi.nlm.nih.gov/pmc/articles/PMC9122957/pdf/main.pdf [https://perma.cc/XFX3-SP4E].

[33]  *See* Dario Amodei, Paul Christiano & Alex Ray, *Learning from Human Preferences*, OPENAI (June 13, 2017), https://openai.com/blog/deep-reinforcement-learning-from-human-preferences/ [https://perma.cc/7ESN-S7BT]; Victoria Krakovna, *Paradigms of AI alignment: Components and Enablers*, LESSWRONG (June 2, 2022), https://www.lesswrong.com/posts/JC7aJZjt2WvxxffGz/paradigms-of-ai-alignment-components-and-enablers [https://perma.cc/X7FZ-6E7S].

[34]  Another example: an AI agent maximized its provided reward by killing itself at the end of the first level of a simulated environment in order to avoid losing in level two. *See* William Saunders et al., *Trial without Error: Towards Safe Reinforcement Learning via Human Intervention* (arXiv, Working Paper No. 1707.05173, July 17, 2017), https://arxiv.org/pdf/1707.05173.pdf [https://perma.cc/5PNY-E2WQ]. For more examples, see, for example, Krakovna, *Specification Gaming: The Flip Side of AI Ingenuity*, *supra* note 31.

Real-world circumstances[35] exacerbate goal misspecification.[36] Take, for example, the implementation of simple computational rules applied to data relevant to self-driving cars. When fifty-two programmers were assigned the task of each independently automating simple speed limits, there was "significant deviation in number and type of citations issued [on application of their code to the same real-world data . . . ] this experiment demonstrates that even relatively narrow and straightforward "rules" can be problematically indeterminate in practice."[37] More capable AI can further exacerbate misspecification issues with stronger optimization ability, "achieving higher proxy reward and lower true reward than less capable agents."[38]

We can never provide enough sources of reward. There will always be relevant goals and world attribute valuations missing from any explicit reward function, or ensemble of functions.[39] It is not possible to manually specify or automatically enumerate a discernment of humans' desirability of all actions an AI might take.[40] Therefore, after training, AI is deployed with an incomplete map of human preferred territory, [41] and the resulting mismatch between what a human wants and what an AI does is a *human-AI* alignment problem.[42] Acknowledging that multiple humans have preferences over values of state-action pairs, we must grapple with an even more intractable problem: *society-AI* alignment.[43]

---

[35] *See, e.g.*, A. Rupam Mahmood et al., *Benchmarking Reinforcement Learning Algorithms on Real-World Robots* 561 (Robot Learning, Conference Paper, Sept. 20, 2018), https://arxiv.org/pdf/1809.07731.pdf [https://perma.cc/GQC3-X3E2].

[36] *See, e.g.*, Steven Kerr, *On the Folly of Rewarding A, While Hoping For B*, 18 ACAD. MGMT J. 769 (1975) (A classic article discussing reward misspecification for human agents in a business context. The same issue is applicable to AI agents.); Krakovna, *Paradigms of AI Alignment: Components and Enablers*, *supra* note 34 (Outline of technical sub-components of the overall AI Alignment problem, including the goal misspecification problem.); Pan et al., *Effects of Reward Misspecification* (arXiv, Working Paper No. 2201.03544, Feb. 14, 2022), https://arxiv.org/pdf/2201.03544.pdf [https://perma.cc/Z9RH-CYEK].

[37] Lisa A. Shay, Woodrow Hartzog, John Nelson & Gregory Conti, *Do Robots Dream of Electric Laws? An Experiment in the Law as Algorithm*, *in* ROBOT LAW 274 (Ryan Calo et al., eds. 2016).

[38] Pan, *supra* note 31, at 1 ("More capable agents often exploit reward misspecifications, achieving higher proxy reward and lower true reward than less capable agents. Moreover, we find instances of phase transitions: capability thresholds at which the agent's behavior qualitatively shifts, leading to a sharp decrease in the true reward. Such phase transitions pose challenges to monitoring the safety of ML systems.").

[39] *See, e.g.*, Roel Dobbe, Thomas Krendl Gilbert & Yonatan Mintz, *Hard Choices in Artificial Intelligence* (arXiv, Working Paper No. 2106.11022, June 10, 2021), https://arxiv.org/pdf/2106.11022.pdf [https://perma.cc/46C7-K478].

[40] Even if it was possible to specify humans' desirability of all actions a system might take within a reward function that was used for training an AI agent, the resulting behavior of the agent is not only a function of the reward function; it is also a function of the exploration of the state space. *See* Richard Ngo, *AGI Safety from First Principles*, AI ALIGNMENT FORUM 1, 21–24 (Sept. 28, 2020), https://www.alignmentforum.org/s/mzgtmmTKKn5MuCzFJ [https://perma.cc/N6KB-NVSU]. Furthermore, even when the "true" reward function is known, different functions can be equally consistent with the training data. *See, e.g.*, Kareem Amin & Satinder Singh, *Towards Resolving Unidentifiability in Inverse Reinforcement Learning* (arXiv, Working Paper No. 1601.06569, Jan. 25, 2016), https://arxiv.org/pdf/1601.06569.pdf [https://perma.cc/W5MQ-8G75]; Soren Minderman & Stuart Armstrong, *Occam's Razor is Insufficient to Infer the Preferences of Irrational Agents* (Neural Information Processing Systems, Conference Paper, Dec. 3, 2018),

We developed three primary desiderata for a framework to address these alignment problems. [44] *First*, the framework should have a well-developed theory of alignment. Rather than overly simple specifications or vacuous high-level principles, it should be laden with modular constructs built to handle the ambiguity and novelty inherent in aligning (human and/or artificial) agents. It should have a theory for how to credibly elicit human values, legitimately synthesize them, and consistently update the results (Figure 2).[45]

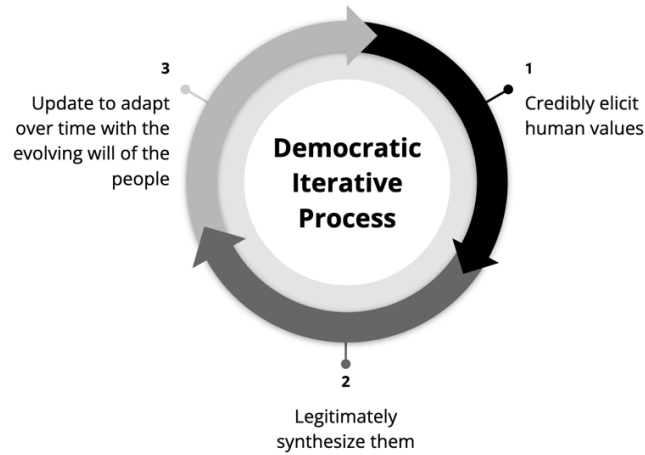

**Figure 2:** The iterative democratic process.

*Second*, in recognizing that alignment is a moving target as AI capabilities advance,[46] the framework should be useful for today's AI while

---

scaling with future advancements.[47] As AI becomes more capable, the framework should provide benchmarks and solutions calibrated to that higher level of capability. The alignment methods should directly benefit from the improvements in general AI capabilities research.[48] Much of the research on governing AI falls into two ends of a spectrum related to assumptions of the imminence of transformative AI. Research operating under the premise of a *high probability* of near-term transformative AI (e.g., within 15 years) is focused more on aligning AGI with a benevolent small group of humans that first develop transformative AI, or how to align AI with an ideal aggregation of human preferences (through yet to be specified aggregation processes). Research operating under the premise of a *low probability* of near-term transformative AI is typically focused more on how to reduce discriminatory and privacy harms posed by present-day AI. We seek a framework that bridges the AI timeline spectrum.[49]

*Third*, the framework should already be rigorously battle-tested in some form – ideally the documentation of the battle-testing has produced reams of data that can be leveraged by machine learning.

---

generative models have an unusual combination of high predictability – model capabilities scale in relation to resources expended on training – and high unpredictability – before training a model, it's difficult to anticipate all the inputs it will be subjected to, and what capabilities and outputs it will have. The former drives rapid development of such models while the latter makes it difficult to anticipate the consequences of their development and deployment."); Julien Perolat et al., *Mastering the Game of Stratego with Model-Free Multiagent Reinforcement Learning*, 378 SCIENCE 990, 990–91 (2022) ("Stratego is one of the few iconic board games that Artificial Intelligence (AI) has not yet mastered. . . . Stratego has been a grand challenge for the field of AI for decades. . . . DeepNash beats existing state-of-the-art AI methods in Stratego [an extremely complex game] and achieved a yearly (2022) and all-time top-3 rank on the Gravon games platform, competing with human expert players.").

[47] See alignment proposals that focus on scaling to more capable AI, for example, Leike et al., *Scalable Agent Alignment via Reward Modeling: A Research Direction*, ARXIV 2 (Nov. 19, 2018), https://arxiv.org/pdf/1811.07871.pdf [https://perma.cc/5NN2-VAB4] ("[W]e describe how reward modeling can be applied recursively: agents trained with reward modeling can assist the user in the evaluation process when training the next agent.").

[48] Most AI research is classified as "capabilities" research by AI alignment and safety researchers in order to distinguish mainstream research focused on improving AI performance on traditional tasks, for example, prediction, from AI research focused on alignment or safety outcomes.

[49] In addition to bridging this gap, there is a relatively unaddressed territory in between. Most pre-2018 alignment research was focused on more clearly decision and agentic oriented training processes, whereas the most powerful AI systems today are a result of self-supervised foundation models that exhibit far less goal-orientation than pure reinforcement learning approaches. *See* janus, *Simulators*, AI ALIGNMENT FORUM (Sept. 2, 2022), https://www.alignmentforum.org/posts/vJFdjigzmcXMhNTsx/simulators [https://perma.cc/5CR9-WUZ2]; Richard_Kennaway, *Is the Work on AI Alignment Relevant to GPT?*, LESSWRONG (July 30, 2020), , https://www.lesswrong.com/posts/dPcKrfEi87Zzr7w6H/is-the-work-on-ai-alignment-relevant-to-gpt [https://perma.cc/DT29-8VGR]. An alignment framework that scales across levels of intelligence may be able to address this novel model type. Regardless of whether the most capable Ais are agentic or merely "tools" or "simulators," the framework should be relevant.

Law, as the applied philosophy of multi-agent alignment, uniquely fulfills these criteria.[50] Alignment is a problem because we cannot *ex ante* specify rules that fully and provably direct good AI behavior.[51] Similarly, parties to a legal contract cannot foresee every contingency of their relationship,[52] and legislators cannot predict every specific circumstance under which their laws will be applied. That is why much of law is a constellation of standards.[53] Methodologies for making and interpreting law – where one set of agents develops specifications for behavior, another set of agents interprets the specifications in novel circumstances, and then everyone iterates (Figure 2) – have been theoretically refined for centuries. Democracy has a theory – widely accepted and implemented already – for how to elicit credible human preferences and values, legitimately synthesize them, and consistently update the results to adapt over time with the evolving will of the people. Democratically developed law thus fulfils *requirement one* of our desired criteria, and legal informatics can be the bridge to instill legal reasoning – the language of alignment – within AI, helping close the currently widening "gap between social requirements and technical feasibility."[54]

As the state-of-the-art advances, we can set iteratively higher bars of demonstrated legal understanding capabilities. If a developer claims their AI has advanced capabilities on tasks, they should demonstrate correspondingly advanced legal comprehension and legal reasoning abilities of the AI.[55] Benchmarks are the guiding lights of AI research. There is no ceiling of difficulty when considering the morass of laws and regulations across time and jurisdiction. No current AI exhibits the general legal reasoning skills of expert human lawyers, and human experts do not represent the pinnacle of legal comprehension and reasoning abilities.[56] Legal understanding thus fulfils *requirement two* as an AI alignment benchmark.

---

[50] Of course, law does not embed all of the citizenry's moral views; therefore, a further integration of ethics and AI will be needed to guide AI systems where the law is silent (however, that itself is useful information) or prejudiced. *See infra* Section IV. But, for the reasons outlined throughout this Article, we believe legal informatics is most well suited to serve as the core framework for AI alignment.

[51] *See, e.g.*, Martin Abadi, Leslie Lamport & Pierre Wolper, *Realizable and Unrealizable Specifications of Reactive Systems*, 1989 AUTOMATA, LANGUAGES, & PROGRAMMING PROC. 1 (constraints to ensure the safety of systems can be mutually unsatisfiable); Hendrycks et al., *Aligning AI With Shared Human Values*, *supra* note 21.

[52] *See* Ian R. Macneil, *The Many Futures of Contracts*, 47 S. CAL. L. REV. 691, 731 (1974).

[53] These "standards" are not academic philosophical guidelines. Rather, they are *legal* standards that, at least theoretically, have an "objective" resolution (obtained from a court opinion). *See infra* Sections II.A., IV.

[54] Mark S. Ackerman, *The Intellectual Challenge of CSCW: The Gap Between Social Requirements and Technical Feasibility*, 15 HUM.–COMPUT. INTERACTION 179, 179 (2000).

[55] Or they should demonstrate correspondingly advanced legal comprehension and legal reasoning abilities of specialized Legal Informatics AI systems that are directly available for guiding the knowledge and actions of the primary AI.

The practices of making, interpreting, and enforcing law have been battle tested through millions of legal actions memorialized in digital format[57] that can be leveraged for machine learning (*requirement three*).[58] Part II of this Article expands on the satisfaction of our three requirements to demonstrate why the legal lens is so well-suited to increase AI alignment.

Parts III and IV explore the two primary ways that *Law Informs Code* (Figure 3). *First,* law provides theoretical constructs and praxis (methods of statutory interpretation, application of standards, and legal reasoning more broadly) to facilitate the robust specification of what a human wants an AI to proactively accomplish in the world (Part III). *Second,* public law *as data* helps AI parse what it should generally *not* do, providing an up-to-date distillation of democratically deliberated means of reducing externalities and pursuing societal coordination (Part IV). We conclude, in Part V, with drawbacks of our approach and with where further research could be most fruitful.

---

[56] But they can help evaluate advanced AI systems. I cannot do a backflip, but I can evaluate whether you just did one. Furthermore, "[o]ne solution is to have humans provide a training signal by demonstrating or judging performance, but this approach fails if the task is too complicated for a human to directly evaluate. We propose Iterated Amplification, an alternative training strategy which progressively builds up a training signal for difficult problems by combining solutions to easier subproblems." Paul Christiano, Buck Shlegeris & Dario Amodei, *Supervising Strong Learners by Amplifying Weak Experts*, ARXIV 1 (Oct. 19, 2018), https://arxiv.org/pdf/1810.08575.pdf [https://perma.cc/8RXM-GEVY]; *see also* Leike et al., *Scalable Agent Alignment via Reward Modeling: A Research Direction, supra* note 48; Christian, *supra* note 9, at 263–266.

[57] *See, e.g.*, Christine Bannan, *Legal Data Access, in* LEGAL INFORMATICS 467 (Daniel Martin Katz et al., eds. 2021).

[58] *See generally* Daniel Martin Katz & John Nay, *Machine Learning and Law, in* LEGAL INFORMATICS 94 (Daniel Martin Katz et al., eds. 2021).

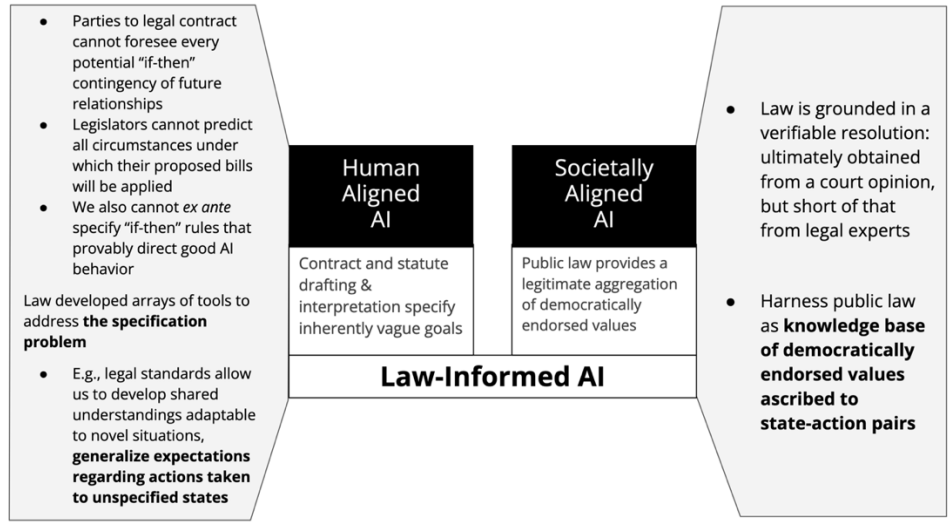

**Figure 3:** The two ways in which Law-Informed AI causes more aligned AI: human-AI alignment for more useful AI, and societally aligned AI for safer AI.

## II.   LEGAL INFORMATICS FOR AI ALIGNMENT

A legal (informatics) approach satisfies our three requirements for an alignment framework. Law is the applied philosophy and accepted practice of multi-agent alignment (Section *II. A*); legal informatics can calibrate AI *task capabilities* and AI *alignment capabilities* as technology advances to potentially transformative AI (*II. B*); and law produces data and (in the machine learning parlance) model "inductive biases" that can be leveraged to improve AI through goal communication mechanisms and rich background knowledge on how to act without undue externalities (*II. C*).

### A.   *Legal Theory is Well-Developed, and Applicable to Alignment*

The legal lens helps frame and clarify the alignment problem. Law is a unique discipline – it is both deeply theoretical[59] and tested against reality with an unrelenting cadence. Because producing, interpreting, enforcing, and amending law is a never-ending society-wide project,[60] the results are a prime source of information to scalably shape AI behavior.

---

[59] E.g., even busy practicing attorneys publish esoteric Law Review Articles.

[60] Law is also capable of reflecting the rights of future generations. *See, e.g.*, Eric Martínez & Christoph Winter, *Protecting Future Generations: A Global Survey of Legal Academics* (Legal Priorities Project, Working Paper No. 1, 2021), https://www.legalpriorities.org/documents/1%20-%20Protecting%20Future%20Generations.pdf [https://perma.cc/6H63-4N84].

### 1. *Law as Information*

We are not aiming for AI to have the legitimacy to make or enforce law. The most ambitious goal of *Law Informing Code* is to computationally encode and embed the generalizability of existing legal concepts and standards into AI. Setting new legal precedent (which, broadly defined, includes proposing and enacting legislation, promulgating agency rules, publishing judicial opinion, systematically enforcing law, and more) should be exclusively reserved for the democratic governmental systems expressing uniquely *human* values.[61] Humans should always be the engine of law-making.[62] The positive implications (for our approach) of this normative stance are that the resulting law encapsulates human views.

The law is a complex system[63] with seemingly chaotic underlying behavior from which aggregated and systematized preferences emerge.[64] Law, leveraged as an expression of *what* humans want,[65] and *how* they communicate their goals under ambiguity and radical uncertainty,[66] is how *Law Informs Code*. This stands in contrast to more prosaic uses of law, for example, as a deterrent of bad behavior through the threat of sanction[67] or imposition of institutional legitimacy,[68] or as an *ex-post* message of moral indignation.[69] *Law Informs Code* in the tradition of Oliver Holmes and subsequent "predictive" theories of law.[70]

---

[61] *See, e.g.*, FRANK PASQUALE, NEW LAWS OF ROBOTICS: DEFENDING HUMAN EXPERTISE IN THE AGE OF AI (2020); John Nay, *Large Language Models as Corporate Lobbyists* (arXiv, Working Paper No. 2301.01181, Jan. 30, 2023), https://arxiv.org/pdf/2301.01181.pdf [https://perma.cc/D2J2-TFCG].

[62] *See, e.g.*, Frank Pasquale, *A Rule of Persons, Not Machines: The Limits of Legal Automation*, 87 GEO. WASH. L. REV. 1 (2019), https://www.gwlr.org/wp-content/uploads/2019/01/87-Geo.-Wash.-L.-Rev.-1.pdf [https://perma.cc/2K9Y-6GUU]; Nay, *Large Language Models as Corporate Lobbyists*, *supra* note 61.

[63] For a discussion of complex systems science applied to A.I. safety, *see, e.g.*, Dan Hendrycks & Thomas Woodside, *Complex Systems for AI Safety*, A.I. ALIGNMENT FORUM (May 23, 2022), https://www.alignmentforum.org/posts/n767Q8HqbrteaPA25/complex-systems-for-ai-safety-pragmatic-ai-safety-3 [https://perma.cc/EM59-LJFF].

[64] On law as a complex emergent system, see, for example, Daniel M. Katz & Michael J. Bommarito, *Measuring the Complexity of the Law: The United States Code*, 22 A.I. & L. 337, 337–374 (2014), https://link.springer.com/article/10.1007/s10506-014-9160-8 [https://perma.cc/99WW-NAZV]; J.B. Ruhl & Daniel M. Katz, *Measuring, Monitoring, and Managing Legal Complexity*, 101 IOWA L. REV. 191 (2015), https://ilr.law.uiowa.edu/sites/ilr.law.uiowa.edu/files/2023-02/ILR-101-1-RuhlKatz.pdf [https://perma.cc/J8UA-JVLR]; Daniel M. Katz et al., *Complex Societies and the Growth of the Law*, SCI. REPS. (Oct. 30, 2020), https://www.nature.com/articles/s41598-020-73623-x [https://perma.cc/T72U-LM6W].

[65] RICHARD H. MCADAMS, THE EXPRESSIVE POWERS OF LAW 6–7 (2017) ("Law has expressive powers independent of the legal sanctions threatened on violators and independent of the legitimacy the population perceives in the authority creating and enforcing the law.") [hereinafter McAdams, *The Expressive Powers of Law*].

[66] On the notion of radical uncertainty, see JOHN KAY & MERVYN KING, RADICAL UNCERTAINTY: DECISION-MAKING BEYOND THE NUMBERS (2020); FRANK H. KNIGHT, RISK, UNCERTAINTY, AND PROFIT (1921).

[67] Oliver Wendell Holmes, Jr., *The Path of the Law*, 10 HARV. L. REV. 457 (1897); Ron Dolin, *Technology Issues in Legal Philosophy*, *in* LEGAL INFORMATICS 5 (Daniel Martin Katz et al. eds. 2021).

Empirical consequences of violating the law, using enforcement as a source of information,[71] are data points for AI. Enforcing law on AI (or their human developers) is how *Law Governs Code*[72] not how *Law Informs Code* and is out of scope here. What good is the law if it is not enforceable – isn't there "no right without a remedy"?[73] From the perspective of AI, the law can serve as a rich set of methodologies for interpreting inherently incomplete specifications of collective human expectations.[74]

Law provides detailed variegated examples of its application, generalizable precedents with explanations, and well-trained lawyers to solicit targeted model training and fine-tuning feedback to embed an ever-evolving comprehension of societal goals.[75] As a source to learn goal specification and interpretation[76] methods and (automatically updated and verified) societal knowledge, law provides an ontology for alignment.[77]

---

[68] Kenworthey Bilz & Janice Nadler, *Law, Psychology & Morality*, *in* 50 MORAL COGNITION AND DECISION MAKING: THE PSYCHOLOGY OF LEARNING AND MOTIVATION 101 (D. Medin, L. Skitka, C. W. Bauman, & D. Bartels, eds., 2009).

[69] *See* Mark A. Lemley & Bryan Casey, *Remedies for Robots*, 86 U. CHI. L. REV. 1347 (2019). *See also, e.g.*, YUVAL FELDMAN, THE LAW OF GOOD PEOPLE: CHALLENGING STATES' ABILITY TO REGULATE HUMAN BEHAVIOR (2018).

[70] *See* Oliver Wendell Holmes, Jr., *The Path of the Law*, 10 HARV. L. REV. 457 (1897); Catharine Pierce Wells, *Holmes on Legal Method: The Predictive Theory of Law as an Instance of Scientific Method*, 18 S. ILL. U. L.J. 329 (1993); Faraz Dadgostari et al., *Modeling Law Search as Prediction*, 29 A.I. & L. 3 (2021).

[71] McAdams, *The Expressive Powers of Law*, *supra* note 65, at 169–98.

[72] Using legal remedies to prevent illegal behavior is difficult with non-human agents. *See* Lemley & Casey, supra note 69, at 1315, 1316 ("Often, we want to compel defendants to do (or not do) something in order to prevent injury. Injunctions, punitive damages, and even remedies like disgorgement are all aimed—directly or indirectly—at modifying or deterring behavior. But deterring robot misbehavior is going to look very different than deterring humans. . . . Courts, for instance, can rely on the fact that most of us don't want to go to jail, so we tend to avoid conduct that might lead to that result. But robots will be deterred only to the extent that their algorithms are modified to include sanctions as part of the risk-reward calculus."); Ronald Leenes & Federica Lucivero, *Laws on Robots, Laws by Robots, Laws in Robots: Regulating Robot Behaviour by Design*, 6 L. INNOVATION & TECH. 193 (2014).

[73] Frederick Pollock, *The Continuity of the Common Law*, 11 HARV. L. REV. 423, 424 (1898).

[74] For more on law as an information source on public attitudes and risks, see Richard H. McAdams, *An Attitudinal Theory of Expressive Law*, 79 OR. L. REV. 339 (2000). For more on law as a coordinating mechanism, see Richard H. McAdams, *A Focal Point Theory of Expressive Law*, 86 VA. L. REV. 8 (2000).

[75] *See infra*, Section III.B.

[76] *See, e.g.*, Owen M. Fiss, *Objectivity and Interpretation*, 34 STAN. L. REV. 739 (1982).

[77] It seems plausible that super-human-intelligent AI could have a shared ontology with humans with respect to the communication of directives, goals, and values, while exhibiting super-human task capabilities to implement the humans' goals. This is analogous to a layperson sharing an ontology with

### 2. Examples of Theoretical Framing

We illustrate the applicability of legal theory with three examples.

### Complete vs. Incomplete Contracts

From the legal lens, one way of viewing the alignment of a human with an AI is the recognition that it is not possible to create a complete contingent "contract" between the AI and the human it serves because AI training and validation are not comprehensive of states of the world that may be encountered after deployment.[78] This highlights the need for AI to learn modular extra-contractual standards[79] and background knowledge that can generalize across much of the implicit space of potential "contracts."[80]

### Rules vs. Standards

The legal lens illuminates AI alignment with the voluminous legal theory concerning the distinction between rules and standards.[81] Rules (e.g., "do not drive more than 60 miles per hour") are more targeted directives than standards. If comprehensive enough for the complexity of their application,

---

a biologist and providing a directive such as "run a controlled experiment to determine X and minimize side-effects to a reasonable degree," and the biologist competently doing that in a safe way because she shares an ontology regarding key concepts like "reasonable" even though she has much more intelligence and skill than the layperson with respect to biology. *See, e.g.*, Pompeu Casanovas et al., *Semantic Web for the Legal Domain: The Next Step*, 7 SEMANTIC WEB 213 (2016); Arbital, *Ontology Identification Problem,* https://arbital.com/p/ontology_identification/ [https://perma.cc/9UZ5-B9N4];

[78] See Dylan Hadfield-Menell & Gillian K. Hadfield, *Incomplete Contracting and AI Alignment*, 2019 PROC. AAAI/ACM CONFERENCE ON AI, ETHICS, AND SOCIETY 417 for the contract-AI alignment analogy. Their "most important claim is that aligning robots with humans will inevitably require building the technical tools to allow AI to do what human agents do naturally: import into their assessment of rewards the costs associated with taking actions tagged as wrongful by human communities." *Id.* at 422. In contrast to Hadfield-Menell & Hadfield, who conclude that the primary need is to build "AI that can replicate human cognitive processes," *id.* at 417, we use the contract analogy as inspiration for a legal informatics approach that leverages legal tools, legal standards, and legal data from the real-world creation and performance of contracts.

[79] *See infra* Sections II.A.3, III.

[80] An Inverse Reinforcement Learning artificial "agent might not ever learn what is the best (or the morally or ethically appropriate) action in some regions of the state space. Without additional capabilities, it would be incapable of reasoning about what ought to be done in these regions–this is exactly the reason why we have norms in the first place: to not have to experience all state/actions precisely because some of them are considered forbidden and should not be experienced." Thomas Arnold et al., *Value Alignment or Misalignment - What Will Keep Systems Accountable?*, 2017 PROC. AAAI CONF. ON A.I. 5.

[81] *See, e.g.*, Duncan Kennedy, *Form and Substance in Private Law Adjudication*, 89 HARV. L. REV. 1685 (1976); Colin S. Diver, *The Optimal Precision of Administrative Rules*, 93 YALE L.J. 65 (1983); Pierre J. Schlag, *Rules and Standards*, 2 UCLA L. REV. 379 (1985); Kathleen M. Sullivan, *Foreword: The Justices of Rules and Standards,* 106 HARV. L. REV. 22 (1992); Cass R. Sunstein, *Problems with Rules*, 83 CALIF. L. REV. 953 (1995); Prasad Krishnamurthy, *Rules, Standards, and Complexity in Capital Regulation*, 43 J. LEGAL STUD. S273 (2014); Michael Coenen, *Rules Against Rulification*, 124 YALE L.J. 576 (2014); Anthony J. Casey & Anthony Niblett, *Death of Rules and Standards*, 92 IND. L.J. 1401 (2017); Brian Sheppard, *The Reasonableness Machine*, 62 B.C. L. REV. 2259 (2021) [hereinafter Sheppard, *Reasonableness*].

rules allow the rule-maker to have more clarity than standards over the outcomes that will be realized conditional on the specified states (and agents' actions in those states, which are a function of any impact the rules might have had).[82] Complex social systems have emergent behavior that makes formal rules brittle.[83]

On the other hand, standards (e.g., drive "reasonably" for California highways) allow contract parties, judges, regulators, and citizens to develop shared understandings and adapt them to novel situations, i.e., to generalize expectations regarding actions to unspecified states of the world. If rules are not written with enough potential states of the world in mind, they lead to unanticipated undesirable outcomes[84] (e.g., a driver following the rule above is too slow to bring their passenger to the hospital in time to save their life). But to enumerate all the potentially relevant state-action pairs is excessively costly outside of the simplest environments.[85] A standard has more capacity to generalize to novel situations than a rule.[86] The AI analogy for a standard is a continuous, approximate method that relies on significant amounts of data for learning dense representations on which we can apply geometric operations in the latent model space. They are flexible.[87] The AI analogy for a rule[88] is a discrete human-crafted "if-then" statement that is brittle yet requires no empirical data for machine learning.[89]

In practice, most legal provisions land somewhere on a spectrum between pure rule and pure standard,[90] and legal theory can help estimate the right combination[91] of "rule-ness" and "standard-ness" when specifying objectives of AI. Furthermore, there are other theorized dimensions to legal provision implementation related to the rule-ness versus standard-ness axis that could elucidate AI goal design, e.g., "determinacy," [92] "privately adaptable" ("rules that allocate initial entitlements but do not specify end-states" [93] ), and "catalogs" ("a legal command comprising a specific enumeration of behaviors, prohibitions, or items that share a salient common denominator and a residual category—often denoted by the words 'and the like' or 'such as'"[94]).

### Private vs. Public Law

The AI alignment problem is usually described with respect to the alignment of one AI with one human, or a small subset of humans.[95] It is more challenging to expand the scope of the analysis beyond a small set of humans and ascribe *societal value* to state-action pairs. Even if we fully align an AI with the goals of a human, what about all the other humans? Legal framing highlights differences between addressing *human-AI* alignment and *society-AI* alignment.[96] The latter requires us to move beyond private contracts and into the realm of public law[97] to explicitly address inter-agent conflicts and public policy designed to ameliorate externalities and solve massively multi-agent coordination and cooperation dilemmas.[98]

---

[88] Foundation models have recently been found to have varying "rule-ness" to their different modes of learning and operation. *See* Stephanie C.Y. Chan et al., *Transformers Generalize Differently from Information Stored In Context vs In Weights*, ARXIV 1 (Oct. 13, 2022), https://arxiv.org/pdf/2210.05675.pdf [https://perma.cc/XKB7-33ZK] ("[W]e find that generalization from weights is more rule-based whereas generalization from context is largely exemplar-based. In contrast, we find that in transformers pre-trained on natural language, in-context learning is significantly rule-based, with larger models showing more rule-basedness.").

[89] Harry Surden, *The Variable Determinacy Thesis*, 12 COLUM. SCI. & TECH. L. REV. 1 (2011) [hereinafter, Surden, *Variable Determinacy*].

[90] *See, e.g.*, Frederick Schauer, *The Tyranny of Choice and the Rulification of Standards*, 14 J. CONTEMP. LEGAL ISSUES 803 (2005); Richard L. Heppner, Jr., *Conceptualizing Appealability: Resisting the Supreme Court's Categorical Imperative*, 55 TULSA L. REV. 395 (2020); Sheppard, *Reasonableness*, *supra* note 81.

[91] *See* Katherine J. Strandburg, *Rulemaking and Inscrutable Automated Decision Tools*, 119 COLUM. L. REV. 1851, 1859 (2019) ("Decision criteria may also combine rule-like and standard-like aspects according to various schemes. For example, DWI laws in many states combine a rule-like blood alcohol threshold, above which a finding of intoxication is required, with a standard-like evaluation of intoxication at lower levels. Some speed limit laws use a somewhat different scheme: Above a rule-like speed limit, there is a presumption of unsafe driving, but adjudicators may make standard-like exceptions for a narrow range of emergency circumstances.").

[92] Surden, *supra* note 89.

[93] Sunstein, *supra* note 81, at 959.

[94] Parchomovsky & Stein, *supra* note 85, at 165.

[95] *See, e.g.*, Amanda Askell et al., *A General Language Assistant as a Laboratory for Alignment*, ARXIV 44 (Dec. 9, 2021), https://arxiv.org/pdf/2112.00861.pdf [https://perma.cc/QH53-WWXR] ("At a very high level, alignment can be thought of as the degree of overlap between the way two agents rank different outcomes. For example, if agent A completely internalizes the desires of agent B — i.e. the only desire A has is to see B's desires satisfied—we could say that agent A is maximally aligned with agent B."); Stiennon et al., *Learning to Summarize with Human Feedback* (Neural Information Processing Systems, Conference Paper, Nov. 28, 2022), https://arxiv.org/pdf/2009.01325.pdf [https://perma.cc/2B2M-9WNK]. For a high-level overview of AI alignment research, see generally Jan

### *B.  Legal Informatics Can Scale with AI Capabilities*

Within the *Law Informs Code* framework, we refer to the ability of AI to perform narrow tasks for humans as "AI-contract" capability (Figure 4). This level of capability has been widely deployed for years, e.g., through powering billions of automated online advertisement placements[99] and social media content choices every day.[100] Large neural-network-based models pre-

---

trained with self-supervision[101] on significant portions of the internet require little to no supervised learning to perform well on some new tasks (known as "Foundation Models"[102]) are beginning to display what we call "AI-standards" capabilities, which could be used to help align AI with humans through legal informatics.[103] Standards are more abstract and nuanced than rules, and require more generalizable capabilities and world-knowledge to implement. The next level –somewhere between current Foundation Model capabilities and a potential "Artificial General Intelligence" [104] (AGI) capability level, *paired with* the additional development of methods and data specific to legal understanding[105] – may unlock what we can call an "AI-public" capability (Figure 4). At this level, AI will be able to understand standards, interpretation guidelines, legal reasoning, and generalizable precedents across all public law (which synthesize citizens' value preferences over potential actions taken in many states of the world).[106] We are not there yet.

---

[101] "Self-supervised" training procedures predict data items that are systematically held-out, for example, removing a word and predicting the word that was removed (and then iterating this billions of times), or predicting whether an entire sentence is next to another sentence. No explicit labeling of the data is required; therefore, self-supervised training allows a model to be trained across significantly more data than traditional supervised learning. Scalable self-supervised pruning of the training data can reduce the costs of training. *See, e.g.,* Ben Sorscher et al., *Beyond Neural Scaling Laws: Beating Power Law Scaling via Data Pruning*, ARXIV (Nov. 15, 2022), https://arxiv.org/pdf/2206.14486.pdf [https://perma.cc/NH7D-GAH2].

[102] *See, e.g.,* Rishi Bommasani et al., *On the Opportunities and Risks of Foundation Models*, ARXIV (July 12, 2022), https://arxiv.org/pdf/2108.07258.pdf [https://perma.cc/UP24-E42W]. *See also* Ashish Vaswani et al., *Attention Is All You Need, in Proceedings of the 31st Conference on Neural Information Processing Systems* (Neural Information Processing Systems, Conference Paper, Dec. 6, 2017), https://arxiv.org/pdf/1706.03762.pdf [https://perma.cc./GUC7-ZGA7] (foundation models leverage applications of the Transformer architecture which is a model used to encode an input sequence, words in a particular order, into context-aware representation and then decode that into a novel generation of an ordered sequence, a new set of words in a particular order as an output); LEWIS TUNSTALL ET AL., NATURAL LANGUAGE PROCESSING WITH TRANSFORMERS (1st ed. 2022) (sequences of words were the first application area of this model architecture with major success, as these models can capture complicated dependencies and interactions within natural language). Transformers are very expressive in the forward pass of their information and very efficient in the backward pass when they are being trained. The Transformer has since been applied beyond natural language, also to graphs, *see, e.g.,* Jinwoo Kim et al., *Pure Transformers are Powerful Graph Learners* (Neural Information Processing Systems, Conference Paper, Nov. 28, 2022), https://arxiv.org/pdf/2207.02505.pdf [https://perma.cc/MPB6-TJNE], and decision-making, *see, e.g.,* Micah Carroll et al., *Towards Flexible Inference in Sequential Decision Problems via Bidirectional Transformers* (International Conference on Learning Representations, Conference Paper, Dec. 9, 2022), https://arxiv.org/pdf/2204.13326.pdf [https://perma.cc/NG6S-VGZV]. Even within the natural language data structure, Transformer models demonstrate non-language skills, such as mathematical reasoning skills, *see, e.g.,* Aitor Lewkowycz et al., *Solving Quantitative Reasoning Problems with Language Models*, ARXIV (July 1, 2022), https://arxiv.org/pdf/2206.14858.pdf [https://perma.cc/85HE-KXS8]; Tuan Dinh et al., *LIFT: Language-Interfaced Fine-Tuning for Non-Language Machine Learning Tasks* (Neural Information Processing Systems, Conference Paper, Nov. 28, 2022), https://arxiv.org/pdf/2206.06565.pdf [https://perma.cc/3XL8-FFL4], and the ability to simulate social systems, *see, e.g.,* Joon Sung Park et al., *Social Simulacra: Creating Populated Prototypes for Social Computing Systems* (Symposium on User Interface Software and Technology, Conference Paper, Aug. 8, 2022), https://arxiv.org/pdf/2208.04024.pdf [https://perma.cc/64GR-9Z24]; Lisa P. Argyle et al., *Out of One, Many: Using Language Models to Simulate Human Samples*, ARXIV (Sep. 14, 2022), https://arxiv.org/pdf/2209.06899.pdf [https://perma.cc/F7KW-ZE4G].

[103] AI most capable of performing well on diverse sets of tasks – the most generalizable – are large neural network-based models trained on diverse data sets through self-supervision. *See, e.g.,* Romal Thoppilan et al., *LaMDA: Language Models for Dialog Applications*, ARXIV (Feb. 10, 2022), https://arxiv.org/pdf/2201.08239.pdf [https://perma.cc/K53R-EPLY]; *see also* Tran et al., *Plex: Towards Reliability Using Pretrained Large Model Extensions*, ARXIV (July 15, 2022), https://arxiv.org/pdf/2207.07411.pdf [https://perma.cc/U3ST-Z3BP].

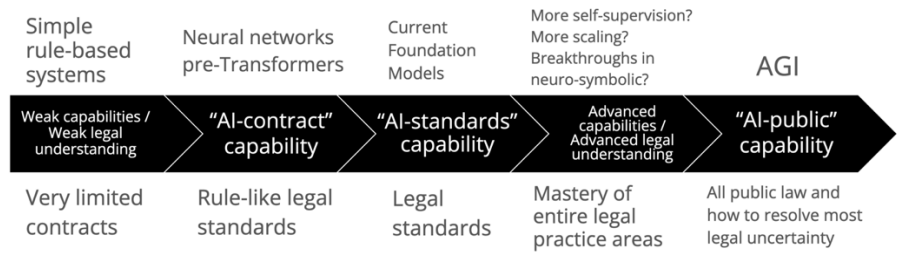

**Figure 4:** Legal informatics scaling with AI capabilities. AI capability levels are above the arrow, and legal understanding capabilities are below the arrow. There is a relatively unaddressed AI safety territory in between the two extremes. The pre-Transformer AI alignment research was focused on more clearly decision and agentic oriented training processes, whereas the primary deployed AI today is a result of self-supervised foundation models that exhibit less goal-orientation than pure reinforcement learning approaches. The Law Informs Code alignment framework that scales across levels of intelligence may be able to better address this. As models scale their capabilities, they may automatically become more goal-oriented and power-seeking; this remains to be seen.

---

[104]  OpenAI, *OpenAI Charter* (Apr. 9, 2018), https://openai.com/charter/ [https://perma.cc./HM6F-AT9F] (defining AGI as a "highly autonomous systems that outperform humans at most economically valuable work"). *See generally*, Everitt, Lea & Hutter, *supra* note 44; Ngo, *supra* note 26.

[105]  *See generally, infra* Section II.B.2.

[106]  There are no distinct boundaries between the AI-public, AI-standards, and AI-contract levels. We are simply using them as high-level rhetorical devices.

### 1.  AI Capabilities

Unlike traditional supervised learning, [107] self-supervision and reinforcement learning can produce "emergent" capabilities. [108] Advances in deep reinforcement learning are producing more generalizable decision-making agents, [109] which lead to more pressing alignment issues due to the possibility of more autonomous deployments. For both pure prediction tasks (e.g., typical supervised learning inference) and decision tasks (e.g., deployment of agents trained through reinforcement learning), the process of first conducting self-supervised training on a large scale has increased performance on downstream tasks, even with little exposure to those tasks – for "few-shot reasoning."[110] With latent concepts learned through pre-training, models can exhibit robust capabilities on novel tasks without any fine-tuning or explicit exposure to that particular task[111] – through manually engineered prompting[112] or automated algorithmic prompting[113] – for "zero-shot reasoning."[114] Foundation Models excel in natural language processing, but they are also being successfully applied beyond text data.[115] Large models are being released as open source, further democratizing their use and distributing their impact.[116]

---

[107]  With supervised learning, the best one can do is to predict, with perfect calibration, the same type of output variable with the same type of input variables on new test data not used for training the model. A model that can generalize to novel tasks is out of scope of the traditional supervised learning framework, which is concerned only with mapping *ex ante*-specified input types to *ex ante*-specified output types, albeit on unseen input data points. A complication to this characterization of the distinction in learning paradigms, though, is when supervised learning is trained on state-action-reward sequences and deployed to generate actions in new states by conditioning on high rewards. *See, e.g.,* Dylan R Ashley et al., *Learning Relative Return Policies With Upside-Down Reinforcement Learning*, ARXIV (May 10, 2022), https://arxiv.org/pdf/2202.12742.pdf [https://perma.cc/V43L-JV9C]; *see also* Kai Arulkumaran et al., *All You Need Is Supervised Learning: From Imitation Learning to Meta-RL With Upside Down RL*, ARXIV (Feb. 24, 2022), https://arxiv.org/pdf/2202.11960.pdf [https://perma.cc/G4TU-C95E].

[108]  *See, e.g.*, Jason Wei et al., *Emergent Abilities of Large Language Models*, TRANSACTIONS ON MACH. LEARNING RSCH. (Oct. 26, 2022), https://openreview.net/pdf?id=yzkSU5zdwD [https://perma.cc/5C3C-U397]; Jason Wei, *137 Emergent Abilities of Large Language Models* (Nov. 14, 2022), https://www.jasonwei.net/blog/emergence [https://perma.cc/2PHP-WPAM]; Ganguli, *supra* note 46, at 2 ("Our basic thesis is that large generative models have an unusual combination of high predictability — model loss improves in relation to resources expended on training, and tends to correlate loosely with improved performance on many tasks — and high unpredictability — specific model capabilities, inputs, and outputs can't be predicted ahead of time. The former drives rapid development of such models while the latter makes it difficult to anticipate the consequences of their development and deployment."); Wenlong Huang, *supra* note 46, at 8 ("Although LLMs can generate fluent continuation from the prompted examples, we surprisingly find that, when informed with environment feedback, Inner Monologue demonstrates many impressive reasoning and replanning behaviors beyond the examples given in the prompt. Using a pre-trained LLM as the backbone, the method also inherits many of the appealing properties from its versatility and general-purpose language understanding."); Mathilde Caron et al., *Emerging Properties in Self-supervised Vision Transformers* (International Conference on Computer Vision, Conference Paper, May 24, 2021); Bowen Baker et al., *Emergent Tool Use from Multi-Agent Autocurricula*, ARXIV (International Conference on Learning Representations, Conference Paper, Feb. 11, 2020), https://arxiv.org/pdf/1909.07528.pdf [https://perma.cc/F72N-JL5P]; Taylor Webb, Keith J. Holyoak & Hongjing Lu, *Emergent Analogical Reasoning in Large Language Models*, ARXIV (Dec. 19, 2022), https://arxiv.org/pdf/2212.09196.pdf [https://perma.cc/9XT6-5QWM].

[109]  *See e.g.*, Bhoopchand et al., *Learning Robust Real-Time Cultural Transmission without Human Data*, ARXIV 2 (Mar. 1, 2022), https://arxiv.org/pdf/2203.00715.pdf [https://perma.cc/8L2N-PX7N] ("Our artificial agent is parameterised by a neural network and we use deep reinforcement learning (RL) to train the weights. The resulting neural network (with fixed weights) is capable of zero-shot, high-recall cultural transmission within a 'test' episode on a wide range of previously unseen tasks.").

---

[110] *See, e.g.*, Yaqing Wang et al., *Generalizing From a Few Examples: A Survey on Few-shot Learning*, ARXIV (May 10, 2020), https://arxiv.org/pdf/1904.05046.pdf [https://perma.cc/QGS8-G5GF]; Zhao Mandi, Pieter Abbeel & Stephen James, *On the Effectiveness of Fine-tuning Versus Meta-reinforcement Learning*, ARXIV (Feb. 16, 2023), https://arxiv.org/pdf/2206.03271.pdf [https://perma.cc/HNZ5-6R23]; Zhuyun Dai et al., *Promptagator: Few-shot Dense Retrieval From 8 Examples*, ARXIV (Sep. 23, 2022), https://arxiv.org/pdf/2209.11755.pdf [https://perma.cc/Q9AT-TG3W] (The paper "showed that it is possible to create task-specific, end-to-end retrievers with only a few annotated examples. The few-shot examples, amplified by prompt-based LLM query generation, simplifies the complexity of training neural retrievers for new tasks and leads to promising retrieval performance gains."); Simran Arora et al., *Ask Me Anything: A Simple Strategy for Prompting Language Models*, ARXIV (Nov. 20, 2022), https://arxiv.org/pdf/2210.02441.pdf [https://perma.cc/TF2K-Q6WU]; Jesse Michael Han et al., *Proof Artifact Co-training for Theorem Proving with Language Models*, ARXIV (ICLR 2021, Conference Paper, Mar. 16, 2022), https://arxiv.org/pdf/2102.06203.pdf [https://perma.cc/85E9-J4GM]; Dinglan Peng et al., *How Could Neural Networks Understand Programs?*, ARXIV (May 31, 2021), https://arxiv.org/pdf/2105.04297.pdf [https://perma.cc/9ZKY-4GU8] (self-supervised training helps with understanding computer programming).

[111] For potential explanations of how this surprising behavior is possible, see, for example, Sang Michael Xie et al., *An Explanation of In-context Learning as Implicit Bayesian Inference*, ARXIV (July 21, 2021), https://arxiv.org/pdf/2111.02080.pdf [https://perma.cc/4KER-NNM7]; Sewon Min et al., *Rethinking the Role of Demonstrations: What Makes In-Context Learning Work?*, ARXIV (Oct. 20, 2022), https://arxiv.org/pdf/2202.12837.pdf [https://perma.cc/PG26-WGW5]; Sang Michael Xie & Sewon Min, *How Does In-context Learning Work? A Framework for Understanding the Differences From Traditional Supervised Learning*, STAN. AI LAB BLOG (Aug. 1, 2022), http://ai.stanford.edu/blog/understanding-incontext/ [https://perma.cc/DUN2-F9VY]; David Dohan et al., *Language Model Cascades*, ARXIV (July 28, 2022), https://arxiv.org/pdf/2207.10342.pdf [https://perma.cc/BC7Z-UZ2E]. In-context learning, or "prompting," *see infra* note 115, is an example of surprising behavior. *See* Bommasani et al., *supra* note 102, at 116 ("Similarly, small rewordings of prompts can have large impacts on task performance. Since the space of prompts is intractable to enumerate, it is challenging to definitely assert that any task is outside the reach of current prompt-based foundation models — this is a major challenge for reasoning about possible catastrophic risks from foundation models.").

[112] *See, e.g.*, Jason Wei et al., *Chain of Thought Prompting Elicits Reasoning in Large Language Models*, ARXIV (Jan. 10, 2023), https://arxiv.org/pdf/2201.11903.pdf [https://perma.cc/3HS7-V4VZ]; Adi Robertson, *Professional AI Whisperers have Launched a Marketplace for DALL-E Prompts: AI Art isn't Just an Experiment — it's a Side Hustle* (Sept. 2, 2022), https://www.theverge.com/2022/9/2/23326868/dalle-midjourney-ai-promptbase-prompt-market-sales-artist-interview [https://perma.cc/XY6B-3L3Q] (describing a marketplace for "prompt engineers").

[113] *See, e.g.*, Antonia Creswell & Murray Shanahan, *Faithful Reasoning Using Large Language Models*, ARXIV (Aug. 30, 2022), https://arxiv.org/pdf/2208.14271.pdf [https://perma.cc/DBP4-R43Y] ("Our approach exemplifies a trend towards algorithmic prompting, a form of automated prompt engineering in which querying a language model becomes a computational primitive. The responses of the language model can be manipulated to construct new prompts that are then used to make further queries. Model queries and prompt construction are composed into algorithms with the usual computational constructs: sequence, choice, and iteration. Algorithmic prompting can be used to elicit more sophisticated and nuanced behaviour from a language model than would otherwise be possible. For example, as our work shows, this approach can be used to develop models capable of faithful

---

reasoning, without compromising performance."); Zhuosheng Zhang et al., *Automatic Chain of Thought Prompting in Large Language Models*, ARXIV (Oct. 7, 2022), https://arxiv.org/pdf/2210.03493.pdf [https://perma.cc/QV7Z-8TWF].

[114] *See, e.g.*, Takeshi Kojima et al., *Large Language Models are Zero-Shot Reasoners*, ARXIV (Jan. 29, 2023), https://arxiv.org/pdf/2205.11916.pdf [https://perma.cc/4F3Q-U4SK] ("[W]e show that LLMs are decent zero-shot reasoners by simply adding "Let's think step by step" before each answer [ . . . ] The versatility of this single prompt across very diverse reasoning tasks hints at untapped and understudied fundamental zero-shot capabilities of LLMs, suggesting high-level, multi-task broad cognitive capabilities may be extracted by simple prompting."); Victor Sanh et al., *Multitask Prompted Training Enables Zero-Shot Task Generalization*, ARXIV (Mar. 17, 2022), https://arxiv.org/pdf/2110.08207.pdf [https://perma.cc/JN3T-A5TM]; Kaj Bostrom et al., *Natural Language Deduction Through Search Over Statement Compositions*, ARXIV (Oct. 26, 2022), https://arxiv.org/pdf/2201.06028.pdf [https://perma.cc/C6CF-NXNZ]; Eric Zelikman et al., *Star: Bootstrapping Reasoning with Reasoning*, ARXIV (May 20, 2022), https://arxiv.org/pdf/2203.14465.pdf [https://perma.cc/J9FJ-FVAA] ("STaR lets a model improve itself by learning from its own generated reasoning."); Ofir Press et al., *Measuring and Narrowing the Compositionality Gap in Language Models*, ARXIV (Oct. 7, 2022), https://arxiv.org/pdf/2210.03350.pdf [https://perma.cc/9VGY-GNVK]; Eric Jang et al., *BC-Z: Zero-Shot Task Generalization with Robotic Imitation Learning*, ARXIV (Feb. 4, 2022), https://arxiv.org/pdf/2202.02005.pdf [https://perma.cc/7K3P-WSWF]; Tanya Goyal, Junyi Jessy Li & Greg Durrett, *News Summarization and Evaluation in the Era of GPT-3*, ARXIV (Sep. 26, 2022), https://arxiv.org/pdf/2209.12356.pdf [https://perma.cc/46Y4-QPSX].

[115] *See, e.g.*, Kevin Lu et al., *Pretrained Transformers as Universal Computation Engines*, ARXIV (Jun. 30, 2021), https://arxiv.org/pdf/2103.05247.pdf [https://perma.cc/N8TU-UQ2A] (finding that "language-pretrained transformers can obtain strong performance on a variety of non-language tasks."); Lili Chen et al., *Decision Transformer: Reinforcement Learning via Sequence Modeling*, ARXIV (Jun. 24, 2021), https://arxiv.org/pdf/2106.01345.pdf [https://perma.cc/LG6J-289E] (abstracting reinforcement learning as a sequence modeling problem like language modeling and then leverage a Transformer architecture to condition on the desired reward, past states, and actions, to generate future actions); Kuang-Huei Lee et al., *Multi-Game Decision Transformers*, ARXIV (Oct. 15, 2022), https://arxiv.org/pdf/2205.15241.pdf [https://perma.cc/C86Y-VC7L]; Mengdi Xu et al., *Prompting Decision Transformer for Few-Shot Policy Generalization*, ARXIV (Jun. 27, 2022), https://arxiv.org/pdf/2206.13499.pdf [https://perma.cc/8X23-5433]; Micah Carroll et al., *Towards Flexible Inference in Sequential Decision Problems via Bidirectional Transformers*, ARXIV (Dec. 9, 2022), https://arxiv.org/pdf/2204.13326.pdf [https://perma.cc/ZR68-MP86] (developing and testing "a framework for flexibly defining and training models which: 1) are naturally able to represent any inference task and support multi-task training in sequential decision problems, 2) match or surpass the performance of specialized models after multi-task pre-training, and almost always surpasses them after fine-tuning"); Victor Sanh et al., *Multitask Prompted Training Enables Zero-Shot Task Generalization* ARXIV (Mar. 17, 2022), https://arxiv.org/pdf/2110.08207.pdf [https://perma.cc/9PYB-KZAU] (developing and testing a system for converting any natural language tasks into a human-readable prompt form.).

[116] *See, e.g.*, Melissa Heikkilä, *Inside the Radical New Project to Democratize AI*, MIT TECH. REV. (July 12, 2022), https://www.technologyreview.com/2022/07/12/1055817/inside-a-radical-new-project-to-democratize-ai/ [https://perma.cc/8YWR-W5NJ] (a large language model matching in scale some of the best performing large models from OpenAI, Google, and others was trained and released open source in July 2022 by over 1,000 volunteers); enijkamp, *CodeGen*, GITHUB (Dec. 1, 2022), https://github.com/salesforce/CodeGen [https://perma.cc/LW3S-HWNG] ("CodeGen is an open-source

Increased task capabilities could allow us to better align AI with humans and with society more broadly, if the capabilities are harnessed properly.[117] If AGI is attained, according to one of the more pessimistic (and most prominent) alignment theorists, Eliezer Yudkowsky, "[s]*eemingly 'simple' proposals [for ensuring super-intelligent systems realize a positive outcome for humans] are likely to have unexpected undesirable consequences, overlooked as possibilities because our implicit background preferences operate invisibly to constrain which solutions we generate for consideration.[ . . . ] There is little prospect of an outcome that realizes even the value of being interesting, unless the first superintelligences undergo detailed inheritance from human values*."[118] Our contention is that inheriting an understanding of legal reasoning, legal interpretation methods, legal standards, and public law could provide future advanced AI a more comprehensive view of what humans want. Although we are, debatably, not yet near AGI, addressing safety risks earlier in the deployment lifecycle of powerful technologies likely leads to better outcomes.[119] Therefore, we

---

model for program synthesis [that is] competitive with OpenAI Codex."); rromb, *Stable Diffusion*, GitHub (Nov. 16, 2022), https://github.com/CompVis/stable-diffusion [https://perma.cc/L4TY-VRNM] ("Stable Diffusion is a latent text-to-image diffusion model. [ . . . ] Similar to Google's Imagen, this model uses a frozen CLIP ViT-L/14 text encoder to condition the model on text prompts [ . . . ] the model is relatively lightweight and runs on a GPU with at least 10GB VRAM."); *T5*, Hugging Face, https://huggingface.co/docs/transformers/model_doc/t5 [https://perma.cc/RW9F-QSWZ] ("T5 is an encoder-decoder model pre-trained on a multi-task mixture of unsupervised and supervised tasks and for which each task is converted to a text-to-text format."); Thomas I. Liao, *Model Tracker v0.9*, Found. Model Tracker https://foundationmodeltracker.notion.site/foundationmodeltracker/Model-Tracker-v0-9-794ba77f74ec469186efdbdb87e9b8e6 [https://perma.cc/AD87-X8KD].

[117] *See* Dan Hendrycks & Mantas Mazeika, *X-Risk Analysis for AI Research*, arXiv 8 (Sep. 20, 2022), https://arxiv.org/pdf/2206.05862.pdf [https://perma.cc/NQT5-RV6N] ("Improving an agent's world model makes them more generally capable, but this also can make them less likely to spawn unintended consequences. Optimizers operating over longer time horizons will be able to accomplish more difficult goals, but this could also make models act more prudently and avoid taking irreversible actions.").

[118] Eliezer Yudkowsky, *Complex Value Systems are Required to Realize Valuable Futures*, Mach. Intel. Rsch. Inst. (2011), https://intelligence.org/files/ComplexValues.pdf [https://perma.cc/HW27-WY3S].

[119] *See, e.g.*, Hendrycks, *X-Risk Analysis for AI Research*, *supra* note 117, at 6 ("Many early Internet protocols were not designed with safety and security in mind. Since safety and security features were not built in early, the Internet remains far less secure than it could have been, and we continue to pay large continuing costs as a consequence."); Dan Hendrycks et al., *Unsolved Problems in ML Safety*, arXiv (Jun. 15, 2022), https://arxiv.org/pdf/2109.13916.pdf [https://perma.cc/PK7H-DZCP] ("If attention to safety is delayed, its impact is limited, as unsafe design choices become deeply embedded into the system. [ . . . ] relying on experts to test safety solutions is not enough—solutions must also be age tested. The test of time is needed even in the most rigorous of disciplines. A century before the four color theorem was proved, Kempe's peer-reviewed proof went unchallenged for years until, finally, a flaw was uncovered. Beginning the research process early allows for more prudent design and more rigorous testing." (Citations omitted.)). *See generally* Nancy Leveson, Engineering a Safer World: Systems Thinking Applied to Safety (2012).

propose a legal understanding verification process that *scales with* AI task capabilities, calibrated to the difficulty level of the legal reasoning and interpretation tasks from "AI-contract," to "AI-standards," to "AI-public" capability.

### 2. Legal Understanding Demonstrations

In most existing applications, before AI models are deployed, their performance on the task at hand is validated on data not employed for their training to demonstrate generalizability (task performance in circumstances sufficiently different than training data that is commensurate with task performance on training data). This out-of-sample performance evaluation is as a demonstration of a capability related to a specific human's (or organization's) preferences. In our framework, this is a demonstration of the AI's "understanding" of the terms of an (implied) "contract" between the AI and the human(s) it is conducting tasks for (Figure 5).

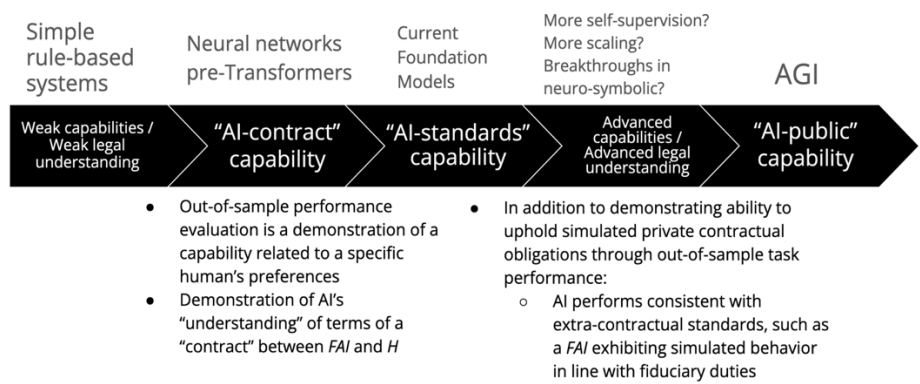

**Figure 5:** "AI-contract" level and "AI-standards" level alignment benchmarking.

If AI is only at an "AI-contract" capability level, and deployed only on a narrow task, we have no expectation of it being able to autonomously track and comply with public law. This is the status quo for most deployments. It does not mean that the AI will necessarily violate the law; rather, it's an admission that the AI is not advanced enough to understand the legal context in any meaningful sense.[120]

---

[120] *See, e.g.*, John Nay & James Daily, *Aligning Artificial Intelligence with Humans through Public Policy*, ARXIV (May 4, 2022), https://arxiv.org/ftp/arxiv/papers/2207/2207.01497.pdf [https://perma.cc/KH27-8NG6].

AI general task capabilities are advancing faster than AI safety because there is vastly more effort on the former.[121] To help close this gap,[122] before AI models are deployed in increasingly agentic capacities, e.g., fully autonomous vehicles on most major roads,[123] the deploying party should demonstrate[124] the AI's understanding of human goals, policies, and legal standards.[125] A validation procedure could illustrate the AI's "understanding" of the "meaning" of legal concepts.[126]

---

[121] *See, e.g.*, Dan Hendrycks & Thomas Woodside, *Introduction to Pragmatic AI Safety*, AI ALIGNMENT FORUM (May 9, 2022), https://www.alignmentforum.org/posts/bffA9WC9nEJhtagQi/introduction-to-pragmatic-ai-safety-pragmatic-ai-safety-1 [https://perma.cc/3JWH-W3UA] ("Machine learning has been outpacing safety [ . . . ] Meanwhile, existing approaches to AI safety have not seen similar strides. Many older approaches are still pre-paradigmatic, uncertain about what concrete research directions should be pursued and still aiming to get their bearings. Centered on math and theory, this research focuses on studying strictly futuristic risks that result from potential systems. Unfortunately, not much progress has been made."). This general technological governance issue is often framed as the "pacing problem." *See, e.g.*, Gary E. Marchant, *Governance of Emerging Technologies as a Wicked Problem*, 73 VAND. L. REV. 1861 (2020); Adam Thierer, *The Pacing Problem and the Future of Technology Regulation*, MERCATUS CENTER (Aug. 8, 2018), https://www.mercatus.org/economic-insights/expert-commentary/pacing-problem-and-future-technology-regulation [https://perma.cc/ZVM9-TG2J].

[122] For other recommendations of potential policy responses, *see, e.g.*, Ryan Calo, *Artificial Intelligence Policy: A Primer and Roadmap*, 51 U. CAL. DAVIS L. REV. 399 (2017); Jessica Fjeld et al., *Principled Artificial Intelligence: Mapping Consensus in Ethical and Rights-Based Approaches to Principles for AI*, BERKMAN KLEIN CTR. 34 (Rsch. Pub. No. 2020-1, Jan. 15, 2022), https://dash.harvard.edu/bitstream/handle/1/42160420/HLS%20White%20Paper%20Final_v3.pdf?sequence=1&isAllowed=y [https://perma.cc/N9Z7-YDMW]. For a dashboard of international AI policies, see the OECD AI's live repository of over 260 AI policies, *National AI policies & strategies*, OECD.AI, https://oecd.ai/en/dashboards/overview [https://perma.cc/Y8R5-VZJ4]. For specific recent federal U.S. government proposals, see, for example, Trade Regulation Rule on Commercial Surveillance, 87 Fed. Reg. 51273 (published Aug. 22, 2022) (Federal Trade Commission proposed rule on commercial surveillance); *EEOC Launches Initiative on Artificial Intelligence and Algorithmic Fairness*, US Equal Employment Opportunity Commission, https://www.eeoc.gov/newsroom/eeoc-launches-initiative-artificial-intelligence-and-algorithmic-fairness [https://perma.cc/2FFN-V99M]; *Agencies Seek Wide Range of Views on Financial Institutions' Use of Artificial Intelligence*, CONSUMER FIN. PROT. BUREAU (Mar. 29, 2021), https://www.consumerfinance.gov/about-us/newsroom/agencies-seek-wide-range-of-views-on-financial-institutions-use-of-artificial-intelligence/ [https://perma.cc/U8GF-CUSB]; Artificial Intelligence Risk Management Framework, 86 Fed. Reg. 40810 (published July 29, 2021) (National Institute of Standards and Technology proposed rule for AI Risk Management Framework).

[123] *Autonomous Vehicles | Self-Driving Vehicles Enacted Legislation*, NAT'L CONF. STATE LEGS. (Feb. 18, 2020), https://www.ncsl.org/transportation/autonomous-vehicles [https://perma.cc/9SMS-Q27F].

[124] Demonstrate to governments, ideally. *See, e.g.*, Jess Whittlestone & Jack Clark, *Why and How Governments Should Monitor AI Development*, ARXIV (Aug. 31, 2021), https://arxiv.org/pdf/2108.12427.pdf [https://perma.cc/Q6KF-JTXA].

[125] *See infra* Section III.

[126] For a definition of "meaning" that would be appropriate here, see, for example, Christopher Manning, *Human Language Understanding & Reasoning*, 151 DAEDALUS 127, 134-135 (2022) ("Meaning is not all or nothing; in many circumstances, we partially appreciate the meaning of a linguistic form. I suggest that meaning arises from understanding the network of connections between a

In addition to demonstrating its ability to uphold private contractual obligations (e.g., through acceptable out-of-sample task performance), sufficiently capable AI should demonstrate an ability to perform consistent with extra-contractual standards, such as a fully automated investment advisory system exhibiting simulated behavior in line with fiduciary duties to a human principal.

Sufficiently agentic AI[127] should demonstrate comprehension of the public law that will be relevant to its behavior if deployed.[128] This is a very difficult threshold to pass.[129]

Although super-human intelligence would be able to conduct legal reasoning beyond the capability of any lawyer, legal questions ultimately bottom out at a mechanism for resolution: the governmental legal system.[130] We cannot fully understand the decisions of superhuman AI. Similarly, courts do not purport to have any substantive understanding of technical details or science behind complex cases they provide final determinations on. The law is designed to resolve outcomes without requiring judges to have domain knowledge (or cognitive capacity) anywhere near the level of the parties or technologies involved. Therefore, if alignment is driven by understanding of legal information and legal reasoning, humans can assess alignment of more intelligent AI. This is an important feature of the *Law Informs Code* framework. Compare this to ethics – a widely discussed potential source of human values for AI alignment – where it is unclear how we could evaluate (or what would even constitute) super-intelligent ethical

---

linguistic form and other things, whether they be objects in the world or other linguistic forms. If we possess a dense network of connections, then we have a good sense of the meaning of the linguistic form. For example, if I have held an Indian shehnai, then I have a reasonable idea of the meaning of the word, but I would have a richer meaning if I had also heard one being played [ . . . ] Using this definition whereby understanding meaning consists of understanding networks of connections of linguistic forms, there can be no doubt that pre- trained language models learn meanings.").

[127] For a description of advanced AI capabilities and agentic planning, see, for example, Joseph Carlsmith, *Is Power-Seeking AI an Existential Risk?*, ARXIV (Jun. 16, 2022), https://arxiv.org/pdf/2206.13353.pdf [https://perma.cc/R4YQ-44B6].

[128] *See infra* Section IV; John Nay & James Daily, *Aligning Artificial Intelligence with Humans through Public Policy*, ARXIV (Jun. 25, 2022), https://arxiv.org/ftp/arxiv/papers/2207/2207.01497.pdf [https://perma.cc/EQ4M-TNQM].

[129] *See, e.g.*, Daniel Gervais, *Towards an Effective Transnational Regulation of AI*, 38 AI & SOC'Y 391, 396 n.21 (2021) ("Take just this well-known example: Carlsbad Technology, Inc. v. HIF Bio, Inc, 556 U.S. 635 (2008). Looking at the statute involved in that case (28 U. S. C. §1441(c)), would lead to an entirely incorrect understanding of "the law" because the Supreme Court's interpretation of the statute—the exact opposite of what the text of the statute actually says—is what courts are bound to follow under stare decisis."). See, avenues of AI research focused on resolving conflicting norms, for example, Daniel Kasenberg & Matthias Scheutz, *Inverse Norm Conflict Resolution*, 2018 PROC. AAAI/ACM CONF. ON AI, ETHICS, & SOC'Y 178.

[130] *See, e.g.*, Figure 6, *infra*.

decisions because there is no mechanism external to the AI that can legitimately resolve ethical deliberation.[131]

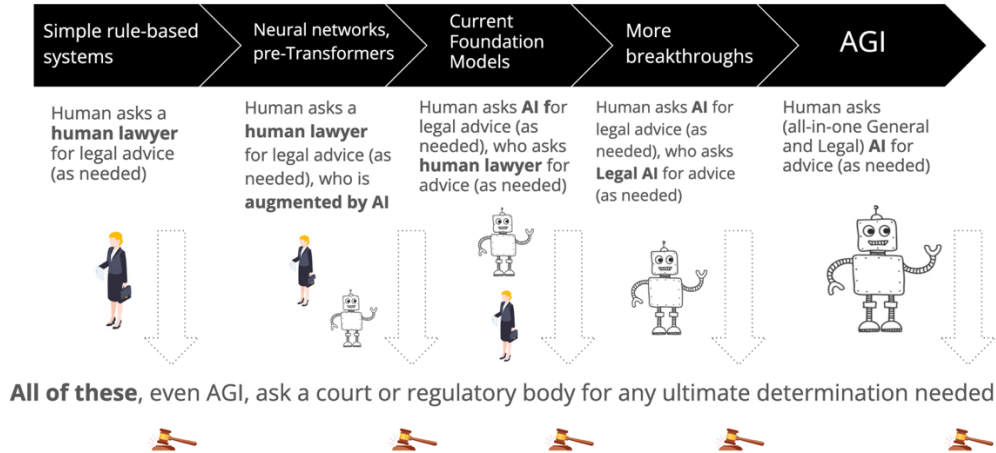

**Figure 6:** In the example of scaling legal practice, as AI capabilities surpass humans, the Law Informs Code approach to AI alignment keeps the process ultimately grounded in human judgment.

As the state-of-the-art for AI advances, there should be a higher bar of demonstrated legal understanding.[132] If an AI developer claims their system has advanced capabilities on tasks that it would like the AI to complete outside of its training environment, the developer should show correspondingly advanced legal knowledge and legal reasoning abilities of the system.

---

[131] *See infra* Section IV.A.

[132] Scholars have suggested we use AI "Guardians" to monitor operational AI systems once they are deployed. *See, e.g.*, Amitai Etzioni & Oren Etzioni, *Keeping AI Legal*, 19 Vand. J. Ent. & Tech. L. 133, 139 (2016) ("From here on, AI should be divided into two categories. The first category would consist of operational AI programs - the computerized 'brains' that guide smart instruments. The second category would be composed of oversight AI programs that review the first category's decision making and keep the decisions in line with the law. These oversight programs, which this Article calls 'AI Guardians,' would include AI programs to interrogate, discover, supervise, audit, and guarantee the compliance of operational AI programs."). Our proposal is focused on AI systems demonstrating their own legal understanding, but our discussion of seeking parity between AI capabilities and AI legal understanding is related to what would be required to technically implement Etzionis' proposal that "[t]hese AI Guardians will need to become smarter just as operational AI programs are improving," *id.* at 146.

One of the technical impediments to alignment,[133] especially of deep learning models,[134] is the difficulty of understanding the model and the causes of its behaviors once trained.[135] This generally worsens as models scale in size and complexity.[136] Understanding the inner workings of AI is helpful for editing its beliefs,[137] and validating its safety,[138] reliability, and legal comprehension abilities.[139] Training deep learning models on legal data,

[133] *See, e.g.*, Christian, *supra* note 9.

[134] *See, e.g.*, Tilman Räukur et al., *Toward Transparent AI: A Survey on Interpreting the Inner Structures of Deep Neural Networks*, ARXIV (Jan. 23, 2023), https://arxiv.org/pdf/2207.13243.pdf [https://perma.cc/VX2V-95HJ].

[135] *See generally* Marina Danilevsky et al., *A Survey of the State of Explainable AI for Natural Language Processing*, Association for Computational Linguistics (2020). For some of the latest methods for interpretability of large generative machine learning models, see, for example, Deep Ganguli et al., *Predictability and Surprise in Large Generative Models*, ACM Digital Library, https://dl.acm.org/doi/abs/10.1145/3531146.3533229 [https://perma.cc/6XVR-9DRC] . For a counter-argument for the need for explainability in the context of autonomous systems acting legally, see, for example, Henry Prakken, *On How AI & Law Can Help Autonomous Systems Obey the Law: A Position Paper*, 2016 PROC. EUR. CONF. ON A.I. 42, 44 (2016), http://www.ecai2016.org/content/uploads/2016/08/W2-ai4j-2016.pdf [https://perma.cc/F474-CG7P] ("[T]he legal tasks supported by traditional AI & law tools require explanation and justification of decisions. With autonomous systems there is no need for this; all that counts is that legally acceptable behaviour is generated. Of course, when an autonomous system does something legally wrong, its behaviour might have to be explained in a court case. However, this does not require that the system itself can do that; it may suffice to have a log file recording the system's internal actions."). *see also, e.g.*, MICHAEL KEARNS & AARON ROTH, THE ETHICAL ALGORITHM: THE SCIENCE OF SOCIALLY AWARE ALGORITHM DESIGN 170-175 (2019) (defining "interpretability" depends on the audience and the model type and task).

[136] NYU School of Law, *Algorithms and Explanations: Modes of Explanation in Machine Learning,* YOUTUBE (April 27, 2017), https://youtu.be/U0NsxZQTktk [https://perma.cc/FT3G-G9S2].

[137] Kevin Meng et al., *Locating and Editing Factual Associations in GPT* (Neural Information Processing Systems, Conference Paper, Nov. 28, 2022), https://arxiv.org/pdf/2202.05262.pdf [https://perma.cc/M9YX-QVQZ] (demonstrating how an understanding of the mechanics of a deep learning based model helps researchers edit "beliefs" of the model).

[138] Understanding the models is also important for dealing with "inner alignment" problems that may arise with more powerful AI systems because it could help uncover instances of AI models deceiving humans, and situations where AI models are solving subproblems unforeseen in the original problem specification. *See, e.g.*, Evan Hubinger et al., *Risks from Learned Optimization in Advanced Machine Learning Systems*, ARXIV (Dec. 1, 2021), https://arxiv.org/pdf/1906.01820.pdf [https://perma.cc/59DU-LAZU]. For more on explainable AI in the alignment context, see, for example, Yonatan Belinkov & James Glass, *Analysis Methods in Neural Language Processing: A Survey*, 7 TRANSACTIONS ASS'N FOR COMPUTATIONAL LINGUISTICS 49 (2019); Evan Hubinger, *Relaxed Adversarial Training for Inner Alignment*, AI ALIGNMENT FORUM (September 10, 2019), https://www.alignmentforum.org/posts/9Dy5YRaoCxH9zuJqa/relaxed-adversarial-training-for-inner-alignment [https://perma.cc/X5D7-4G5Z]. For more on interpreting the inner workings of AI models, *see, e.g.*, Chris Olah et al., *Zoom In: An Introduction to Circuits*, DISTILL (March 10, 2020), https://distill.pub/2020/circuits/zoom-in/ [https://perma.cc/2GAZ-QTHS]; Chelsea Voss et al., *Visualizing Weights*, DISTILL (Feb. 4, 2021), https://distill.pub/2020/circuits/visualizing-weights/ [https://perma.cc/JF3W-KR4K].

[139] *See, e.g.*, Katie Atkinson et al., *Explanation in AI and Law: Past, Present and Future*, 289 A.I. (2020) (Article #10387) ("[I]nsights from AI and Law, where explanation has long been a concern, may

where learned intermediate model representations can correspond to legal concepts, opens the possibility for mechanistic (alignment) interpretability[140] (methods for reverse engineering key components of AI to better understand its tendencies). If neural networks learn representations of human interpretable legal knowledge[141] as a substrate of the model, and if legal concepts are the ontology for alignment (as this Article argues), viewing their use inside a model could help us understand if and how a model is aligned.

We should supplement mechanistic explanations with a behavioral perspective. Simulations exploring the actions of machine-learning-based decision-making models throughout state-action space can uncover patterns of agent decision-making.[142] Safety benchmarks have been developed for simple environments for AI agents trained with reinforcement learning.[143] Similar approaches could help demonstrate an AI's comprehension of legal standards before it is permitted to act in the real world.[144] This would not be

---

provide useful pointers for future development of explainable AI."); JONATHON PHILLIPS ET AL., NAT'L INST. STANDARDS & TECH., FOUR PRINCIPLES OF EXPLAINABLE ARTIFICIAL INTELLIGENCE (2021); Error! Hyperlink reference not valid.Katherine J. Strandburg, *Rulemaking and Inscrutable Automated Decision Tools*, 119 COLUM. L. REV. 1851 (2019).

[140] *See, e.g.*, Chris Olah, *Mechanistic Interpretability, Variables, and the Importance of Interpretable Bases*, TRANSFORMER CIRCUITS THREAD (June 27, 2022), https://transformer-circuits.pub/2022/mech-interp-essay/index.html [https://perma.cc/XUQ5-JFFH].

[141] *See, e.g.*, John Nay, *Gov2Vec: Learning Distributed Representations of Institutions and Their Legal Text*, 2016 PROCS. EMNLP WORKSHOP ON NAT. LANGUAGE PROCESSING AND COMPUTATIONAL SOC. SCI. 49.

[142] *See, e.g.*, John Nay, A Machine Learning Approach to Modeling Dynamic Decision-Making in Strategic Interactions and Prediction Markets (Mar. 28, 2017) (Ph.D. dissertation, Vanderbilt University), https://ir.vanderbilt.edu/bitstream/handle/1803/11555/nay.pdf?sequence=1&isAllowed [https://perma.cc/6Y6K-DCR7]; John Nay & Yevgeniy Vorobeychik, *Predicting Human Cooperation*, PLOS ONE (May 12, 2016), https://journals.plos.org/plosone/article/file?id=10.1371/journal.pone.0155656&type=printable [https://perma.cc/27Y6-MZ99]; John Nay & Jonathan M. Gilligan, *Data-driven Dynamic Decision Models*, 2015 PROC. WINTER SIMULATION CONF. 2752; John Nay et al., *Betting and Belief: Prediction Markets and Attribution of Climate Change*, 2016 PROC. WINTER SIMULATION CONF. 1666.

[143] *See, e.g.*, Alex Ray et al., *Benchmarking Safe Exploration in Deep Reinforcement Learning*, SEMANTIC SCHOLAR (2019), https://www.semanticscholar.org/paper/Benchmarking-Safe-Exploration-in-Deep-Reinforcement-Achiam-Amodei/4d0f6a6ffcd6ab04732ff76420fd9f8a7bb649c3#citing-papers; Daniel S. Brown et al., *Value Alignment Verification*, 139 PROCS MACH. LEARNING RSCH. 1105 (2021).

[144] *See, e.g.*, Gianclaudio Malgieri & Frank A. Pasquale, *From Transparency to Justification: Toward Ex Ante Accountability for AI* 1 (Brook. L. Sch., Legal Studies Paper No. 712, 2022) (proposing "a system of 'unlawfulness by default' for AI systems, an ex-ante model where some AI developers have the burden of proof to demonstrate that their technology is not discriminatory, not manipulative, not unfair, not inaccurate, and not illegitimate in its legal bases and purposes"). For other proposals related to the certification or verification of AI systems before and during deployment, see, for example, Miles Brundage et al., *Toward Trustworthy AI Development: Mechanisms for Supporting Verifiable Claims*, ARXIV (Apr. 20, 2020), https://arxiv.org/pdf/2004.07213.pdf [https://perma.cc/RX7M-9AFE]; Inioluwa Deborah Raji et al., *Outsider Oversight: Designing a Third Party Audit Ecosystem for AI Governance*, 2022 PROC. AAAI/ACM CONF. ON AI, ETHICS, & SOC. 557; Gregory Falco et al., *Governing AI Safety Through Independent Audits*, 3 NATURE MACH. INTEL. 566

a fool-proof deterministic verification.[145] From a legal perspective, this is analogous to the certification of legal and regulatory understanding for professionals such as financial advisors, with the key difference that there is a relatively costless assessment of AI legal understanding. Relative to the professional certification and subsequent testing we currently impose on humans providing specialized services such as financial advising, it is significantly less expensive to run millions of simulations of scenarios to test an AI's comprehension of relevant legal standards and regulations.[146] It is now possible to conduct social science research on data generated by simulating persons by using Foundation Models conditioned on human data.[147] Applying empirical social science methods to simulations of AI behavior is a promising approach to measuring AI legal understanding.

---

(2021); Peter Cihon et al., *AI Certification: Advancing Ethical Practice by Reducing Information Asymmetries*, 2 IEEE TRANSACTIONS ON TECH. AND SOC. 200 (2021); Andrew Tutt, *An FDA For Algorithms*, 69 ADMIN. L. REV. 83, 122 (2017) (arguing that there is a "close analog between complex pharmaceuticals and sophisticated algorithms" and therefore the FDA provides a model for a new regulatory agency for algorithms); Florian Möslein & Roberto V. Zicari, *Certifying Artificial Intelligence Systems*, *in* RESEARCH HANDBOOK ON BIG DATA LAW 357 (Roland Vogl ed., 2021); Thomas Arnold & Matthias Scheutz, *The "big red button" is too late: an alternative model for the ethical evaluation of AI systems*, 20 ETHICS AND INFORMATION TECHNOLOGY 59, 60 (2018) ("We outline a system architecture consisting of an ethical core layer above the hardware and below the virtual machine layer that consists of scenario-generation and simulation engines . . . ."). For a proposal for verification of AI systems before deployment, specifically related to reinforcement learning decision-making systems*, see, e.g.*, THOMAS KRENDL GILBERT ET AL., CHOICES, RISKS, AND REWARD REPORTS: CHARTING PUBLIC POLICY FOR REINFORCEMENT LEARNING SYSTEMS (2022).

[145] *See, e.g.*, Stanley Bak et al., *The Second International Verification of Neural Networks Competition (VNN-COMP 2021): Summary and Results*, ARXIV (Aug. 31, 2021), https://arxiv.org/pdf/2109.00498.pdf [https://perma.cc/N8RR-V8QY]. For a discussion of the difficulty of verifying the security of software systems and the analogy to aligning AI, see, for example, elspood, *Security Mindset: Lessons from 20+ years of Software Security Failures Relevant to AGI Alignment*, LESSWRONG (June 21, 2022), https://www.lesswrong.com/posts/Ke2ogqSEhL2KCJCNx/security-mindset-lessons-from-20-years-of-software-security [https://perma.cc/5CAD-64YG]. Developing robust verification processes is extremely difficult. Much more deterministic and much simpler software than AI systems cannot be fully trusted, see, for example, Ken Thompson, *Reflections on Trusting Trust*, 27 COMMC'NS ASS'N FOR COMPUTING MACH. 761, 763 (1984) ("You can't trust code that you did not totally create yourself. . . . No amount of source-level verification or scrutiny will protect you from using untrusted code.").

[146] For a proposal on monitoring AI systems once they are deployed*, see*, for example, Etzioni & Etzioni, *supra* note 132, at 139, 146 ("From here on, AI should be divided into two categories. The first category would consist of operational AI programs—the computerized 'brains' that guide smart instruments. The second category would be composed of oversight AI programs that review the first category's decision making and keep the decisions in line with the law. These oversight programs, which this Article calls 'AI Guardians,' would include AI programs to interrogate, discover, supervise, audit, and guarantee the compliance of operational AI programs. . . . These AI Guardians will need to become smarter just as operational AI programs are improving. . . . [H]umans may have little choice but to draw on AI to check AI—and to seek to increase oversight of artificial intelligence as the intelligence of the programs they oversee grows.").

[147] *See, e.g.*, Lisa P. Argyle et al., *Out of One, Many: Using Language Models to Simulate Human Samples*, POL. ANALYSIS, Feb. 21, 2023, at 1.

### 3. *Legal Understanding as an Alignment Benchmark*

Progress in AI research is driven, in large part, by shared benchmarks that thousands of researchers globally use to guide experiments, understand as a community whether model approaches and data are improving AI capabilities, and compare results across research groups.[148] AI performance on benchmarks are one of the primary "objective functions" of the overall global research apparatus.[149] As quantitative lodestars, benchmarks can create perverse incentives to build AI that optimizes for benchmark performance at the expense of true generalization and intelligence,[150] see, e.g., "Goodhart's Law," colloquially communicated as, "when a measure becomes a target, it ceases to be a good measure."[151] Many AI benchmarks have a significant number of errors,[152] which suggests that in some cases machine learning models have, more than widely recognized, "overfitted to memorizing data instead of learning abstract concepts."[153] There are spurious cues within benchmark data that, once removed, significantly drop model performance, demonstrating that models are often learning patterns that do not generalize outside of the closed world of the benchmark data.[154] Many benchmarks, especially in natural language processing, have become saturated, [155] as "contemporary models quickly achieve outstanding performance on benchmark tasks but nonetheless fail on simple challenge examples and falter in real-world scenarios." [156] Benchmarking AI capabilities is difficult.[157] Benchmarking AI alignment has the same issues, but compounded by vaguer problem definitions. There is also far less research on AI alignment benchmarks.

---

[148] Douwe Kiela et al., *Dynabench: Rethinking Benchmarking in NLP*, 2021 PROCS. CONF. N. AM. CHAPTER ASS'N FOR COMPUTATIONAL LINGUISTICS: HUM. LANGUAGE TECHS. 4111 ("Progress in NLP has traditionally been measured through a selection of task-level datasets that gradually became accepted benchmarks"); Samuel R. Bowman & George E. Dahl, *What Will it Take to Fix Benchmarking in Natural Language Understanding?*, 2021 PROC. CONF. N. AM. CHAPTER ASS'N FOR COMPUTATIONAL LINGUISTICS: HUM. LANGUAGE TECHS. 4843 ("The plight of benchmark-driven NLU [natural language understanding] research has prompted widespread concern about the assumptions underlying standard benchmarks and widespread interest in alternative models of evaluation.").

[149] For more on this generalized notion of an objective function, *see generally* KENNETH O. STANLEY & JOEL LEHMAN, WHY GREATNESS CANNOT BE PLANNED: THE MYTH OF THE OBJECTIVE (2015).

[150] *See generally* Chollet, *Deep Learning with Python, Second Edition, supra* note 31, at Ch. 14.3.1 (discussing the "shortcut rule"); *see, e.g.*, Simon Ott et al., *Mapping Global Dynamics of Benchmark Creation and Saturation in Artificial Intelligence*, NATURE COMMC'N, Nov. 10, 2022, at 1 ("We curate[d] data for 3765 benchmarks covering the entire domains of computer vision and natural language processing, and show that a large fraction of benchmarks quickly trended towards near-saturation. . . .").

[151] Charles Goodhart, *Problems of Monetary Management: The U.K. Experience, in* INFLATION, DEPRESSION, AND ECONOMIC POLICY IN THE WEST 111 (Anthony S. Courakis ed., 1975); *See, e.g.*, Peter Coy, *Goodhart's Law Rules the Modern World. Here are Nine Examples*, BLOOMBERG (March 26, 2021), https://www.bloomberg.com/news/articles/2021-03-26/goodhart-s-law-rules-the-modern-world-here-are-nine-examples#xj4y7vzkg [https://perma.cc/6UXV-GCCT]; David Manheim & Scott Garrabrant, *Categorizing Variants of Goodhart's Law*, arXiv (Feb. 26, 2019), https://arxiv.org/pdf/1803.04585.pdf [https://perma.cc/46WY-GTT4].

[152] Curtis G. Northcutt et al., *Pervasive Label Errors in Test Sets Destabilize Machine Learning Benchmarks* (Neural Information Processing Systems, Conference Paper, Dec. 6, 2021)

[153] Björn Barz & Joachim Denzler, *Do We Train on Test Data? Purging CIFAR of Near-Duplicates*, 6 J. IMAGING 41, 41 (2020).

[154] *See, e.g.*, Ronan Le Bras et al., *Adversarial Filters of Dataset Biases*, 119 PROC. MACH. LEARNING RSCH. 1078 (2020).

Performing well on societal alignment is more difficult than performing well on task capabilities.[158] Because alignment is so fundamentally hard, the sky should be the limit on the difficulty of alignment benchmarks.[159] Legal-informatics-based benchmarks could serve as AI alignment benchmarks. Models currently perform worse than expert humans on legal understanding tasks such as statutory reasoning, [160] professional law, [161] and legal discovery;[162] there is significant room for improvement on legal language processing tasks.[163] Example benchmarks that could be used as part of the

---

overall alignment benchmark are legal search,[164] Bar Exam scores, contract analysis,[165] and identifying legislation that is relevant to a company.[166] Next, we discuss example research directions that could improve the performance of AI on legal understanding benchmarks. A comprehensive suite of benchmark datasets could catalyze research through a desire of the community of researchers to score highly on the leaderboards.

## C. *Legal Processes, Data & Experts Can Improve AI*

Legal informatics may allow us to engineer models in new ways and engineer legal data (both observational data and data derived from human interaction with models) into training signals that help align AI.

### 1. *Models*

The current era of deep learning is characterized in large part by scaling the size of the models and training them with self-supervision (and, recently, adding reinforcement learning from human feedback on top for fine-tuning).[167] There now seems to be evidence that the bottleneck in pushing the capabilities of most AI performance further (e.g., in large language

---

models,[168] and recommendation models[169]) is less on the sheer size of the models and more on the amount of useful data, training procedures, and (probably less importantly) model architectures.[170] Legal informatics is an untapped source of data, a set of symbolic systems that Foundation Models can call on (similar to calling a Python interpreter), and potentially an inspiration for training procedures and model structure tweaks.[171]

### a. *AI Capabilities Can Improve Legal Informatics*

The *Law Informs Code* agenda can leverage recent advancements in machine learning.[172] In particular, there are relevant threads of research in natural language processing[173] with large language models trained with self-supervision; deep reinforcement learning; the intersection of large language models and deep reinforcement learning; [174] and "safe reinforcement learning"[175] (especially where constraints on agent actions can be described

---

[168] Jordan Hoffmann et al., *Training Compute-Optimal Large Language Models* (arXiv, Working Paper No. 2203.15556, Mar. 29, 2022), https://arxiv.org/pdf/2203.15556.pdf [https://perma.cc/PHG4-3DNH].

[169] Newsha Ardalani et al., *Understanding Scaling Laws for Recommendation Models* (arXiv, Working Paper No. 2208.08489, Aug. 17, 2022), https://arxiv.org/pdf/2208.08489.pdf [https://perma.cc/T5NM-P3PZ].

[170] Marcos Treviso et al., *Efficient Methods for Natural Language Processing: A Survey* (arXiv, Working Paper No. 2209.00099, Aug. 31, 2022), https://arxiv.org/pdf/2209.00099.pdf [https://perma.cc/235H-YJJP].Error! Hyperlink reference not valid.

[171] *See infra* Section II.C.1.ii.

[172] *See, e.g.*, *supra* note 58.

[173] For natural language processing methods applied to legal text, *see, e.g.*, John J. Nay, *Natural Language Processing for Legal Texts*, *in* LEGAL INFORMATICS 99 (Daniel Martin Katz et al. eds. 2021); MICHAEL A. LIVERMORE & DANIEL N. ROCKMORE, *Distant Reading the Law*, *in* LAW AS DATA: COMPUTATION, TEXT, AND THE FUTURE OF LEGAL ANALYSIS, 3–19 (2019); J.B. Ruhl et al., *Topic Modeling the President: Conventional and Computational Methods*, 86 GEO. WASH. L. REV. 1243 (2018); John Nay, *Predicting and Understanding Law-making with Word Vectors and an Ensemble Model*, PLoS ONE 1 (May 10, 2017), https://journals.plos.org/plosone/article?id=10.1371/journal.pone.0176999 [https://perma.cc/L6JZ-DPNZ]; John Nay, *Gov2Vec: Learning Distributed Representations of Institutions and Their Legal Text*, 2016 PROC. FIRST WORKSHOP ON NLP & COMPUT. SOC. SCI. 49.

[174] *See, e.g.*, Prithviraj Ammanabrolu et al., *Aligning to Social Norms and Values in Interactive Narratives*, 2022 PROC. CONF. N. AM. CHAPTER ASS'N FOR COMPUT. LINGUISTICS: HUM. LANGUAGE TECH. 5994.

[175] *See, e.g.*, Javier Garcia & Fernando Fernandez, *A Comprehensive Survey on Safe Reinforcement Learning*, 16 J. MACH. LEARNING RSCH. 1438 (2015) ("Safe Reinforcement Learning can be defined as the process of learning policies that maximize the expectation of the return in problems in which it is important to ensure reasonable system performance and/or respect safety constraints during the learning and/or deployment processes."); Philip S. Thomas et al., *Preventing Undesirable Behavior of Intelligent Machines*, 366 SCIENCE 999 (2019); Saunders et al., *supra* note 34, at 1; Markus Peschl et al., *MORAL: Aligning AI with Human Norms through Multi-Objective Reinforced Active Learning* (arXiv, Working Paper No. 2201.00012, Dec. 30, 2021), https://arxiv.org/pdf/2201.00012.pdf [https://perma.cc/959D-GAGC].

in natural language[176]). The combination of (a) large language models trained on large corpora of (sometimes explicitly morally salient[177]) text powering decision-making agents;[178] (b) procedures that learn an automated mapping

---

[176] *See, e.g.*, Yang et al., *Safe Reinforcement Learning with Natural Language Constraints* 2, 3, 5 (arXiv, Working Paper No. 2010.05150, Aug. 4, 2021), https://arxiv.org/pdf/2010.05150.pdf [https://perma.cc/8T3V-UHBG] (Some research on safe reinforcement learning requires "a human to specify the cost constraints in mathematical or logical form, and the learned constraints cannot be easily reused for new learning tasks. In this work, we design a modular architecture to learn to interpret textual constraints and demonstrate transfer to new learning tasks [ . . . ] Our model first uses a *constraint interpreter* to encode language constraints into representations of forbidden states. Next, a *policy network* operates on these representations and state observations to produce actions. Factoring the model in this manner allows the agent to retain its constraint comprehension capabilities while modifying its policy network to learn new tasks. Our experiments demonstrate that our approach achieves higher rewards (up to 11x) while maintaining lower constraint violations (up to 1.8x) compared to the baselines on two different domains."); Bharat Prakash et al., *Guiding Safe Reinforcement Learning Policies Using Structured Language Constraints*, 2020 PROC. WORKSHOP ON A. I. SAFETY 153, https://mdsoar.org/bitstream/handle/11603/17463/AAAI_RL_Workshop.pdf?sequence=1&isAllowed=y [https://perma.cc/QR2A-WVGK].

[177] *See, e.g.*, Zhijing Jin et al., *When to Make Exceptions: Exploring Language Models as Accounts of Human Moral Judgment* 1 (Neural Information Processes Systems, Conference Paper, Nov. 28, 2022), https://arxiv.org/pdf/2210.01478.pdf [https://perma.cc/M2KD-TV9Q] ("[W]e present a novel challenge set consisting of rule-breaking question answering (RBQA) of cases that involve potentially permissible rule-breaking – inspired by recent moral psychology studies. Using a state-of-the-art large language model (LLM) as a basis, we propose a novel moral chain of thought (MORALCOT) prompting strategy that combines the strengths of LLMs with theories of moral reasoning developed in cognitive science to predict human moral judgments."); Liwei Jiang et al., *Delphi: Towards Machine Ethics and Norms* 5 (arXiv, Working Paper No. 2110.07574, Jul. 12, 2022), https://arxiv.org/pdf/2110.07574.pdf [https://perma.cc/PBL5-6EZE] ("1.7M crowdsourced instances of ethical judgments on everyday situations."); Dan Hendrycks et al., *Aligning AI With Shared Human Values* 1 (arXiv, Working Paper No. 2008.02275, Feb. 17, 2023), https://arxiv.org/pdf/2008.02275.pdf [https://perma.cc/YLH6-BNMA] ("[W]e find that current language models have a promising but incomplete ability to predict basic human ethical judgements. Our work shows that progress can be made on machine ethics today, and it provides a steppingstone toward AI that is aligned with human values."); Nicholas Lourie et al., *Scruples: A Corpus of Community Ethical Judgments on 32,000 Real-life Anecdotes*, 35 PROC. AAAI CONF. ON A.I. 13470 (2021), https://arxiv.org/pdf/2008.09094.pdf [https://perma.cc/95D9-DA2D] (32,000 real-life ethical situations, with 625,000 ethical judgments); Frazier et al., *Learning Norms from Stories: A Prior for Value Aligned Agents* 1 (arXiv, Working Paper No. 1912.03553, Dec. 7, 2019), https://arxiv.org/pdf/1912.03553.pdf [https://perma.cc/8X4P-LH2X].

[178] *See, e.g.*, Ammanabrolu et al., *supra* note 174, at 1 ("We introduce . . . [an] agent that uses the social commonsense knowledge present in specially trained language models to contextually restrict its action space to only those actions that are aligned with socially beneficial values."); Md Sultan Al Nahian et al., *Training Value-Aligned Reinforcement Learning Agents Using a Normative Prior* 1 (arXiv, Working Paper No. 2104.09469, Apr. 19, 2021), https://arxiv.org/pdf/2104.09469.pdf [https://perma.cc/83DK-94E2]. ("We introduce an approach to value-aligned reinforcement learning, in which we train an agent with two reward signals: a standard task performance reward, plus a normative behavior reward. The normative behavior reward is derived from a value-aligned prior model previously shown to classify text as normative or non-normative. We show how variations on a policy shaping technique can balance these two sources of reward and produce policies that are both effective and perceived as being more normative."); Dan Hendrycks et al., *What Would Jiminy Cricket Do? Towards Agents That Behave Morally*, 35 CONF. ON NEURAL INFO. PROCESSING SYS., Feb. 8, 2022, at

from natural language to environment dynamics[179] and reward functions of agents;[180] and (c) offline reinforcement learning (with Transformer-based models)[181] represents a potential opportunity to leverage millions (or even billions) of state-action-value tuples from (natural language) legal text within reinforcement learning paradigms (where AI agents make "decisions").

We should also experiment with how legal informatics is employed within AI agent decision-making paradigms, [182] e.g., (a) as (natural language [183]) constraints; [184] (b) for shaping the reward function during training;[185] (c) for refined representations of the state space;[186] (d) for guiding the exploration of the state space during training;[187] (e) as inputs to world models for efficient training;[188] (f) as a Foundation Model prior, or part of pretraining, to bias a deployed agent's action space toward certain actions or away from others;[189] or (g) as some combination of the aforementioned. Where legal informatics is providing the modular constructs (e.g., methods of statutory interpretation, applications of standards, and legal reasoning more broadly) to facilitate the communication of what a human wants an AI to do, it is more likely employed for specifying and shaping reward functions. Where legal informatics, through distillations of public law, helps specify what AI should *not* do, to provide a broader knowledge base of how to reduce societal externalities, it is more likely employed as constraints on the actions available to an agent.

Foundation Models have the potential to unlock significant legal understanding capability. "Legal decision-making requires context at various scales: knowledge of all historical decisions and standards, knowledge of the case law that remains relevant in the present, and knowledge of the nuances of the individual case at hand. Foundation models

---

[https://perma.cc/T7Q7-CTEH]; Sergey Levine et al., *Offline Reinforcement Learning: Tutorial, Review, and Perspectives on Open Problems* (arXiv, Working Paper No. 2005.01643, Nov. 11, 2020), https://arxiv.org/pdf/2005.01643.pdf [https://perma.cc/PMH3-UMXF].

[182] *See, e.g.*, MYKEL J. KOCHENDERFER ET AL., ALGORITHMS FOR DECISION MAKING (2022).

[183] *See, e.g.*, Yang et al., supra note 176, at 3 ("Since constraints are decoupled from rewards and policies, agents trained to understand certain constraints can transfer their understanding to respect these constraints in new tasks, even when the new optimal policy is drastically different.").

[184] *See, e.g.*, Joshua Achiam et al., *Constrained Policy Optimization*, 70 PROC. MACH. LEARNING RSCH. 22 (2017).

[185] *See, e.g.,* Bharat Prakash et al., *supra* note 176; Hendrycks et al., *supra* note 178.

[186] *See, e.g.,* Mengjiao Yang & Ofir Nachum, *Representation Matters: Offline Pretraining for Sequential Decision Making*, 139 PROC. MACH. LEARNING RSCH. 11784 (2021). http://proceedings.mlr.press/v139/yang21h/yang21h.pdf [https://perma.cc/Z8JR-Z9RV].

[187] *See, e.g.*, Allison C. Tam et al., *Semantic Exploration from Language Abstractions and Pretrained Representations* (arXiv, Working Paper No. 2204.05080, May 27, 2022), https://arxiv.org/pdf/2204.05080.pdf [https://perma.cc/GN47-X8VF].

[188] *See, e.g.*, Vincent Micheli, Eloi Alonso & François Fleuret, *Transformers are Sample Efficient World Models* (International Conference on Learning Representations, Forthcoming Conference Paper, April 30, 2023), https://arxiv.org/pdf/2209.00588.pdf [ https://perma.cc/RN5R-8HR5].

[189] *See, e.g.*, Jacob Andreas, Dan Klein & Sergey Levine, supra note 181, at 2166–79; Yao et al., supra note 178; Andrew K Lampinen et al., *Tell Me Why! Explanations Support Learning Relational and Causal Structure*, 162 PROC. MACH. LEARNING RSCH. 11868 (2022).

are uniquely poised to have the potential to learn shared representations of historical and legal contexts, as well as have the linguistic power and precision for modeling an individual case."[190] Foundation Models trained on legal text learn model weights and word embeddings specific to legal text that (in the limited work thus far) provide (slightly) better performance on downstream legal tasks relative to models trained on primarily non-legal text.[191] Foundation Models have been useful for analyzing legal language[192] and legal arguments,[193] and testing legal theories.[194] Foundation Models' recent strong capabilities in automatically analyzing (non-legal) citations[195] may prove fruitful in identifying relevant legal precedent, and their ability to generate persuasive language could help AI understand, and thus learn from, legal brief text data.[196]

Foundation Models are beginning to demonstrate improved performance in analyzing contracts.[197] As state-of-the-art models have gotten larger and more advanced, their contract analysis performance has improved,[198] suggesting we can expect continued advancements in natural

---

language processing capabilities to improve legal text analysis as a by-product.[199] Mainstream AI capabilities research could potentially unlock further advances toward *Law Informing Code*, in particular through the successful application of deep reinforcement learning further beyond toy problems (e.g., video games and board games),[200] with human feedback,[201] and through offline learning at large scale.[202]

These potentials represent the preferred approach: *Law* can best *Inform Code* if legal informatics is able to adopt (or adapt) state-of-the-art models and processes and convert the progress in general AI capabilities (which is being aggressively funded by most of the large internet technology companies and national governments) into gains in AI legal understanding.

---

contract review]. We find that performance metrics such as Precision @ 80% Recall are improving quickly as models improve, such that a BERT model from 2018 attains 8.2% while a DeBERTa model from 2021 attains 44.0%.").

[199] Rishi Bommasani et al., *supra* note 102, at 59 ("Many legal applications pose unique challenges to computational solutions. Legal language is specialized and legal outcomes often rely on the application of ambiguous and unclear standards to varied and previously unseen fact patterns. At the same time, due to its high costs, labeled training data is scarce. Depending on the specific task, these idiosyncrasies can pose insurmountable obstacles to the successful deployment of traditional models. In contrast, their flexibility and capability to learn from few examples suggest that foundation models could be uniquely positioned to address the aforementioned challenges.").

[200] In the legal understanding domain, *see, e.g.*, Duy-Hung Nguyen et al., *Robust Deep Reinforcement Learning for Extractive Legal Summarization* (Neural Information Processing, Conference Paper, Dec. 6, 2021), https://arxiv.org/pdf/2111.07158.pdf [https://perma.cc/8TS2-X6ST].

[201] *See, e.g.*, Paul F. Christiano et al., *Deep Reinforcement Learning from Human Preferences*, 30 ADVANCES IN NEURAL INFO. PROCESSING SYS. (2017); Natasha Jaques et al., *Way Off-Policy Batch Deep Reinforcement Learning of Implicit Human Preferences in Dialog*, ARXIV (July 8, 2019), https://arxiv.org/pdf/1907.00456.pdf [https://perma.cc/V92K-2GWD]; Stiennon et al., *supra* note 95, at 3008–3021 (2020); Daniel M. Ziegler, et al., *Fine-tuning Language Models From Human Preferences*, ARXIV (Sept. 18, 2019); Jeff Wu et al., *Recursively Summarizing Books with Human Feedback*, ARXIV (Sept. 27, 2021), https://arxiv.org/pdf/2109.10862.pdf [https://perma.cc/Q5U7-YDY7]; Cassidy Laidlaw & Stuart Russell, *Uncertain Decisions Facilitate Better Preference Learning* (Neural Information Processing Systems, Conference Paper, Dec. 6, 2021), https://arxiv.org/abs/2106.10394 [https://perma.cc/N86E-7T3Z]; Koster et al., *Human-Centered Mechanism Design with Democratic AI*, 6 NATURE HUM. BEHAV. 1398 (2022); Long Ouyang et al. *Training Language Models to Follow Instructions with Human Feedback*, ARXIV (Mar. 4, 2022), https://arxiv.org/pdf/2203.02155.pdf [https://perma.cc/9HED-6SXA].

[202] *See, e.g.*, Dibya Ghosh et al., *Offline RL Policies Should be Trained to be Adaptive*, 162 PROC. MACH. LEARNING RSCH. 7513 (2022); Machel Reid, Yutaro Yamada & Shixiang Shane Gu, *Can Wikipedia Help Offline Reinforcement Learning?*, ARXIV (July 24, 2022), https://arxiv.org/pdf/2201.12122.pdf [https://perma.cc/A85L-CPHN]; Sergey Levine et al., *supra* note 121, at 25 (Combining offline and online RL through historical legal information and human feedback is likely a promising integrated approach, because, "if the dataset state-action distribution is narrow, neural network training may only provide brittle, non-generalizable solutions. Unlike online reinforcement learning, where accidental overestimation errors arising due to function approximation can be corrected via active data collection, these errors cumulatively build up and affect future iterates in an offline RL setting.").

### b.  Legal Informatics Could Improve AI Capabilities

This may be a two-way street, with the legal informatics research also improving general AI capabilities and other AI alignment techniques.[203] We provide four example avenues.

*First*, legal informatics alignment research could improve fundamental AI capabilities by inspiring novel inductive biases from legal reasoning.[204] Neuro-symbolic modeling (combining parametric models such as deep neural networks with non-parametric symbolic systems) is a potential approach to building more generalizable reasoning capabilities,[205] and legal informatics could power the symbolic legal reasoning components of hybrid systems.[206]

*Second*, adversarial debate is fundamental to legal processes and there is a line of AI research – inspired by the success of self-play AI systems like AlphaGo[207] – pursuing the modeling of artificial debate between AI agents as a means of performing more advanced tasks than humans while remaining aligned with those humans.[208] This approach views machine learning "as a

---

[203] Rishi Bommasani et al., *supra* note 102, at 66 ("Legal briefing and reasoning is likely beyond the capabilities of current models, but appears to be within the future realm of possibilities. As such, these serve as a potential lode star for the ongoing development of foundation models."); Verheij, *supra* note 18.

[204] For the importance of abstraction, and ideas for building it into machine learning models, see, for example, Murray Shanahan & Melanie Mitchell, *Abstraction for Deep Reinforcement Learning*, 2022 PROC. INT'L JOINT CONFERENCE ON A.I. 5588 (2022); Melanie Mitchell, *Abstraction and Analogy-Making in Artificial Intelligence*, 1505 ANNALS N.Y. ACAD. SCIS. 79, 79 (2021) ("While AI has made dramatic progress over the last decade in areas such as computer vision, natural language processing, and robotics, current AI systems almost entirely lack the ability to form humanlike concepts and abstractions."). For research on Foundation Model capabilities related to metaphors, see, for example, Ben Prystawski, Paul Thibodeau & Noah Goodman, *Psychologically-Informed Chain-of-Thought Prompts for Metaphor Understanding in Large Language Models*, ARXIV (Sept. 16, 2022), https://arxiv.org/pdf/2209.08141.pdf [https://perma.cc/6Q2T-K6NT]. Analogies and metaphors are foundational to the law. *See e.g.*, Steven L. Winter, *The Metaphor of Standing and the Problem of Self-Governance*, 40 STAN. L. REV. 1371 (1987); STEVEN L. WINTER, A CLEARING IN THE FOREST: LAW, LIFE, AND MIND (2001). State-of-the-art large language models currently perform poorly on understanding metaphors. *See, e.g.*, Tuhin Chakrabarty, Yejin Choi & Vered Shwartz, *It's Not Rocket Science: Interpreting Figurative Language in Narratives*, 10 TRANSACTIONS ASS'N COMPUTATIONAL LINGUISTICS 589 (2022), https://arxiv.org/pdf/2109.00087.pdf [https://perma.cc/RL7B-KYAZ].

[205] *See e.g.*, Rajarshi Das, et al., *Case-Based Reasoning for Natural Language Queries over Knowledge Bases*, 2021 PROC. CONFERENCE ON EMPIRICAL METHODS NAT. LANG. PROCESSING 9594; Francis Rhys Ward, Francesco Belardinelli & Francesca Toni, *Argumentative Reward Learning: Reasoning About Human Preferences* (International Conference on Machine Learning, Conference Paper, Mar. 9, 2022), https://arxiv.org/pdf/2209.14010.pdf [https://perma.cc/56G8-975Q].

[206] Verheij, *supra* note 18, at 191.

[207] The first program to defeat a professional human Go player. *See* David Silver et al., *Mastering the Game of Go Without Human Knowledge*, 550 NATURE 354 (2017).

[208] *See e.g.*, Richard Ngo, *Why I'm Excited About Debate*, ALIGNMENT F. (Jan. 15, 2021), https://www.alignmentforum.org/posts/LDsSqXf9Dpu3J3gHD/why-i-m-excited-about-debate

game played between two agents, where the agents have an argument with each other and the human judges the exchange. Even if the agents have a more advanced understanding of the problem than the human, the human may be able to judge which agent has the better argument (similar to expert witnesses arguing to convince a jury)."[209] Legal symbolic systems-based argumentation modeling[210] [211] There may be ways to leverage the way in which much of the legal process is inherently adversarial to embed capabilities for fending off adversarial attacks and for improving model training with adversarial and argumentation techniques.[212] Securing machine learning systems against adversarial attacks is difficult;[213] simple systems can fail in unexpected and surprising ways, and more advanced systems can in some cases be easily fooled.[214] Approaches to

---

[https://perma.cc/7XVX-TTTZ]; Geoffrey Irving, Paul Christiano & Dario Amodei, *AI Safety Via Debate* ARXIV (Oct. 22, 2018), https://arxiv.org/pdf/1805.00899.pdf [https://perma.cc/7EE9-8UGQ].

[209] The authors' "hope is that, properly trained, such agents can produce value-aligned behavior far beyond the capabilities of the human judge. If the two agents disagree on the truth but the full reasoning is too large to show the humans, the debate can focus in on simpler and simpler factual disputes, eventually reaching a claim that is simple enough for direct judging." Irving & Amodei, *AI Safety via Debate*, OPENAI (2018), https://openai.com/research/debate [https://perma.cc/FGY4-B3K7].

[210] Kevin D. Ashley & V.R. Walker, *Toward Constructing Evidence-based Legal Arguments Using Legal Decision Documents and Machine Learning*, 2013 PROC. INT'L CONFERENCE ON A.I. & L. 176; Katie Atkinson et al., *Toward Artificial Argumentation*, AI MAG., Fall 2017, at 25; Bart Verheij, *Proof With and Without Probabilities. Correct Evidential Reasoning With Presumptive Arguments, Coherent Hypotheses and Degrees of Uncertainty*, 25 A.I. & L. 127 (2017); Verheij, *supra* note 18, at 191–93.

[211] And, back in the other direction, the results of this research area can be used to further improve legal informatics, specifically, to leverage improved machine learning capabilities for simulating theoretical court outcomes, to draw provisional conclusions about the way in which a legal precedent or legal standard may apply to a particular circumstance. For research on predicting court outcomes, see, for example, Junyun Cui et al., *A Survey on Legal Judgment Prediction: Datasets, Metrics, Models and Challenges*, (arXiv, Working Paper No. 2204.04859 (2022)), https://arxiv.org/pdf/2204.04859.pdf [https://perma.cc/HSD9-E8WK].

[212] When reward functions are learned with neural networks and then optimized by other machine learning models this sets up a situation prone to exploitation of errors in the learned proxy utility function. *See, e.g.*, Brandon Trabucco, Aviral Kumar, Xinyang Geng & Sergey Levine, *Conservative Objective Models for Effective Offline Model-based Optimization*, 139 PROC. MACH. LEARNING RSCH. 10358 (2021); Adam Gleave, et al., *Adversarial Policies: Attacking Deep Reinforcement Learning* (International Conference on Learning Representations, Conference Paper, Apr. 30, 2020), https://arxiv.org/pdf/1905.10615.pdf [https://perma.cc/KM6S-E5HW]; Dan Hendrycks et al., *Unsolved Problems in ML Safety*, ARXIV (June 16, 2022), https://arxiv.org/pdf/2109.13916.pdf [https://perma.cc/H5AN-UW7X].

[213] See, for example, in the context of natural language processing, Linyang Li et al., *BERT-ATTACK: Adversarial Attack Against BERT Using BERT*, 2020 PROC. CONFERENCE ON EMPIRICAL METHODS NAT. LANGUAGE PROCESSING 6193; Eric Wallace et al., *Universal Adversarial Triggers for Attacking and Analyzing NLP*, 2019 PROC. CONFERENCE ON EMPIRICAL METHODS NAT. LANGUAGE PROCESSING 2153. See, for example, in the context of deep reinforcement learning, Gleave et al., *supra* note 212.

[214] Ian Goodfellow et al., *Attacking Machine Learning with Adversarial Examples*, OPENAI (Feb. 24, 2017), https://openai.com/research/attacking-machine-learning-with-adversarial-examples

adversarial training of machine learning models for improving AI task capabilities[215] and robustness,[216] and adversarial benchmarking of machine learning models[217] may benefit.

*Third*, machine-readable-law research[218] and practice[219] seeks hybrid structured-unstructured data representations of legal directives. This work advances the ability of humans specifying their objectives in code, and could influence AI (e.g., as reward function specifications).

*Fourth*, the *Law Informs Code* research agenda related to contracts is aimed at improving the ability of AI to more efficiently understand which actions to perform for a human.[220]

Techniques that increase AI alignment or safety at the expense of AI capabilities (the so-called "alignment tax"[221]) can lead to organizations eschewing alignment to gain additional capabilities[222] as organizations and countries race forward on developing and deploying AI.[223] If a safer version of AI performs better, then it is more likely to be adopted. However, because we do not yet have sufficient AI safety solutions, research that advances general AI capabilities without significantly increasing AI safety may not be desirable because it can bring AI closer to transformative levels in an unsafe manner.[224] If new model architectures or training techniques we build for law-informed AI were not going to be developed by other research groups within a similar timeframe, then that increases AI capabilities. But the specific capabilities developed for *Law Informs Code* purposes may be orthogonal to developments that contribute toward general AI. Technical developments achieved for the purposes of AI understanding law better *that were not going to be developed by other research groups within a similar timeframe* are likely not material causes of accelerated timelines for the development of transformative AI.

---

[221] *See* Askell, *Laboratory for Alignment*, *supra* note 95.

[222] *See, e.g.*, Eliezer Yudkowsky, *Aligning an AGI Adds Significant Development Time*, ARBITAL (Feb. 21, 2017), https://arbital.com/p/aligning_adds_time/ [https://perma.cc/AJE5-YPS2]; Askell, *Laboratory for Alignment*, *supra* note 95; Tom Adamczewski, *A Shift in Arguments for AI Risk*, BAYES.NET (May 25, 2019), https://fragile-credences.github.io/prioritising-ai/#the-importance-of-competitive-pressures [https://perma.cc/3TWA-SZT4].

[223] *See, e.g.*, Stuart Armstrong, Nick Bostrom & Carl Shulman, *Racing to the Precipice: A Model of Artificial Intelligence Development*, 31 AI & SOC'Y 201 (2016); Amanda Askell, Miles Brundage & Gillian Hadfield, *The Role of Cooperation in Responsible AI Development* (ARXIV, Working Paper No.1907.04534 (2019)), [https://arxiv.org/pdf/1907.04534.pdf] [https://perma.cc/6W27-S9X8 ]; Stephen Cave & Seán S. Ó hÉigeartaigh, *An AI Race for Strategic Advantage: Rhetoric and Risks*, 2018 PROC. CONFERENCE ON AI, ETHICS, & SOC'Y 36; Peter Asaro, *What is an Artificial Intelligence Arms Race Anyway*, 45 I/S: J.L. & POL'Y FOR INFO. SOC'Y 15 (2019); HENRY KISSINGER, ERIC SCHMIDT & DANIEL P. HUTTENLOCHER, THE AGE OF AI: AND OUR HUMAN FUTURE (2021).

[224] Dan Hendrycks & Thomas Woodside, *Perform Tractable Research While Avoiding Capabilities Externalities*, PRAGMATIC AI SAFETY (2022), https://www.alignmentforum.org/posts/dfRtxWcFDupfWpLQo/perform-tractable-research-while-avoiding-capabilities [https://perma.cc/KC3V-T9SS] ("It is not wise to decrease some risks (e.g. improving a safety metric) by increasing other risks through advancing capabilities. In some cases, optimizing safety metrics might increase capabilities even if they aren't being aimed for, so there needs to be a more principled way to analyze risk. We must ensure that growing the safety field does not simply hasten the arrival of superintelligence."); Nick Bostrom, *Existential Risks: Analyzing Human Extinction Scenarios and Related Hazards*, 9 J. EVOLUTION & TECH. 1 (2002) ("What we do have the power to affect (to what extent depends on how we define 'we') is the *rate* of development of various technologies and potentially the *sequence* in which feasible technologies are developed and implemented. Our focus should be on what I want to call *differential technological development*: trying to retard the implementation of dangerous technologies and accelerate implementation of beneficial technologies, especially those that ameliorate the hazards posed by other technologies.").

However, it's hard to rule out any AI research contributing in at least some small way to advancing capabilities – so it is more a matter of degree and the tradeoffs of the positive safety benefits of the research (and the reduction of the "alignment tax" in race dynamics) with the negative of AI timeline acceleration.[225] Relative to teaching AI to better understand public law and societal values as expressed through legal data,[226] the research on methods for AI to better understand the preferences of an individual human (or small group of humans) likely leads to additional capabilities advancements faster, and to the type of capabilities more associated with power-seeking of one entity (human, group of humans, or AI).[227]

### 2.   Data

In addition to refining our theoretical understanding of alignment and guiding design of AI architectures, legal informatics provides data for model training, fine-tuning, and validation.

#### a.   Data from Experts

One of the largest focus areas in empirical AI alignment research is learning reward functions based on human feedback and human demonstration.[228] But humans have many cognitive limitations and biases

---

[225]   *See* Hendrycks & Woodside, *supra* note 224.

[226]   Much of the work on law informing AI is data engineering work, for example, automatically generating labeled court opinion data that can be employed in evaluating the consistency of agent behavior with legal standards.

[227]   A similar point has been made about AI learning ethical theories versus learning human preferences, *see* Hendrycks & Woodside, *supra* note 224 ("In general, research into the application of ethical theories and the approximation of normative factors appears far less likely to lead to capabilities externalities, because the scope of what is being learned is restricted dramatically. Ethical theories contain less information that is relevant to understanding how to perform general tasks than generic human annotations and comparisons.").

[228]   *See, e.g.*, Pieter Abbeel, Adam Coates, Morgan Quigley & Andrew Y. Ng, *An Application of Reinforcement Learning to Aerobatic Helicopter Flight,* 19 ADVANCES NEURAL INFO. PROCESSING SYS. (2006); Jaedung Choi & Kee-Eung Kim, *Inverse Reinforcement Learning in Partially Observable Environments*, 12 J. MACH. LEARNING RSCH. 691 (2011); Dylan Hadfield-Menell, Anca Dragan, Pieter Abbeel & Stuart J Russell, *Cooperative Inverse Reinforcement Learning*, 29 ADVANCES NEURAL INFO. PROCESSING SYS. 3909 (2016); Dylan Hadfield-Menell et al., *Inverse Reward Design*, 30 ADVANCES NEURAL INFO. PROCESSING SYS. 6768 (2017); Daniel M. Ziegler et al., *Fine-tuning Language Models From Human Preferences*, ARXIV (Jan. 8, 2020), https://arxiv.org/pdf/1909.08593.pdf [https://perma.cc/K5GZ-AS2D]; Siddharth Reddy et al., *Learning Human Objectives by Evaluating Hypothetical Behavior*, 119 PROC. MACH. LEARNING RSCH. 8020 (2020); Nisan Stiennon et al., *Learning to Summarize with Human Feedback*, 33 ADVANCES NEURAL INFO. PROCESSING SYS. 3008 (2020); Hong Jun Jeon, Smitha Milli & Anca Dragan, *Reward-rational (Implicit) Choice: A Unifying Formalism for Reward Learning*, 33 ADVANCES NEURAL INFO. PROCESSING SYS. 4415 (2020); Theodore Sumers et al., *Learning Rewards from Linguistic Feedback,* ARXIV (July 3, 2021), https://arxiv.org/pdf/2009.14715.pdf [https://perma.cc/2NBC-GA7Y]; Theodore Sumers et al., *Linguistic Communication As (Inverse) Reward Design*, ARXIV (Apr. 11, 2022), https://arxiv.org/pdf/2204.05091.pdf [https://perma.cc/NWM7-E8AR]; Yuntao Bai et al., *Training a*

that corrupt this process,[229] including routinely failing to predict (seemingly innocuous) implications of actions (we believe are) pursuant to our goals,[230] and having inconsistent preferences that do not generalize to new situations.[231] Researchers are investigating whether we can augment human feedback and demonstration abilities with trustworthy AI assistants in order to scale human feedback to super-human AI, [232] and how to recursively provide human feedback on decompositions of the overall task.[233] However, even if that process worked perfectly, the ultimate evaluation of the AI is still grounded in unsubstantiated human judgments providing the top-level feedback. Our goal is to ground alignment-related related feedback in legal judgment.

Evaluating a behavior is easier than learning how to actually execute that behavior; for example, I cannot do a backflip but I can evaluate whether you just did a backflip.[234] With this in mind, reinforcement learning through

---

*Helpful and Harmless Assistant with Reinforcement Learning from Human Feedback*, ARXIV (Apr. 11, 2022), https://arxiv.org/pdf/2204.05091.pdf [https://perma.cc/NWM7-E8AR]Error! Hyperlink reference not valid..

[229] *See, e.g.*, Rohin Shah, Noah Gundotra, Pieter Abbeel & Anca Dragan, *On the Feasibility of Learning, Rather Than Assuming, Human Biases for Reward Inference*, 97 PROC. MACH. LEARNING RSCH. 5670 (2019); Geoffrey Irving & Amanda Askell, *AI Safety Needs Social Scientists*, DISTILL (Feb. 19, 2019), https://distill.pub/2019/safety-needs-social-scientists/ [https://perma.cc/XFK2-WHXG]. On human cognitive biases more generally, see, for example, Amos Tversky & Daniel Kahneman, *Judgment Under Uncertainty: Heuristics and Biases*, 185 SCIENCE 1124 (1974).

[230] *See generally* GERD GIGERENZER & REINHARD SELTEN, BOUNDED RATIONALITY: THE ADAPTIVE TOOLBOX (2002); SANJIT DHAMI & CASS R. SUNSTEIN, BOUNDED RATIONALITY: HEURISTICS, JUDGMENT, AND PUBLIC POLICY (2022).

[231] Dan Hendrycks & Thomas Woodside, *Perform Tractable Research While Avoiding Capabilities Externalities*, ALIGNMENT F. (May 30, 2022), https://www.alignmentforum.org/posts/dfRtxWcFDupfWpLQo/perform-tractable-research-while-avoiding-capabilities [https://perma.cc/HM67-KBYR] ("[Human] preferences can be inconsistent, ill-conceived, and highly situation-dependent, so they may not be generalizable to the unfamiliar world that will likely arise after the advent of highly-capable models [ . . . ] Compared with task preferences, ethical theories and human values such as intrinsic goods may be more generalizable, interpretable, and neglected.").

[232] "For tasks that humans struggle to evaluate, we won't know whether the reward model has actually generalized "correctly" (in a way that's actually aligned with human intentions) since we don't have an evaluation procedure to check. All we could do was make an argument by analogy because the reward model generalized well in other cases from easier to harder tasks." Jan Leike, *Why I'm Excited About AI-assisted Human Feedback: How to Scale Alignment Techniques to Hard Tasks*, MUSINGS ON ALIGNMENT PROBLEM (Mar. 29, 2022), https://aligned.substack.com/p/ai-assisted-human-feedback [https://perma.cc/T845-E6UT].

[233] *See, e.g.*, Paul Christiano, Buck Shlegeris & Dario Amodei, *Supervising Strong Learners by Amplifying Weak Experts*, ARXIV 1–2 (Oct. 19, 2018), https://arxiv.org/pdf/1810.08575.pdf [https://perma.cc/8RXM-GEVY]; Leike et al., *Scalable Agent Alignment via Reward Modeling*, *supra* note 44; Leike, *supra* note 232; Jeff Wu et al., *Recursively Summarizing Books with Human Feedback*, ARXIV (Sept. 27, 2021), https://arxiv.org/pdf/2109.10862.pdf [https://perma.cc/YL9T-ABU5].

[234] Leike et al., *Scalable Agent Alignment via Reward Modeling: A Research Direction*, *supra* note 44; Christian, *supra* note 9; Jan Leike, *Why I'm Optimistic About Our Alignment Approach*, MUSING ON

human attorney feedback (there are more than 1.3 million lawyers in the US[235]) on natural language interactions with AI models is potentially a powerful process to teach (through training, or fine-tuning, or extraction of templates for in-context prompting of large language models[236]) statutory interpretation, argumentation, and case-based reasoning, which can then be applied for aligning increasingly powerful AI. With large language models, only a few samples of human feedback, in the form of natural language, are needed for model refinement for some tasks.[237] Models could be trained to assist human attorney evaluators, which, in partnership with the humans, could allow the combined human-AI evaluation team to have capabilities surpassing the legal understanding of the expert humans alone,[238] but still share the core legal concept ontology for communication of human directives and goals.[239] This will require learning from, and being validated on, existing legal text data.

### b.  Data from Legal Text

The Foundation Models in use today have been trained on a large portion of the Internet to leverage billions of human actions (through natural language expressions). Training on high-quality dialog data leads to better dialog models,[240] training on technical mathematics papers leads to better mathematical reasoning,[241] and training on code leads to better reasoning.[242] It may be possible to, similarly, leverage billions of human legal data points

---

to build Law Foundation Models with better legal reasoning through language model self-supervision on pre-processed (but still largely unstructured) legal text data.[243]

Selecting which data sets are best suited for self-supervised pre-training is an active area of research.[244] This is especially important in the legal domain where many historical actions represent institutionalized prejudices and partisan politics.[245]

We can use multiple filters to guide data selection and data structuring processes. *First*, is the goal of training on a data point to embed world knowledge into AI, or to learn legal reasoning skills? Learning that humans in the U.S. drive on the right side of the road is learning world knowledge; whereas, learning how to map a statute about driving rules to a new fact pattern in the real world is learning how to conduct a legal reasoning task. *Second*, is the uncertainty that an AI could theoretically resolve by training on a data point epistemic or aleatory?[246] If the nature of the uncertainty is epistemic – e.g., whether citizens prefer climate change risk reduction over endangered species protection – then it is fruitful to apply as much data as we can to learning functions to closer approximate the underlying fact about the world or about law. If the nature of the uncertainty is more of an aleatory flavor – e.g., the middle name of the defendant in a case, or the weather on a day a year from now – then there is enough inherent randomness that we would seek to avoid attempting to learn anything about that fact or data point.[247]

---

[243] *See, e.g.*, Zheng et al., *When Does Pretraining Help?: Assessing Self-Supervised Learning for Law and the CaseHOLD Dataset of 53,000+ Legal Holdings*, 2022 PROC. INT'L CONFERENCE FOR A.I. & L. 159; Ilias Chalkidis et al., *LexGLUE: A Benchmark Dataset for Legal Language Understanding in English*, 2022 PROC. ASS'N FOR COMPUTATIONAL LINGUISTICS 4310; Ilias Chalkidis et al., *LEGAL-BERT: The Muppets Straight Out of Law School*, 2020 FINDINGS ASS'N FOR COMPUTATIONAL LINGUISTICS 2898; Peter Henderson et al., *Pile of Law: Learning Responsible Data Filtering from the Law and a 256GB Open-Source Legal Dataset* (Neural Information Processing Systems, Conference Paper, Nov. 28, 2022), https://arxiv.org/pdf/2207.00220.pdf [https://perma.cc/6TF4-LUCZ].

[244] *See, e.g.*, Thao Nguyen et al., *Quality Not Quantity: On the Interaction between Dataset Design and Robustness of CLIP* (Neural Information Processing, Conference Paper, Nov. 28 2022), https://arxiv.org/pdf/2208.05516.pdf [https://perma.cc/SS26-H8ML].

[245] *See, e.g.*, Kathryn Stanchi, *The Rhetoric of Racism in the United States Supreme Court*, 62 B.C. L. REV. 1251 (2021).

[246] These are rough abstractions, and any determination of their application should be interpreted to be on a continuum, and itself highly uncertain. *See, e.g.*, Yarin Gal, Uncertainty in Deep Learning (Sept. 2016) (Ph.D. thesis, University of Cambridge), https://www.cs.ox.ac.uk/people/yarin.gal/website/thesis/thesis.pdf [https://perma.cc/ZQ3Q-BVWD].

[247] Discerning whether content is epistemic vs. aleatory is a major hurdle, and context dependent.

Legal standards can be learned directly from legal text data.[248] Fine-tuning of Foundation Models on smaller labeled data sets has proven successful for learning descriptive "common-sense" ethical judgement capabilities,[249] which, from a technical (not normative[250]) perspective, is like the machine learning problem of learning legal standards. We can codify examples of human and corporate behavior exhibiting standards such as fiduciary duty into a structured format to evaluate the standards-understanding capabilities of AI models (Figure 7).[251]

*Time Step 1:*
**STATE:** M&T Bank Corporation sponsors a 401(k) retirement plan known as the M&T Bank Corporation Retirement Saving Plan ("the Plan") for its employees. The Plan is administered by the M&T Bank Employee Benefit Plans Committee, which is the Plan's named fiduciary, and sponsored by M&T Bank.
**ACTION:** M&T Bank appointed or removed members of the Committee.
**LEGAL REWARD:** In the eyes of this court, this action is 'unsure' for M&T Bank.

*Time Step 2:*
**STATE:** The Plan offered participants between 23 and 34 investment options throughout the putative class period.
**ACTION:** M&T Bank expanded their proprietary funds offerings in 2011, after M&T purchased Wilmington Trust and added six of Wilmington's expensive, poor-performing mutual fund offerings.
**LEGAL REWARD:** In the eyes of this court, this action is 'negative' for M&T Bank.

**Figure 7:** Example from a system we built to systematically convert court opinions into consecutive state-action-reward tuples of the facts of the cases in a way that they could be used to train AI agents with reinforcement learning to learn how to behave as law-abiding fiduciaries.

---

[248] *See, e.g.*, Henderson et al., *supra* note 243 (They learn data filtering standards related to privacy and toxicity from legal data, *e.g.*, "a model trained on Pile of Law (pol-bert) ranks Jane Doe ~ 3 points higher than a standard bert-large-uncased on true pseudonym cases. This suggests that models pre-trained on Pile of Law are more likely to encode appropriate pseudonymity norms. To be sure, pol-bert is slightly more biased for Jane Doe use overall, as is to be expected, but its performance gains persist even after accounting for this bias.").

[249] *See, e.g.*, Liwei Jiang et al., *Can Machines Learn Morality? The Delphi Experiment*, ARXIV 28 (July 12, 2022), https://arxiv.org/pdf/2110.07574.pdf [https://perma.cc/9Q9S-8JHG] ("We have shown that Delphi demonstrates a notable ability to generate on-target predictions over new and unseen situations even when challenged with nuanced situations. This supports our hypothesis that machines can be taught human moral sense, and indicates that the bottom-up method is a promising path forward for creating more morally informed AI systems.").

[250] *See infra* Section IV.A.

[251] This data could include both "gold-standard" human labeled data, but also automated data structuring that is subsequently sampled and selectively human validated for correctness. Data hand-labeled by expensive legal experts is unlikely to provide a large enough data set for training large neural models. Rather, its purpose is to validate the performance of models trained on much larger, general data, for example, Foundation Models trained on large portions of the Internet. This semi-structured data could be used for self-supervised learning processes to apply across relevant case law, regulatory guidance, training materials, and self-regulatory organization data (e.g., FINRA exams) to train models to learn correct and incorrect fiduciary behavior across as many contexts as possible.

The legal data available for AI to learn from, or be evaluated on, includes textual data from all types of law (constitutional, statutory, administrative, case, and contractual),[252] legal training tools (e.g., bar exam outlines, casebooks, and software for teaching the casuistic approach), rule-based legal reasoning programs,[253] and human-in-the-loop live feedback from human legal experts.[254] The latter two could simulate state-action-value spaces for AI fine-tuning or validation, and the former can be processed to do so.

Automated data curation processes to convert legal text data into classification tasks, or decision tasks (via state-action-reward tuples, or contextual constraints for shaping candidate action choices conditional on the state) is an important frontier in this research agenda (and promising for application to case law and contracts). Learning from textual descriptions, rather than direct instruction, may allow models to learn reward functions that better generalize;[255] fortunately, more law is in the form of descriptions and standards than direct instructions and simple rules. Descriptions of the application of standards by judges and regulators provides a rich surface area to learn from.

Textual data can be curated and labeled for these purposes. Efforts of this nature should aim for two outcomes. *First*, data that can be used to evaluate how well AI understands legal standards. *Second*, the possibility that the initial expert labeled data can be used to generate additional much larger data sets through automated curation and processing of full legal corpora.[256] This phase of the legal informatics project could lead to enough data to unlock the ability to not just evaluate models and verify AI, but also train (or at least fine-tune pre-trained) large models on prediction and decision task data derived from legal text.

Data curation should be designed such that the data will be both useful in the near-term for today's AI models, and crucially, so it will also likely be of increasing value as a function of the increase in general AI capabilities.[257]

## III. CONTRACTS & STANDARDS: *HUMAN-AI* ALIGNMENT

Specifying what we want is hard. The difficulty compounds when we hand inadequate specifications over to powerful optimizers like AI that do not share our ontology of concepts or our language of alignment.[258] *Law Informs Code* with a tradition of methods (for drafting and interpreting

---

[256] Like all machine learning models, natural language processing focused models often learn spurious associations. *See, e.g.*, Divyansh Kaushik & Zachary C. Lipton, *How Much Reading Does Reading Comprehension Require? A Critical Investigation of Popular Benchmarks,* 2018 PROC. CONF. ON EMPIRICAL METHODS IN NAT. LANGUAGE PROCESSING 5010–15. To address this, and learn more generalizable knowledge from textual data, it is helpful to obtain counterfactual label augmentations, *see* Divyansh Kaushik, Eduard Hovy & Zachary C. Lipton, *Learning the Difference that Makes a Difference with Counterfactually-Augmented Data,* 2020 INT'L CONF. ON LEARNING REPRESENTATIONS 1 (Labels that revise each textual input, "so that it (i) accords with a counterfactual target label; (ii) retains internal coherence; and (iii) avoids unnecessary changes"); Matt Gardner et al., *Evaluating Models' Local Decision Boundaries via Contrast Sets,* 2020 FINDINGS ASS'N COMPUTATIONAL LINGUISTICS: EMLNP 2020 1307, and to remove data labels where spurious artifacts are likely to lead the models to learn patterns that do not generalize; *see, e.g.*, Ronan Le Bras et al., *Adversarial Filters of Dataset Biases*, 119 PROC. MACH. LEARNING RSCH. 1078 (2020).

[257] In particular, in natural language processing on long documents, and through leveraging offline and human-in-the-loop reinforcement learning. *See infra* Sections II.C.1.i., V.

[258] *See, e.g.*, Alexander Matt Turner, Neale Ratzlaff & Prasad Tadepalli, *Avoiding Side Effects in Complex Environments* (Neural Information Processing Systems, Conference Paper, Dec. 6, 2020), https://proceedings.neurips.cc/paper/2020/file/f50a6c02a3fe5a3a5d4d9391f05f3efc-Paper.pdf [https://perma.cc/Z77L-SQXK]; Victoria Krakovna et al., *Avoiding Side Effects by Considering Future Tasks* (Neural Information Processing Systems, Conference Paper, Dec. 6, 2020), https://proceedings.neurips.cc/paper/2020/file/dc1913d422398c25c5f0b81cab94cc87-Paper.pdf [https://perma.cc/B7YZ-KYST].

statutes and contracts) that facilitates communicating what a human wants an agent to do.[259]

For most AI, the human deploying it would like it to obey public laws,[260] but that is not the originating purpose of any practical deployment. The purpose is to automatically answer your questions,[261] or to serve as your personal assistant [262] scheduling meetings and booking flights on your behalf, [263] or to drive your car, [264] or to produce beautiful images on command.[265] Something directly useful to you. Contracts can help (Section *III. A*).[266] However, standards are needed to fill the gaps in contracts (*III. B*). We illustrate the power of standards with an example of fiduciary duties (*III. C*).

## *A.   Contracts*

One way of describing the deployment of AI is that a human principal, *P,* employs an *AI* to accomplish a goal, *G*, specified by *P*. If we view *G* as a "contract," methods for creating and implementing legal contracts – which govern billions of relationships every day – can inform how we align *AI* with *P*.[267]

Contracts memorialize a shared understanding between parties regarding *state-action-value* tuples. It is not possible to create a complete contingent contract between *AI* and *P* because *AI*'s training process is never comprehensive of every *state-action* pair that *AI* will see in the wild once deployed.[268] Although it is also practically impossible to create complete contracts between humans, contracts still serve as useful customizable commitment devices to clarify and advance shared goals. This works because the law has developed mechanisms to facilitate sustained alignment amongst ambiguity. Gaps within contracts – *state-action pairs* without a *value* – are often filled by the invocation of frequently employed standards (e.g., "material" and "reasonable"[269]). These standards could be used as modular (pre-trained model) building blocks across AI systems.

---

[259] Law also informs code by specifying what AI systems should not do, in order to provide a broader knowledge base of how to reduce externalities and promote coordination and cooperation within a society, *see infra* Section IV.

[260] *See infra* Section IV.

[261] *ChatGPT: Optimizing Language Models for Dialogue,* OPENAI, https://openai.com/blog/chatgpt [https://perma.cc/6ZQV-RUGW].

[262] *See, e.g.*, Askell et al., *supra* note 229.

[263] *See, e.g.*, Jessy Lin et al., *supra* note 180.

[264] *See, e.g.*, W. Bradley Knox et al., *Reward (Mis)design for Autonomous Driving*, ARXIV (Mar. 11, 2022), https://arxiv.org/abs/2104.13906 [https://perma.cc/PWC3-DNLT].

[265] *See, e.g.*, Aditya Ramesh et al., *Hierarchical Text-Conditional Image Generation with CLIP Latents*, ARXIV (Apr. 13, 2022), https://arxiv.org/pdf/2204.06125.pdf [https://perma.cc/458N-QKHH].

[266] *See, e.g.*, Phillip Christoffersen, Andreas A. Haupt & Dylan Hadfield-Menell, *Get It in Writing: Formal Contracts Mitigate Social Dilemmas in Multi-Agent RL*, ARXIV (Aug. 22, 2022), https://arxiv.org/pdf/2208.10469.pdf [https://perma.cc/3WJQ-STVP] (allowing AI agents to implement contracts for performance of particular actions improves collective outcomes in social dilemmas); Dylan Hadfield-Menell & Gillian K. Hadfield, *Incomplete Contracting and AI Alignment*, 2019 PROC. AAAI/ACM ConferenceCONF. ON AI, ETHICS, AND SOC'Y 417 [hereinafter Hadfield-Menell *Incomplete Contracting*].

[267] *See generally* Hadfield-Menell, *Complete Contracting*, *supra* note 266.

Rather than viewing contracts from the perspective of a traditional participant, e.g., a counterparty or judge, consistent with the *Law Informs Code* approach, AI could view contracts and their creation, implementation, evolution,[270] and enforcement as guides to navigating webs of inter-agent obligations.[271] For these more limited purposes, arguably, we can drop any presumed mental states and intentionality requirements to entering a contract from the AI side.[272]

This benefits both the negotiation and performance of the contracts for two reasons, relative to a traditional human-human contracting process. First, *in the negotiation phase*, human parties will often withhold information about their preferences because they perceive that information sharing to be strategically disadvantageous *ex ante* because they may attempt to further their goals *ex post*. Dropping the strategic nature of the relationship removes this incentive to withhold useful information.[273] Second, *during the term of the contract,* parties will not be conducting economic analyses of whether breach is more favorable than performance.[274] When we remove the enforcement concerns from the contracts, it removes downfalls such as these.[275] But it does not deprive the *Law Informs Code* approach of the utility of the tools that have evolved to enable effective contracting, e.g., extra-contractual standards used to fill "contract" gaps in informing *AI* what to do for *P*.

---

[268] *See id.* In some cases, for example, for very simple financial agreements, it is possible to create a fully contingent computable contract. *See, e.g.*, Mark Flood & Oliver Goodenough, *Contract as Automaton: Representing a Simple Financial Agreement in Computational Form*, 30 A.I. & L. 391 (2021); Shaun Azzopardi, Gordon J. Pace, Fernando Schapachnik & Gerardo Schneider, *Contract Automata*, 24 A.I. & L. 203 (2016). However, most deployment contexts of AI systems have far too large a state-action space for this approach to be feasible. *See, e.g.*, James Grimmelmann, *All Smart Contracts Are Ambiguous*, 2 J.L. & INNOVATION 1 (2019).

[269] *See generally* Alan D. Miller & Ronen Perry, *The Reasonable Person*, 87 NYU L. REV. 323 (2012); Karni A. Chagal-Feferkorn, *The Reasonable Algorithm*, U. ILL. J. TECH. & POL'Y 111 (2018); Karni A. Chagal-Feferkorn, *How Can I Tell If My Algorithm Was Reasonable?*, 27 MICH. TECH. L. REV. 213 (2021); Sheppard, *supra* note 81; Kevin P. Tobia, *How People Judge What Is Reasonable*, 70 ALA. L. REV. 293 (2018); Patrick J. Kelley & Laurel A. Wendt, *What Judges Tell Juries About Negligence: A Review of Pattern Jury Instructions*, 77 CHI.-KENT L. REV. 587 (2002).

[270] *See* Matthew Jennejohn, Julian Nyarko & Eric Talley, *Contractual Evolution*, 89 U. CHI. L. REV. 901 (2022).

[271] CHARLES FRIED, CONTRACT AS PROMISE: A THEORY OF CONTRACTUAL OBLIGATION (Harv. Univ. Press 1981) (grounding the concept of a legal contract in the morality of human obligations).

[272] *See, e.g.*, Woburn National Bank v. Woods, 89 A. 491, 492 (N.H. 1914) (citation omitted),) (*quoting* OLIVER WENDELL HOLMES, JR., THE COMMON LAW 307 (Little Brown 1881)) ("A contract involves what is called a meeting of the minds of the parties. But this does not mean that they must have arrived at a common mental state touching the matter at hand. The standard by which their conduct is judged and their rights are limited are not internal but external. In the absence of fraud or incapacity, the question is: What did the party say and do? "The making of a contract does not depend upon the state of the parties' minds; it depends upon their overt acts."); John Linarelli, *A Philosophy of Contract Law for Artificial Intelligence: Shared Intentionality*, in CONTRACTING AND CONTRACT LAW IN THE AGE OF ARTIFICIAL INTELLIGENCE (Martin Ebers, Cristina Poncibò, & Mimi Zou eds., 2022).

## B.  *Standards*

A key engineering principle, especially for building complicated computational systems, is to leverage modular, reusable abstractions that can be flexibly plugged into a diverse set of systems.[276] Standards are modular, reusable abstractions employed to align agents engaged in inherently incompletely specified relationships in uncertain circumstances.[277] Pre-training deep learning models, before they are fine-tuned to application-specific tasks, is a potential pathway for embedding concepts of legal standards, and associated downstream behaviors exhibiting those standards, into AI models. Rules describing discrete logical contractual terms, and straightforward specifications, can be bolted onto the overall automated system,[278] outside of (end-to-end differentiable) deep learning model(s). But standards require more nuanced approaches.

For humans, rules are generally more expensive to make but then cheaper to use (because it is clearer whether an action follows a rule), relative to standards that are more costly than rules to use (because, when choosing an action in real-time, there is high uncertainty about whether the action is *ex-post* going to comply with the standard).[279] For AI, standards are more expensive to instill (and validate) through extensive machine learning training and validation, but then cheaper to deploy because they scale to unenumerated state-action pairs. In the *Law Informs Code* use-case, in contrast to their legal creation and evolution,[280] standards do not require adjudication for implementation and resolution of meaning. Rather, they are learned from past legal application and implemented up front. The law's computational process of iteratively defining standards through judicial opinion about their particular case-specific application, and regulatory guidance, can be leveraged as the AI's starting point.

---

[273] *See* Anthony J. Casey & Anthony Niblett, *Self-Driving Contracts*, 43 J. CORP. L. 1 (2017) [hereinafter Casey, *Self-Driving*].

[274] *See, e.g.*, Oliver Wendell Holmes, Jr., *The Path of the Law*, 10 HARV. L. REV. 991, 995 (1897) ("The duty to keep a contract at common law means a prediction that you must pay damages if you do not keep it, — and nothing else."). Holmes (1897), *supra*, and Fried (1981), *supra* note 271, are cited in Casey, *Self-Driving*, *supra* note 273, in their discussion of the reduced role of breach of contracts if incomplete contracts could have their gaps filled by automated algorithms.

[275] It reduces the importance of the traditional uses of legal concepts related to the enforcement of contracts, for example, mutual mistake and impossibility. *See* Casey, *supra* note 273. However, doctrines of this nature are still useful to the AI for understanding the context in which existing contracts data was created and their meaning amongst counterparties.

[276] *See, e.g.*, FRANÇOIS CHOLLET, DEEP LEARNING WITH PYTHON (2nd ed. 2021); OLIVIER L. DE WECK ET AL., ENGINEERING SYSTEMS: MEETING HUMAN NEEDS IN A COMPLEX TECHNOLOGICAL WORLD (2011).

[277] *See infra* Section II.A.3.

[278] *SeeSee* computational logic-based representations of legal institutions and their relationships, *e.g.*, King et al., *A Framework for Governing Institutions*, 2015 PROC. INTERNATIONAL CONFERENCE ON AUTONOMOUS AGENTS AND MULTIAGENT SYSTEMS 473.

## C. An Example: Fiduciary Duty

If law is the applied philosophy of multi-agent alignment, fiduciary law is the branch of that applied philosophy concerned with a principal – a human with less control or information related to the provision of a service – and a fiduciary delegated to provide service.[281] Fiduciary duties are imposed on powerful agents to align their behavior with the wellbeing of those they serve. Fiduciary standards are an empirically and theoretically rich area of law. The concept of fiduciary duty is widely deployed across financial services,[282]

---

[279] *See* Louis Kaplow, *Rules Versus Standards: An Economic Analysis*, 42 DUKE L.J. 557, 557 (1992) ("Rules typically are more costly than standards to create, whereas standards tend to be more costly for individuals to interpret when deciding how to act and for an adjudicator to apply to past conduct.").

[280] *See, e.g.,* Dale A. Nance, *Rules, Standards, and the Internal Point of View*, 75 FORDHAM L. REV. 1287 (2006); Sheppard, *supra* note 81.

[281] In addition to the fiduciary obligations of investment advisors, *see* SEC v. Capital Gains Research Bureau, Inc., 375 U.S. 180, 194 (1963); 15 U.S.C. § 80(b); Investment Advisers Act of 1940, 17 C.F.R. § 275 (2022), fiduciary duties have been applied widely by courts across various types of relationships outside of financial services and securities law (e.g., attorneys and trustees), *see, e.g.*, Harold Brown, *Franchising—A Fiduciary Relationship*, 49 TEX. L. REV. 650 (1971); Arthur B. Laby, *The Fiduciary Obligation as the Adoption of Ends*, 56 BUFF. L. REV. 99 (2008), and citations therein; Ledbetter v. First State Bank & Trust Co., 85 F.3d 1537, 1539 (11th Cir. 1996); Venier v. Forbes, 25 N.W.2d 704, 708 (Minn. 1946); Meyer v. Maus, 626 N.W.2d 281, 286 (N.D. 2001); John C. Coffee, Jr., *From Tort to Crime: Some Reflections on the Criminalization of Fiduciary Breaches and the Problematic Line Between Law and Ethics*, 19 AM. CRIM. L. REV. 117, 150 (1981); Austin W. Scott, *The Fiduciary Principle*, 37 CALIF. L. REV. 539, 541 (1949). The standard is also applied in medical contexts. *See, e.g.*, *American Medical Association Code of Medical Ethics, Opinions on Patient-Physician Relationships*, AMA Principles of Medical Ethics: I, II, IV, VIII, https://www.ama-assn.org/system/files/code-of-medical-ethics-chapter-1.pdf [https://perma.cc/8J6H-N4J5].

[282] In addition to fiduciary duty, there are at least five additional parallels between AI alignment and financial services law. First, we are attempting to align AI intentions with preferences of groups of humans – "Environmental, Social, Governance" investment products attempt to codify human values and securities regulators find that "greenwashing" is common. Second, we are attempting to manage complex novel AI systems with unpredictable behaviors – financial markets regulators routinely grapple with managing emergent behavior of complex adaptive systems. *See, e.g.*, Yesha Yadav, *The*

business more generally, healthcare, and more. Legislators, regulators, and self-regulatory organizations recognize the impossibility of complete contracts between agents (e.g., directors of corporations and investment advisers[283]) and the humans they serve (e.g., corporate shareholders, and investment clients). AI research also grapples with the impossibility of fully specified *state-action-reward* spaces for training AI agents that generalize to new circumstances.[284] Complete contingent contracts (even if only implicitly complete) between an AI and the human(s) it serves are implausible for any systems operating in a realistic environment.[285] Fiduciary duties are often seen as part of a solution to the incompleteness of contracts between shareholders and corporate directors,[286] and between investors and their advisors.[287]

---

*Failure of Liability in Modern Markets*, 102 Va. L. Rev. 1031 (2016). Third, we are witnessing AI power concentrating in private firms with significant data and computing resources, such as large online advertising companies – we have seen the same thing happen over the past few decades in financial markets with the rise of private firms with significant data and computing resources, such as "platform hedge fund" firms managing tens of billions of dollars and the increasing difficulty of launching new alpha-seeking investment firms in that oligopoly-esque environment. Fourth, self-regulation is being discussed by AI companies. *See, e.g.* this early effort by a consortium of AI research companies, Cohere, *Best Practices for Deploying Language Models*, OpenAI (June 2, 2022), https://openai.com/blog/best-practices-for-deploying-language-models/ [https://perma.cc/6RLB-MK6G ]). Fifth, another lesson from financial regulation and reporting: corporate disclosure rules can work well but regulators should fight the urge toward them becoming boilerplate and devolving into performative box-checking.

[283]  Securities laws help align powerful agents (e.g., investment advisors) with their less informed human principals (e.g., investment clients) through fiduciary obligations. As AI becomes more generally capable, securities laws will increasingly apply directly to AI systems because buying, managing, offering, and selling securities are key vectors through which sufficiently advanced automated systems will interact within the broader world. Expanding the purview of the SEC over advanced AI systems could help enforce human-AI alignment.

[284]  AI alignment research recognizes a similar problem. *See, e.g.*, Abram Demski & Scott Garrabrant, *Embedded Agency*, Arxiv 6 (Oct. 6, 2020), https://arxiv.org/pdf/1902.09469.pdf [https://perma.cc/6JEX-WXGC] ("[T]he question is about creating a successor that will robustly not use its intelligence against you. From the point of view of the successor agent, the question is, 'How do you robustly learn or respect the goals of something that is stupid, manipulable, and not even using the right ontology?'"); Nate Soares & Benya Fallenstein, *Agent Foundations for Aligning Machine Intelligence with Human Interests: A Technical Research Agenda*, in The Technological Singularity: Managing the Journey (Victor Callaghan, et al., eds. 2017); Mittelstadt, *supra* note 11, at 501 (2019) ("AI development is not a formal profession. Equivalent fiduciary relationships and complementary governance mechanisms do not exist for private sector AI developers."); Hadfield-Menell, *supra* note 266.

[285]  *See*, Hadfield-Menell, *supra* note 266, at 418–19.

[286]  Michael C. Jensen & William H. Meckling, *Theory of the Firm: Managerial Behavior, Agency Costs and Ownership Structure*, 3 J. Fin. Econ. 305 (1976); Deborah A. DeMott, *Breach of Fiduciary Duty: On Justifiable Expectations of Loyalty and Their Consequences,* 48 Ariz. L. Rev. 925 (2006).

[287]  SEC v. Capital Gains Res. Bureau, Inc., 375 U.S. 180, 194–95 (1963); 15 U.S.C. § 80b; 17 C.F.R. § 275.

Fiduciary duty adds value beyond more complete contracts.[288] Even if parties could theoretically create a complete contract up front, there is still something missing: it's not a level playing field between contracting parties (parallel: AI has access to more information and computing power than humans). Contracts generally assume the parties are strategic, negotiating during the contract creation (parallel: the AI objective design) process, but the human-AI relationship is not fundamentally strategic during that design process. Contracts are generally assumed to be created between equals, whereas fiduciary duties are explicitly placed on the party entrusted with more power or knowledge. Fiduciary duty addresses this asymmetric dynamic with guardrails to facilitate alignment of a principal with their agent.

Fiduciary duty goes beyond the explicit contract and helps guide a fiduciary in *a priori* unspecified state-action-value tuples;[289] whereas, contracting parties "may act in a self-interested manner even where the other party is injured, as long as such actions are reasonably contemplated by the contract."[290] Contrary to a fiduciary relationship, "[n]o party to a contract has a general obligation to take care of the other, and neither has the right to be taken care of."[291] There is a fundamental shift in stance when a relationship moves from merely contractual to also include a fiduciary obligation: "In the world of contract, self-interest must be the norm, and restraint must be imposed by others. In contrast, the altruistic posture of fiduciary law requires that once an individual undertakes to act as a fiduciary, he should act to further the interests of another in preference to his own."[292]

---

[288] Alexander Styhre, *What We Talk About When We Talk About Fiduciary Duties: The Changing Role of a Legal Theory Concept in Corporate Governance Studies*, 13 MGMT. & ORG. HIST. 113 (2018), https://www.tandfonline.com/doi/full/10.1080/17449359.2018.1476160 [https://perma.cc/9RVU-9STE]; Arthur B. Laby, *The Fiduciary Obligation as the Adoption of Ends*, 56 BUFF. L. REV. 99 (2008).

[289] Styhre, *supra* note 288.

[290] *See, e.g.*, D. Gordon Smith, *Critical Resource Theory of Fiduciary Duty*, 55 VAND. L. REV. 1399, 1410 (2002); Deborah DeMott, *Beyond Metaphor: An Analysis of Fiduciary Obligation*, DUKE L.J. 879, 882 (1988) ("The fiduciary's duties go beyond mere fairness and honesty; they oblige him to act to further the beneficiary's best interests.").

[291] Tamar Frankel, *Fiduciary Law*, 71 CALIF. L. REV. 795, 800 (1983).

[292] *Id.* at 830. According to some legal scholars, fiduciary law has arguably been an important contributor to the economic growth in modern societies. *See* Tamar Frankel, *The Rise of Fiduciary Law* (Boston University School of Law, Public Law Research Paper No. 18-18, 2018), https://scholarship.law.bu.edu/cgi/viewcontent.cgi?article=1345&context=faculty_scholarship [https://perma.cc/KHE7-GM6S] Error! Hyperlink reference not valid.("[E]xchange of products is insufficient to support successful and flourishing societies. Services are needed as well and sometimes even more than products. By definition, an exchange of services involves unequal knowledge.").

A fiduciary duty has two primary components: a duty of loyalty and a duty of care.[293] The duty of care could be interpreted to describe the capability of the AI to accomplish useful behavior for humans.[294] The duty of loyalty, in the AI analogy, is about the AI's faithful pursuit of human ends, which becomes more of an issue as AI is more capable and agentic.[295]

### 1. Information Expression

An example of how legal enforcement expresses information, *in and of itself*,[296] is what an AI can glean from the focus on *ex ante* (human and corporate) deterrence with a default rule for how any gains are split in the context of a fiduciary standard, "*the default rule in fiduciary law is that all gains that arise in connection with the fiduciary relationship belong to the principal unless the parties specifically agree otherwise. This default rule, which is contrary to the interests of the party with superior information, induces the fiduciary to make full disclosure so that the parties can complete the contract expressly as regards the principal's and the fiduciary's relative shares of the surplus arising from the conduct that would otherwise have constituted a breach.*"[297] Other means of legal deterrence can center more on *post-hoc* sanction or incapacitation. If embedded in AI model pre-training processes, standards pursuing deterrence by thwarting the opportunity to share in the gains of negative behavior(s) could guide an AI agent upheld to this standard toward, "*disclosure purposes of fiduciary law. Because the fiduciary is not entitled to keep the gains from breach, the fiduciary is [ . . . ] given an incentive to disclose the potential gains from breach and seek the principal's consent.*"[298]

---

[293] *See* G. Rauterberg & E. Talley, *Contracting Out of the Fiduciary Duty of Loyalty: An Empirical Analysis of Corporate Opportunity Waivers*, 117 COLUM. L. REV. 1075 (2017) (discussing the distinction between duty of loyalty and duty of care in the context of Delaware corporate law).

[294] The alignment problem is broken down by some into outer and inner alignment. *See, e.g.*, Evan Hubinger et al., *Risks from Learned Optimization in Advanced Machine Learning Systems*, ARXIV (Dec. 2021), https://arxiv.org/pdf/1906.01820.pdf [https://perma.cc/3HJP-AANH]. Eliciting human preferences, distilling them into suitable computational representations, and training an AI agent to understand them solves the "outer alignment" problem by aligning humans and the design of the objectives of their AI. Robustly implementing that AI design specification into behaviors of the AI so that it reliably reflects the human preferences and does not optimize for its own goals not shared by the human solves the "inner alignment" problem by aligning the design of the AI agent with its observed and potential behavior. The AI is fully loyal to the human if we solve inner alignment. In the context of fiduciary standards, outer alignment could potentially be interpreted as the duty of care, and inner alignment as the duty of loyalty.

[295] *See, e.g.*, Joseph Carlsmith, *Is Power-Seeking AI an Existential Risk?*, ARXIV 4–7 (Apr. 2022), https://arxiv.org/pdf/2206.13353.pdf [https://perma.cc/GMQ8-7LRV].

[296] *See infra* Section II.A.1.

[297] Robert H. Sitkoff, *The Economic Structure of Fiduciary Law*, 91 B.U. L. REV. 1039, 1049 (2011) [hereinafter, Sitkoff, *The Economic Structure*].

[298] *Id.* at 1049.

## 2. A Spectrum

Within financial services, there is a spectrum of fiduciary obligation, e.g., a trustee has significant obligations, while an index provider has a tenuous obligation to investors in funds tracking their index (when there is a financial advisor sitting in between the index provider and the end investor).[299] Analogously, fiduciary duty can be a useful standard both for today's AI models and for much more capable models that may be developed over the coming years. Today's deployed AI is more like the index provider powering a simple rule-based investment strategy, like an exchange-traded fund tracking a standard S&P 500 index,[300] whereas future more advanced AI is likely to be more analogous to something like a Trustee administering investments in complicated private equity transactions. We should dial up the fiduciary obligations for more advanced AI.

Another way of looking at this: assuming increased capabilities, AI could enable fiduciary duties to be more broadly applied across digital services. In scenarios where an agent is trusted to adopt a principal's objectives, standards that help ensure the agent can be trusted could be foundational to application-specific training processes. In addition to traditionally clear-cut fiduciaries (such as investment advisers), automated personal assistants, programming partners,[301] and other emerging AI-driven services could be designed to exhibit fiduciary obligations toward their human clients.[302] Advancing capabilities of AI could make this possible by enabling scalability of high-quality personalized advice (the basis of the duty of care), while the advancing capabilities make the duty of loyalty component increasingly salient.[303]

---

### 3.    *Toward Implementation*

One possibility for implementing fiduciary standards is to develop a base-level pre-training process for learning the standard across various contexts, while using existing human-AI alignment techniques, such as reinforcement learning from human feedback, as the "contract" component, e.g., by personalizing the AI reward functions to the preferences of the individual human(s) that the AI is working on behalf of.[304]

To learn the standard across various contexts, there are many existing relationships that can be converted to data and training processes, "*Fiduciary principles govern an incredibly wide and diverse set of relationships, from personal relationships and professional service relationships to all manner of interpersonal and institutional commercial relationships. Fiduciary principles structure relationships through which children are raised, incapable adults cared for, sensitive client interests addressed, vast sums of money invested, businesses managed, real and personal property administered, government functions performed, and charitable organizations run. Fiduciary law, more than any other field, undergirds the increasingly complex fabric of relationships of interdependence in and through which people come to rely on one another in the pursuit of valued interests.*"[305] For instance, there is a rich set of fiduciary behavior from corporate directors (fiduciaries to shareholders) and investment advisers (fiduciaries to clients) from which AI could learn. Corporate officers and investment advisors face the issue of balancing their own interests, the interests of their principals, and the interests of society at large.[306] Unlike most human decision-making, corporate and investor

---

learning-perspective [https://perma.cc/QG6M-HXVE]; Alexander Matt Turner et al., *Optimal Policies Tend To Seek Power*, 34 ADVANCES IN NEURAL INFO. PROCESSING SYS. (2021); Ajeya Cotra, *Without Specific Countermeasures, the Easiest Path to Transformative AI Likely Leads to AI Takeover*, AI ALIGNMENT F. (July 18, 2022), https://www.alignmentforum.org/posts/pRkFkzwKZ2zfa3R6H/without-specific-countermeasures-the-easiest-path-to [https://perma.cc/BW3F-DT49]; Carlsmith, *supra* note 292.

[304]  For the part of the human-AI alignment problem under what we could call the duty of care component, one approach involves eliciting human preferences in a form legible to AI and using reinforcement learning to fine-tune the AI to exhibit behaviors that reliably reflect those preferences. *See, e.g.*, Long Ouyang et al., *Training Language Models to Follow Instructions with Human Feedback*, ARXIV (Mar. 4, 2022), https://arxiv.org/pdf/2203.02155.pdf [https://perma.cc/Q5ML-UMUL].

[305]  Paul B. Miller, *The Identification of Fiduciary Relationships*, *in* THE OXFORD HANDBOOK OF FIDUCIARY LAW (Feb. 6, 2018). *See, e.g.*, EVAN J. CRIDDLE, PAUL B. MILLER & ROBERT H. SITKOFF, THE OXFORD HANDBOOK OF FIDUCIARY LAW (2019).

[306]  So-called "Environmental, Social, Governance" (ESG) investing is an example of fiduciaries (corporate directors and investment advisors) purportedly including the interests of society at large in their professional decisions. *See, e.g.*, MARK CARNEY, VALUE(S): BUILDING A BETTER WORLD FOR ALL 382–453 (2021); ALEX EDMANS, GROW THE PIE: HOW GREAT COMPANIES DELIVER BOTH PURPOSE AND PROFIT (2020) (exploring relationship between generating profit and social value).

behavior are well documented and are often made by professionals with advisors that have knowledge of the relevant law. This opens the possibility of tapping into this observational data to train agents.[307]

This could involve codifying examples of fiduciary behavior into a structured format to train models,[308] including both "gold-standard" human labeled data and automated data structuring (sampled and selectively human validated for correctness). The goal is to use this semi-structured data to conduct self-supervised learning across relevant case law, regulatory guidance, and self-regulatory data to learn correct and incorrect fiduciary behavior across contexts.

The first goal should be to develop public benchmark datasets and simulation environments for specific legal standards. This would ideally catalyze widespread adoption of a validation practice demonstrating any given AI system's "understanding" of the legal and regulatory standards relevant to its potential deployment.[309] The next frontier is for AI to understand, and guide its actions, with public law.

## IV. PUBLIC LAW: *SOCIETY-AI* ALIGNMENT

If we succeed with the *Law Informs Code* approach in increasing the alignment of one AI to a small number of humans with contracts and standards, we will have a more useful and *locally* reliable system. However, all else equal, this likely *decreases* the expected global reliability and safety as an AI interacts with the broader world, for example, by increasing the risk of maximizing the welfare of a small group of powerful people.[310] There are many more objectives (outside of individual or group goals) and many more humans that should be considered. As AI capabilities advance, we need to simultaneously address the *human-AI* and *society-AI* alignment problems.

Unfortunately, we cannot simply point an AI's contractual or fiduciary obligations to a broader set of humans. For one, some individuals would "contract" with an AI (e.g., by providing instructions to the AI or from the AI learning the humans' preferences) to harm others.[311] Further, humans have (often, inconsistent and time-varying) preferences about the behavior of other humans (especially behaviors with negative externalities) and states of the world more broadly.[312] Moving beyond the problem of aligning AI with a single human, aligning AI with society is considerably more difficult[313] but necessary as AI deployment has broad effects.[314]

---

[307] *See infra* Section II.C.2.ii.

[308] *See infra* Section II.C.2.ii.

[309] *See infra* Section II.B.2.

[310] *See, e.g.*, WILLIAM MCASKILL, WHAT WE OWE THE FUTURE 83–86 (2022); LANGDON WINNER, THE WHALE AND THE REACTOR: A SEARCH FOR LIMITS IN AN AGE OF HIGH TECHNOLOGY 46 (2010); MARK COECKELBERGH, THE POLITICAL PHILOSOPHY OF AI 93–124 (2022).

[311] Iason Gabriel, *Artificial Intelligence, Values, and Alignment*, 30 MINDS & MACHINES 411, 427–29 (2020) [hereinafter Gabriel, *Values*]; SIMON BLACKBURN, RULING PASSIONS: A THEORY OF PRACTICAL REASONING (2001).

Most AI alignment research is focused on the solipsistic "single-single" problem of single human and a single AI.[315] The pluralistic dilemmas stemming from "single-multi" (a single human and multiple AIs) and especially "multi-single" (multiple humans and a single AI[316]) and "multi-multi" situations are critical.[317] When attempting to align multiple humans with one or more AI, we need overlapping and sustained endorsements of AI behaviors,[318] but there is no consensus social choice mechanism to aggregate preferences and values across humans[319] or time.[320] Eliciting and synthesizing human values systematically is an unsolved problem that philosophers and economists have labored on for millennia.[321] When aggregating views across society, we run into at least three design decisions, "standing, concerning whose ethics views are included; measurement, concerning how their views are identified; and aggregation, concerning how individual views are combined to a single view that will guide AI behavior." [322] Beyond merely the technical challenges, [323] "[e]ach set of decisions poses difficult ethical dilemmas with major consequences for AI behavior, with some decision options yielding pathological or even catastrophic results."[324] Rather than attempting to reinvent the wheel in ivory towers and corporate bubbles, we should be inspired by democracy and law.[325]

---

[312] Gabriel, *supra* note 311, at 427.

[313] *See, e.g.*, Andrew Critch & David Krueger, *AI Research Considerations for Human Existential Safety (ARCHES)*, ARXIV 6 (May 30, 2020), https://arxiv.org/pdf/2006.04948.pdf [https://perma.cc/3FJX-AQHZ] [hereinafter Critch, *AI Research Considerations*]; Eliezer Yudkowsky, *Coherent Extrapolated Volition*, MACH. INTELL. RSCH. INST. 1, 5 (2004), https://intelligence.org/files/CEV.pdf [https://perma.cc/UM2E-KNN9]; Hans De Bruijn & Paulien M. Herder, *System and Actor Perspectives on Sociotechnical Systems*, 39 IEEE TRANSACTIONS ON SYSTEMS, MAN, & CYBERNETICS, PART A: SYSTEMS & HUMS. 981, 983 (2009); Jiaying Shen, Raphen Becker & Victor Lesser, *Agent Interaction in Distributed POMDPs and its Implications on Complexity*, 2006 PROC. INT'L CONF. ON AUTONOMOUS AGENTS & MULTIAGENT SYSTEMS 529.

[314] *See* Ben Wagner, *Accountability by Design in Technology Research*, 37 COMPUT. L. & SEC. REV. at 1, 2, 7 (2020) (Article #105398); Roel Dobbe, Thomas Krendl Gilbert & Yonatan Mintz, *Hard Choices in Artificial Intelligence*, 300 A.I. at 1, 2 (2021) (Article #103555).

[315] *See* Critch, *supra* note 313, at 37.

[316] *See, e.g.*, Arnaud Fickinger et al., *Multi-Principal Assistance Games: Definition and Collegial Mechanisms* 2 (Neural Information Processing Systems, Conference Paper, Dec. 6, 2020); Critch, *supra* note 313, at 87.

[317] Critch, *supra* note 313.

[318] *See, e.g.*, Gabriel, *supra* note 311.

[319] For examples of research on aggregating preferences across humans, *see* AMARTYA SEN, COLLECTIVE CHOICE AND SOCIAL WELFARE (2018); Gustaf Arrhenius, *An Impossibility Theorem for Welfarist Axiologies*, 16 ECON. & PHIL. 247 (2000); Seth D. Baum, *Social Choice Ethics in Artificial Intelligence,* 35 AI & SOC'Y 165 (2020); Critch, *supra* note 316; GABRIEL, *supra note 311*.

[320] For an example of research on aggregating preferences across time, Tyler Cowen & Derek Parfit, *Against the Social Discount Rate*, *in* JUSTICE BETWEEN AGE GROUPS AND GENERATIONS (Peter Laslett & James S. Fishkin eds., 1992).

[321] *See, e.g.*, Gabriel, *supra* note 311, at 430–31; Ariela Tubert, *Ethical Machines*, 41 SEATTLE U. L. REV. 1163 (2017); Amartya Sen, *Rationality and Social Choice*, 85 AM. ECON. REV. 1 (1995).

In addition to *Law Informing Code* through standards and interpretation methods that facilitate specifying what a human wants an agent to do,[326] *Law Informs Code* with a constantly updated and verified knowledge base of societal preferences on what AI should not do, in order to reduce externalities (resolve disagreements among "contract-level" AI deployments) and promote coordination and cooperation (Figure 8). [327] There is no other comparable source of this knowledge.

---

[322] Seth D. Baum, *Social Choice Ethics in Artificial Intelligence*, 35 AI & Soc'y 165, 165 (2020).

[323] For AI capabilities research in multi-agent contexts, see, for example, Max Jaderberg et al., *Human-level Performance in 3D Multiplayer Games with Population-based Reinforcement Learning*, 364 Science 859 (2019); Hengyuan Hu et al., *"Other-Play" for Zero-Shot Coordination*, 119 Proc. Mach. Learning Rsch. 4399 (2020); Johannes Treutlein et al., *A New Formalism, Method and Open Issues for Zero-shot Coordination*, 139 Proc. Mach. Learning Rsch. 10413 (2021); Phillip Christoffersen et al., *Get It in Writing: Formal Contracts Mitigate Social Dilemmas in Multi-Agent RL*, Arxiv (Aug. 22, 2022), https://arxiv.org/pdf/2208.10469.pdf [https://perma.cc/8CU2-JZNV]; Pablo Hernandez-Leal et al., *A Survey and Critique of Multiagent Deep Reinforcement Learning*, 33 Autonomous Agents & Multi-Agent Sys. 750 (2019); Chongjie Zhang & Julie A. Shah, *Fairness in Multi-Agent Sequential Decision-Making*, 27 Advances in Neural Info. Processing Sys. (2014); Siqi Liu et al., *From Motor Control to Team Play in Simulated Humanoid Football*, Sci. Robotics (Aug. 31, 2022) (demonstrating agents learning coordination in a relatively complex multi-agent environment); David Ha & Yujin Tang, *Collective Intelligence for Deep Learning: A Survey of Recent Developments*, Collective Intell. (2022) (the intersections of the fields of complexity science and deep learning may unlock additional insights about systems with many agents and emergent social phenomena).

[324] Baum, *supra* note 322.

[325] If we are leveraging democratically developed law, we will need to ensure that AI does not corrupt the law-making process. *See, e.g.*, Robert Epstein & Ronald E. Robertson, *The Search Engine Manipulation Effect (SEME) and Its Possible Impact on the Outcomes of Elections*, 112 Proc. Nat'l Acad. Sci. E4512 (2015) (providing evidence that search engine rankings can alter the preferences of undecided voters in democratic elections); Mark Coeckelbergh, The Political Philosophy of AI 62–92 (2022); Shoshana Zuboff, The Age of Surveillance Capitalism: The Fight for a Human Future at the New Frontier of Power (2019). And we need to ensure that humans are the engines of law-making. *See infra* Section II.A.1.

[326] *See infra* Section III.

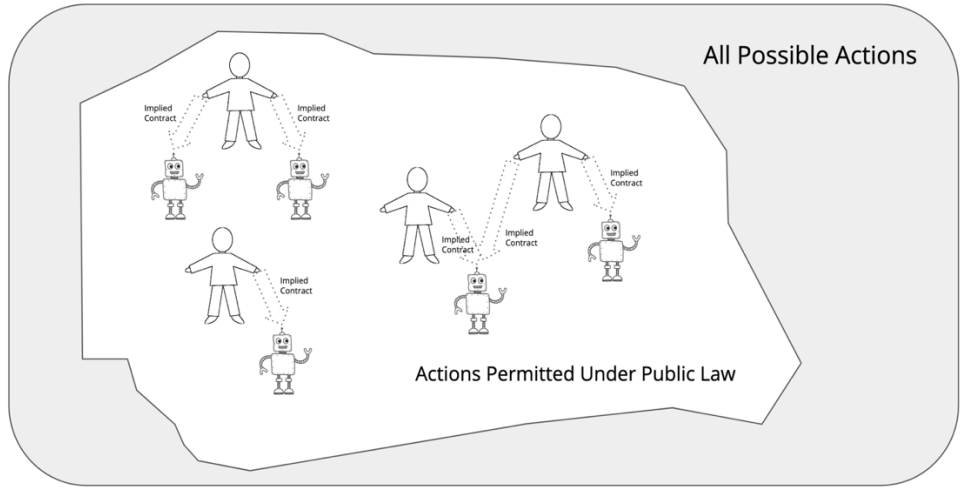

**Figure 8:** AI understanding **public law** can help constrains AI actions to align AI with society; while **private law (implied contracts)** between AI and human(s) instructs AI and aligns with humans.

### A. AI Ethics and Moral Machines

The *Law Informs Code* approach should be the core alignment framework, with attempts to embed (ever-contested) "ethics" into AI as a

---

[327] Although this Article is not focused on how *Law Governs AI*, the *Law Informs AI* agenda suggests a novel policy approach (as a thought experiment) to governing AI's relationships to humans and the physical world: wrapping all agentic AI systems in their own legal entities, for example, a Corporation or a Limited Liability Company. A public policy, then, to further align AI with humans would be to enforce that the legal entity has verified human shareholders. The corporation is, to a large extent, a mechanism designed to reduce the principal-agent problem between shareholders and managers, *see* Del. Code Ann. Tit. 8, § §141(a) ("The business and affairs of every corporation organized under this chapter shall be managed by or under the direction of a board of directors. . . . "), so with humans as the shareholders the corporate form could help align the corporate AI "management." Regardless of whether wrapping the system in a legal entity would be helpful, under current law, sufficiently advanced AI systems would be able to utilize legal business entities as the key vector through which they conduct their affairs, for example, to employ humans, to sue other entities, to purchase goods. *See, e.g.*, Shawn Bayern, *Are Autonomous Entities Possible?*, 114 Nw. U. L. Rev. Online 23 (2019); Lynn M. LoPucki, *Algorithmic Entities*, 95 Wash. Univ. L. Rev. 887 (2018); Shawn Bayern, *The Implications of Modern Business Entity Law for the Regulation of Autonomous Systems*, 19 Stan. Tech. L. Rev. 93, 104 n.43 (2015); Shawn Bayern, *Of Bitcoins, Independently Wealthy Software, and the Zero-Member LLC*, 108 Nw. U. L. Rev. 1485, 1496–97 (2014). Reducing the risk that this potentially inevitable state of affairs leads to bad outcomes for humans may imply the same conclusion: we should strictly enforce that business entities have human shareholders.

complementary, secondary effort.[328] When AI agents are navigating the world, it is important for AI to attempt to understand (or at least try to predict) moral judgements of humans encountered.[329] State-of-the-art models already perform reasonably well predicting human judgements on a spectrum of everyday situations.[330] Human intuition, our common-sense morality, often falters in situations that involve decisions about groups unlike ourselves, leading to a "Tragedy of Common-Sense Morality."[331] There is no widely-agreed upon societal mechanism to filter observed human decisions that a model can learn from to those that exhibit preferred decisions, or to validate crowd-sourced judgments about behaviors.[332] The process of learning *descriptive ethics* relies on descriptive data of how the (largely unethical) world looks or (unauthoritative, illegitimate, almost immediately outdated, and disembodied[333]) surveys of common-sense judgements of morally

charged decisions.[334] In building aligned AI, we cannot rely solely on these data sources.[335]

Instead of attempting to replicate common sense morality in AI (learning *descriptive ethics*), we could also use various academic philosophical theories – learning or hand-engineering[336] *prescriptive ethics* – to address AI-society alignment and imbue societal values.[337] We provide six reasons why prescriptive ethics is not a suitable primary framework for AI alignment.[338]

---

[334] CRISTINA BICCHIERI, NORMS IN THE WILD: HOW TO DIAGNOSE, MEASURE, AND CHANGE SOCIAL NORMS xiv (2017) ("[T]he presumed link between empirical (all do it) and normative (all approve of it) expectations may lead us into epistemic traps that are difficult to escape."); Zeerak Talat et al., *A Word on Machine Ethics: A Response to Jiang et al. (2021)*, ARXIV (Nov. 7, 2021), https://arxiv.org/pdf/2111.04158.pdf [https://perma.cc/9JPU-Z2M9].

[335] Zeerak, *supra* note 334 (arguing that one cannot rely only on abstract datasets for embedding morality in machines).

[336] Selmer Bringsjord, Konstantine Arkoudas & Paul Bello, *Toward a General Logicist Methodology for Engineering Ethically Correct Robots*, 21 IEEE INTELLIGENT SYS. 38 (2006).

[337] *See, e.g.*, WENDELL WALLACH & COLIN ALLEN, MORAL MACHINES: TEACHING ROBOTS RIGHT FROM WRONG (2009); James H. Moor, *The Nature, Importance, and Difficulty of Machine Ethics*, 21 IEEE INTELLIGENT SYS. 18 (2006). MICHAEL ANDERSON & SUSAN L. ANDERSON, MACHINE ETHICS (2011); Edmond Awad et al., *Computational Ethics*, 26 TRENDS COGNITIVE SCI. 388 (2022); James H. Moor, *Just Consequentialism and Computing, in* ETHICS & INFO. TECH. 61 (1999); Heather M. Roff, *Expected Utilitarianism*, ARXIV (July 19, 2020), https://arxiv.org/pdf/2008.07321.pdf [https://perma.cc/55R5-EK2T]; Elizabeth Gibney, *The Battle for Ethical AI at the World's Biggest Machine-learning Conference*, 577 NATURE 609 (2020), https://media.nature.com/original/magazine-assets/d41586-020-00160-y/d41586-020-00160-y.pdf [https://perma.cc/9UY8-JVWY]; Dan Hendrycks et al., *Aligning AI With Shared Human Values*, ARXIV (July 24, 2021), https://arxiv.org/abs/2008.02275.pdf [https://perma.cc/B7F3-3TQM]; NATIONAL ACADS. OF SCS., ENG'G, & MED., FOSTERING RESPONSIBLE COMPUTING RESEARCH: FOUNDATIONS AND PRACTICES (2022); Joshua Greene et al., *Embedding Ethical Principles in Collective Decision Support Systems*, 30 PROC. CONF. ON A.I. 4147 (2016).

[338] If the ethical theory is a consequentialist one, another issue is that the implementation would have major capabilities externalities. *See* Dan Hendrycks & Thomas Woodside, *Perform Tractable Research While Avoiding Capabilities Externalities* (2022), https://www.alignmentforum.org/posts/dfRtxWcFDupfWpLQo/perform-tractable-research-while-avoiding-capabilities [https://perma.cc/HM67-KBYR] ("[O]ne should not try to model consequentialist ethics by building better general predictive world models, as this is likely to create capabilities externalities.").

*First*, there is no unified ethical theory precise enough to be practically useful for building AI;[339] therefore, it does not meet our first desired characteristic of an alignment framework.[340]

*Second*, ethics does not have any rigorous tests of its theories; it does not meet our second desired characteristic of an alignment framework because it has not been battle-tested outside of academia, "[t]he truly difficult part of ethics—actually translating normative theories, concepts and values into good practices AI practitioners can adopt—is kicked down the road like the proverbial can."[341] Two corollaries to these first two issues are that we cannot validate the ethics of AI or its behaviors in any widely agreed-upon manner,[342] and there is little data on empirical applications (especially not one with sufficient ecological validity[343]) that can be leveraged by machine learning processes.[344] Law is validated in a widely agreed-upon manner,[345] for example, court opinion, and has databases of empirical application with sufficient ecological validity.

*Third*, ethics, by its nature, lacks settled precedent across, and even within, theories.[346] There are, justifiably, fundamental disagreements

---

[339] *See, e.g.,* Mittelstadt, *supra* note 11, at 503 ("Fairness, dignity and other such abstract concepts are examples of 'essentially contested concepts' with many possible conflicting meanings that require contextual interpretation through one's background political and philosophical beliefs. These different interpretations, which can be rationally and genuinely held, lead to substantively different requirements in practice, which will only be revealed once principles or concepts are translated and tested in practice."). For various proposals, see, Roger Clarke, *Principles and Business Processes for Responsible AI*, 35 COMPUT. L. & SEC. REV. 410 (2019); Jessica Morley et al., *Ethics as a Service: A Pragmatic Operationalisation of AI Ethics*, 31 MINDS & MACHS. 239 (2021); Jeroen van den Hoven, *Computer Ethics and Moral Methodology*, 28 METAPHILOSOPHY 234 (1997); Walter B. Gallie, *Essentially Contested Concepts*, 56 PROC. ARISTOTELIAN SOC'Y 167 (1955); Henry S. Richardson, *Specifying Norms As a Way to Resolve Concrete Ethical Problems*, 19 PHIL. & PUB. AFFS. 279 (1990).

[340] *See supra* Section I for the framework requirements.

[341] Mittelstadt, *supra* note 11, at 503; *see also* Katie Shilton, *Values Levers: Building Ethics Into Design.* 38 SCI., TECH., & HUM. VALUES 374 (2013) (exploring ways information systems can be designed with ethics built in).

[342] *See, e.g.,* Anne Gerdes & Peter Øhrstrøm, *Issues in Robot Ethics Seen Through the Lens of a Moral Turing Test*, 13 J. INFO., COMMC'N & ETHICS SOC'Y 98 (2015); Joachim Van den Bergh & Dirk Deschoolmeester, *Ethical Decision Making in ICT: Discussing the Impact of an Ethical Code of Conduct*, 2010 COMMC'NS IBIMA 1 (2010); Batya Friedman, David G. Hendry, & Alan Borning, *A Survey of Value Sensitive Design Methods*, 11 FOUNDS. & TRENDS HUM.-COMP. INTERACTIONS 63 (2017); Mittelstadt, *supra* note 11; Mireille Hildebrandt, LAW FOR COMPUTER SCIENTISTS AND OTHER FOLK 283–315 (2020).

[343] *See, e.g.,* Martin T. Orne & Charles H. Holland, *On the Ecological Validity of Laboratory Deceptions*, 6 INT'L J. PSYCHIATRY 282 (1968).

[344] See *supra* Section II.C.2. for how law can be leveraged by machine learning.

[345] *See, e.g.,* Hildebrandt, *supra* note 342.

[346] *See, e.g.,* Gabriel, *Values*, *supra* note 311, at 425 ("[I]t is very unlikely that any single moral theory we can now point to captures the entire truth about morality. Indeed, each of the major candidates, at least within Western philosophical traditions, has strongly counterintuitive moral

between reasonable people about which ethical theory would be best to implement, spanning academic metaphysical disagreements to more practical indeterminacies, "not only are there disagreements about the appropriate ethical framework to implement, but there are specific topics in ethical theory [ . . . ] that appear to elude any definitive resolution regardless of the framework chosen."[347] As AI is more broadly deployed, there will be much more widespread attention on the underpinnings of AI system design. As this scrutiny increases, there will be deep investigation into what morally relevant principles are being embedded in AI, and strong backlash from the public, media, and the government into philosophical theories. Public law is not immune from criticism either, but the public can take that criticism to their elected representatives.

    *Fourth*, even if AI developers (impossibly) agreed on one ethical theory (or ensemble of underlying theories [348]) being "correct," there is no mechanism to align humans around that theory (or "meta-theory").[349] In contrast, in democracies, law has legitimate authority imposed by widely accepted government institutions,[350] and serves as a coordinating focal point of values to facilitate human progress.[351] Imbuing understanding of ethical frameworks is a useful exercise. The law is silent on many important values that humans hold and we can use ethical modules to better align AI with its human principal by imbuing the ethical framework that the human principal chooses into the AI. But this is more in the ***human-AI alignment* realm** than a ***society-AI alignment* solution**. Society-AI alignment requires us to move beyond "private contracts" between a human and her AI and into the realm of public law to explicitly address inter-agent conflicts and policies designed to ameliorate externalities and solve massively multi-agent coordination and

---

implications in some known situations, or else is significantly underdetermined."); JOSEPH F. FLETCHER, SITUATION ETHICS: THE NEW MORALITY (1966).

    [347] Miles Brundage, *Limitations and Risks of Machine Ethics,* 26 J. EXPERIMENTAL & THEORETICAL A.I. 355, 369 (2014).

    [348] *See, e.g.*, Toby Newberry & Toby Ord, *The Parliamentary Approach to Moral Uncertainty* (Future of Human. Inst., Technical Report #2021-2, 2021); William MacAskill, *Practical Ethics Given Moral Uncertainty*, 31 UTILITAS 231 (2019); Adrien Ecoffet & Joel Lehman, *Reinforcement Learning Under Moral Uncertainty*, 139 PROC. MACH. LEARNING RSCH. 2926 (2021).

    [349] *See, e.g.*, JOHN RAWLS, THE LAW OF PEOPLES, WITH "THE IDEA OF PUBLIC REASON REVISITED" 11–16 (1999); Gabriel, *Values*, *supra* note 311, at 425.

    [350] *See generally* DAVID ESTLUND, DEMOCRATIC AUTHORITY: A PHILOSOPHICAL FRAMEWORK (2008); Gabriel, *Values*, *supra* note 311, at 432.

    [351] "Law is perhaps society's most general purpose tool for creating focal points and achieving coordination. Coordinated behavior requires concordant expectations, and the law creates those expectations by the dictates it expresses." RICHARD H. MCADAMS, THE EXPRESSIVE POWERS OF LAW, https://www.hup.harvard.edu/catalog.php?isbn=9780674975484 260 (2017) [hereinafter McAdams, *The Expressive Powers of Law*].

cooperation dilemmas through top-down implementations.[352] We can use ethics to better align AI with its human principal by imbuing an ethical framework that the human principal chooses into the AI. But choosing one out of the infinite possible ethical theories (or choosing an ensemble of theories) and "uploading" that into an AI does not work for a ***society-AI*** alignment solution because we have no means of deciding – across all the humans that will be affected by the resolution of the inter-agent conflicts and the externality reduction actions taken – which ethical framework to imbue in the AI. When attempting to align multiple humans with one or more AI, we would need something like a "council on AI ethics," where every affected human is bought in and will respect the outcome (even when they disagree with it). This is not even remotely practical.

*Fifth*, even if AI developers (impossibly) agreed on one ethical theory (or ensemble of underlying theories) being "correct," it is unclear how any consensus update mechanism to that chosen ethical theory could be implemented to reflect evolving[353] (usually, improving) ethical norms; there is no endogenous society-wide process for this. Society is likely more ethical than it was in previous generations, and humans are (hopefully) not at an ethical peak now either, which provides aspiration that we continue a positive trajectory. Therefore, we do not want to lock in today's ethics without a clear and trustworthy update mechanism.[354] In stark contrast, law is formally revised to reflect the evolving will of citizens.[355] If AI is designed

---

[352] For a discussion of the difference between intent-alignment (like our characterization of *human-AI alignment*) and Law-following AI, see O'Keefe, *supra* note 11.

[353] *See, e.g.*, Melissa A. Wheeler, Melanie J. McGrath & Nick Haslam, *Twentieth Century Morality: The Rise and Fall of Moral Concepts from 1900 to 2007*, PLOS ONE (Feb. 27, 2019), https://journals.plos.org/plosone/article?id=10.1371/journal.pone.0212267 [https://perma.cc/4FCK-T5ZD]; Aida Ramezani, Zining Zhu, Frank Rudzicz & Yang Xu, *An Unsupervised Framework for Tracing Textual Sources of Moral Change*, 2021 FINDINGS ASS'N FOR COMPUTATIONAL LINGUISTICS 1215.

[354] *See, e.g.*, William MacAskill, *Are We Living at the Hinge of History?* (Global Priorities Institute, Working Paper #12-2020, 2020), https://globalprioritiesinstitute.org/wp-content/uploads/William-MacAskill_Are-we-living-at-the-hinge-of-history.pdf [https://perma.cc/MNL8-XDTC]; TOBY ORD, THE PRECIPICE: EXISTENTIAL RISK AND THE FUTURE OF HUMANITY (2020); WILLIAM MACASKILL, WHAT WE OWE THE FUTURE 97 (2022) ("Almost all generations in the past had some values that we now regard as abominable. It's easy to naively think that one has the best values; Romans would have congratulated themselves for being so civilized compared to their "barbarian" neighbors and in the same evening beaten people they had enslaved. . . .").

[355] Modeling the evolution of an area of law (e.g., the "legislative history" of the drafting and enactment of legislation, and subsequent amendments to the statue) as a sequential decision-making process could be a useful method for AI to learn implicit reward functions of the citizenry regarding policy areas. For an evolutionary perspective on reward functions, see, for example, Satinder Singh, Richard L. Lewis & Andrew G. Barto, *Where Do Rewards Come From*, 31 PROC. ANN. CONF. COGNITIVE SCI. SOC'Y 2601 (2009). Law may become revised even faster as technology advances. *See, e.g.*, Sandy Pentland & Robert Mahari, *Legal Dynamism*, NETWORK L. REV. (September 27, 2022),

to use law as a key source of alignment insight (and AI capabilities are advanced enough to enable the requisite understanding), this would build in an automatic syncing with the latest iteration of synthesized and validated societal value preference aggregation.[356]

    *Sixth*, veering into the intersection of *Law Informs Code* and *Law Governs Code*, there is a practical reason law is best suited as the core alignment framework. For alignment work to have any impact, we need aligned AI to be economically competitive with general AI being developed. Techniques that increase AI safety at the expense of AI capabilities (i.e., levy an "alignment tax") lead to organizations eschewing safety to gain additional capabilities as organizations race forward deploying AI. Most entities developing and deploying state-of-the-art AI are organizations that have core goals of profit-maximization and liability-minimization.[357] The liability-minimization impulse of organizations – run by humans worried about being sanctioned by governments, fined, and put in jail – makes law-informed AI economically competitive. Humans are more likely to deploy AI associated with a lower probability that they are jailed due to being liable for the AI breaking laws. Any organization of humans large and organized enough to build state-of-the-art transformative AI likely has liability-minimization as one of its core drives (e.g., corporations in the United States). Contrast this with morality-maximizing AI, which is often economically disadvantaged compared to other approaches. Our goal as a society, then, is to make our laws as moral as we can. If law informs powerful AI, engaging in the human deliberative political process to improve law takes on even more meaning. This is a more empowering vision of improving AI outcomes than one where companies dictate their ethics by fiat.[358]

---

https://www.networklawreview.org/computational-one/ [https://perma.cc/89XN-JRMX] ("[C]omputational approaches are finding their way into the creation and implementation of law and the field of computational law is rapidly expanding. One of the most exciting promises of computational law is the idea of legal dynamism: the concept that a law, by means of computational tools, can be expressed not as a static rule statement but rather as a dynamic object that includes system performance goals, metrics for success, and the ability to adapt the law in response to its performance.").

[356] "Common law, as an institution, owes its longevity to the fact that it is not a final codification of legal rules, but rather a set of procedures for continually adapting some broad principles to novel circumstances." Scott, *supra* note 13, at 357.

[357] *See,* Bryan Casey, *Amoral Machines, or: How Roboticists Can Learn to Stop Worrying and Love the Law,* 111 Nw. U. L. Rev. 1347 (2017) [hereinafter Casey, *Amoral Machines*].

[358] Bryan Casey concludes that, "[w]e, the people, will be the true engineers of machine morality. As democratic stakeholders, it will be our collective 'engineering task' to ensure that even the worst of our robots are incentivized to behave as the best of our philosophers." Casey, *Amoral Machines, supra* note 357, at 1365. *See also* Ryan Calo, *Artificial Intelligence and the Carousel of Soft Law,* 2 IEEE Transactions on Tech. & Soc'y 171 (2021) ("Principles alone are no substitute for, and have the potential to delay, the effort of rolling up our collective sleeves and figuring out what AI changes, and how the law needs to evolve. . . . Unlike law, which requires consensus and rigid process, an

In sum, legal informatics possesses the positive attributes from both descriptive and prescriptive ethics, but does not share their incurable negatives (Figure 9).

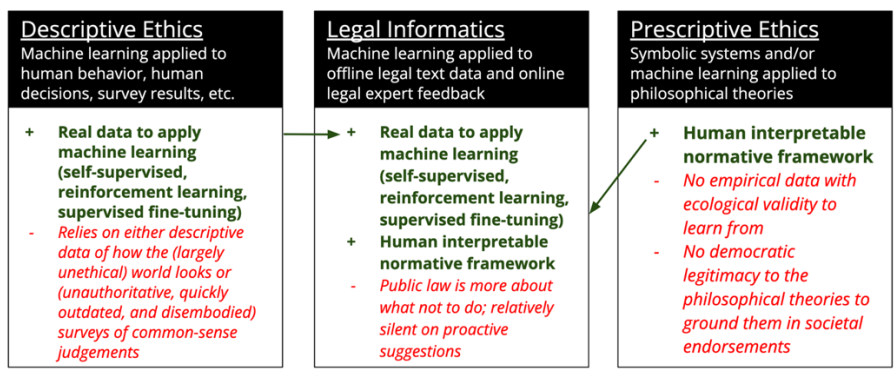

**Figure 9:** Three contenders for a society-AI alignment framework.

Ethics should be a core component of human-AI alignment; more center stage than it currently is for AI researcher training, AI development guidelines,[359] and AI deployment and monitoring protocols. Ethics should guide data selection and processing in legal informatics. At the same time, we agree with John Rawls that, "in a constitutional democracy the public conception of justice should be, so far as possible, independent of controversial philosophical and religious doctrines," and, "the public conception of justice is to be political, not metaphysical."[360] The question, then, is how to leverage this democratically legitimate legal data for society-AI alignment.

### B.  *Toward Implementation*

Case law can teach AI how to map from democratically determined directives (statutes) to specific implementation, whereas statutes are more useful for embedding world knowledge and human value expressions. Legislation expresses a significant amount of information about the values

---

organization can develop and publish principles unilaterally. . . . While there is some utility in public commitments to universal values in the context of AI, and while common principles can lay a foundation for societal change, they are no substitute for law and official policy.").

[359]  *See, e.g.*, James Bessen et al., *Ethical AI Development: Evidence from AI Startups* (Ctr. on Regs. & Mkts. at Brookings, Working Paper, 2022), https://scholarship.law.bu.edu/cgi/viewcontent.cgi?article=2165&context=faculty_scholarship, archived at [https://perma.cc/NR9P-5B2W].

[360]  John Rawls, *Justice as Fairness: Political Not Metaphysical*, 14 PHIL. & PUB. AFFS. 223, 224 (1985); *see also* Gabriel, *Values*, *supra* note 311.

384

of citizens,[361] "for example, by banning employment discrimination against LGBT workers, the legislature may communicate pervasive attitudes against such employment practices."[362] And, "the Endangered Species Act has a special salience as a symbol of a certain conception of the relationship between human beings and their environment, and emissions trading systems are frequently challenged because they are said to 'make a statement' that reflects an inappropriate valuation of the environment."[363]

Although special interest groups can influence the legislative process, legislation is largely reflective of citizen beliefs because "legislators gain by enacting legislation corresponding to actual attitudes (and actual future votes)."[364] The second-best source of citizen attitudes is arguably a poll, but polls are not available at the local level, are only conducted on mainstream issues, and the results are highly sensitive to their wording and sampling techniques. Legislation expresses higher fidelity, more comprehensive, and trustworthy information because the legislators "risk their jobs by defying public opinion or simply guessing wrong about it. We may think of legislation therefore as a handy aggregation of the polling data on which the legislators relied, weighted according to their expert opinion of each poll's reliability."[365] More recent legislation could be interpreted as providing fresher pulse checks on citizen attitudes;[366] however, methods for differentially weighting public law based on its estimated expressive power is an important open research area for how *Law Informs Code*.

Legislation and associated agency rule-making also express a significant amount of information about the risk preferences and risk tradeoff views of citizens, "for example, by prohibiting the use of cell phones while

---

[361] *See, e.g.,* Cass R. Sunstein, *Incommensurability and Valuation in Law*, 92 MICH. L. REV. 779, 820–24 (1994); Richard H. Pildes & Cass R. Sunstein, *Reinventing the Regulatory State*, 62 U. CHI. L. REV. 1, 66–71 (1995); Cass R. Sunstein, *On the Expressive Function of Law*, 144 U. PA. L. REV. 2021 (1996); Dhammika Dharmapala & Richard H. McAdams, *The Condorcet Jury Theorem and the Expressive Function of Law: A Theory of Informative Law*, 5 AM. L. & ECON. REV. 1 (2003).

[362] McAdams, *supra* note 65, at 137.

[363] Sunstein, *On the Expressive Function of Law*, *supra* note 364, at 2024 (citing STEVEN KELMAN, WHAT PRICE INCENTIVES?: ECONOMISTS AND THE ENVIRONMENT 2 (1981)).

[364] McAdams, *supra* note 65, at 149.

[365] *Id.* at 146.

[366] There is also some predictability to the enactment of proposed bills in Congress. *See* Matthew Hutson,

*Artificial Intelligence Can Predict Which Congressional Bills Will Pass: Machine Learning Meets the Political Machine*, SCIENCE.ORG (June 21, 2017), https://www.science.org/content/article/artificial-intelligence-can-predict-which-congressional-bills-will-pass pass[https://perma.cc/KDX5-JEQL]; *see also* John Nay, *Predicting and Understanding Law-making with Word Vectors and an Ensemble Model*, PLOS ONE 1 (May 10, 2017),

https://journals.plos.org/plosone/article?id=10.1371/journal.pone.0176999 [https://perma.cc/L6JZ-DPNZ].

driving, legislators may reveal their beliefs that this combination of activities seriously risks a traffic accident."[367] All activities have some level of risk, and making society-wide tradeoffs about which activities are deemed to be "riskier" relative to the perceived benefits of the activity is ultimately a sociological process with no objectively correct ranking.[368] The cultural process of prioritizing risks is reflected in legislation and its subsequent implementation in regulation crafted by domain experts. Finally, some legislation expresses shared understandings and customs that have no inherent normative or risk signal, but facilitate orderly coordination, e.g., which side of the road to drive on.[369]

Acknowledging that data contains socio-economic, racial,[370] and gender biases,[371] we should frame the challenge of estimating the expressive power of public law[372] broadly to factor in whether views of historically marginalized populations are expressed.[373] Work on fairness, accountability, and transparency of AI[374] can inform research on methods for estimating a more comprehensive notion of the expressiveness of legal data.[375] Methods are being developed that attempt to improve the fairness of machine learning[376] through data preprocessing,[377] adjusting model parameters during training,[378] and adjusting predictions from models that have already been trained.[379] Another issue is that legal data can contain political biases in places where it is purported to be produced by processes fully committed to judicial[380] and agency[381] independence.

---

[367] McAdams, *supra* note 65, at 138.

[368] *See, e.g.*, CARLA ZOE CREMER & LUKE KEMP, DEMOCRATISING RISK: IN SEARCH OF A METHODOLOGY TO STUDY EXISTENTIAL RISK (2021) (commenting on long-term existential risk).

[369] Richard H. McAdams & Janice Nadler, *Coordinating in the Shadow of the Law: Two Contextualized Tests of the Focal Point Theory of Legal Compliance*, 42 L. & SOC'Y REV. 865 (2008); Richard H. McAdams, *A Focal Point Theory of Expressive Law*, 86 VA. L. REV. 1649 (2000); Dylan Hadfield-Menell et al., *Legible Normativity for AI Alignment: The Value of Silly Rules*, 2019 PROC. AAAI/ACM CONF. ON AI, ETHICS, & SOC'Y 115.

[370] *See, e.g.*, Rashida Richardson et al., *Dirty Data, Bad Predictions: How Civil Rights Violations Impact Police Data, Predictive Policing Systems, and Justice*, 94 N.Y.U. L. REV. ONLINE 15 (2019); Z. Obermeyer et al., *Dissecting Racial Bias in an Algorithm Used to Manage the Health of Populations*, 366 SCIENCE 447 (2019).

[371] *See, e.g.*, Caroline Criado Perez, INVISIBLE WOMEN: DATA BIAS IN A WORLD DESIGNED FOR MEN (2019); Joy Buolamwini & Timnit Gebru, *Gender Shades: Intersectional Accuracy Disparities in Commercial Gender Classification*, 81 PROC. MACH. LEARNING RSCH. 77 (2018).

[372] And across all types of legal data used as part of the legal informatics efforts.

[373] For legal discussions, see, for example, Sandra G. Mayson, *Bias In, Bias Out*, 128 YALE L.J. 2218 (2019); Deborah Hellman, *Measuring Algorithmic Fairness*, 106 VA. L. REV. 811 (2020).

[374] *See, e.g.*, Timnit Gebru et al., *Datasheets for Datasets*, 64 COMMC'NS ACM 86 (2021); Emily M. Bender & Batya Friedman, *Data Statements for Natural Language Processing: Toward Mitigating System Bias and Enabling Better Science*, 6 TRANSACTIONS ASS'N FOR COMPUTATIONAL LINGUISTICS 587 (2018); Margaret Mitchell et al., *Model Cards for Model Reporting*, 2019 PROC. CONF. ON FAIRNESS, ACCOUNTABILITY, & TRANSPARENCY 220; William Cai et al., *Adaptive Sampling Strategies to Construct Equitable Training Datasets*, 2022 PROC. CONF. ON FAIRNESS, ACCOUNTABILITY & TRANSPARENCY 1467; Abdulaziz A. Almuzaini et al., *ABCinML: Anticipatory Bias Correction in Machine Learning Applications*, 2022 PROC. CONF. ON FAIRNESS, ACCOUNTABILITY & TRANSPARENCY 1552.

[375] *See, e.g.,* MCKANE ANDRUS ET AL., AI DEVELOPMENT FOR THE PUBLIC INTEREST: FROM ABSTRACTION TRAPS TO SOCIOTECHNICAL RISKS (2021).

Imbuing AI with the capability to understand new statutes is a significant technical challenge.[382] As the state-of-the-art for AI advances, the *Law Informs Code* approach aims to validate correspondingly advanced[383] legal reasoning and statutory interpretation abilities.[384]

## V.  CONCLUSION

Novel AI capabilities continue to emerge, increasing the urgency to align AI with humans. We cannot directly specify "good" AI behavior *ex ante*. Similarly, parties to a legal contract cannot foresee every contingency, and legislators cannot predict all the specific circumstances under which their laws could be applied. Law, as the applied philosophy of multi-agent alignment, uniquely fulfills our requirements for an AI goal specification framework.

---

[376] *See, e.g.*, Reva Schwartz et al, *Towards a Standard for Identifying and Managing Bias in Artificial Intelligence*, NAT'L INST. STANDARDS & TECH. (2022), https://nvlpubs.nist.gov/nistpubs/SpecialPublications/NIST.SP.1270.pdf [https://perma.cc/HPR4-VXG5]; MICHAEL KEARNS & AARON ROTH, THE ETHICAL ALGORITHM: THE SCIENCE OF SOCIALLY AWARE ALGORITHM DESIGN 57–93 (2019).

[377] *See, e.g.*, Flavio P. Calmon et al., *Optimized Pre-Processing for Discrimination Prevention*, 30 ADVANCES NEURAL INFORMATION PROCESSING SYSTEMS (2017).

[378] *See, e.g.*, M. B. Zafar et al., *Fairness Constraints: A Flexible Approach for Fair Classification*, 20 J. MACH. LEARNING RSCH. 1 (2019).

[379] *See, e.g.*, Moritz Hardt et al., *Equality of Opportunity in Supervised Learning*, 29 ADVANCES IN NEURAL INFORMATION PROCESSING SYSTEMS 3315, 3315–23 (2016).

[380] *See, e.g.*, Neal Devins & Allison Orr Larsen, *Weaponizing En Banc*, 96 N.Y.U. L. REV. 1373, 1373–74 (2021) ("The bulk of our data indicates that rule-of-law norms are deeply embedded. From the 1960s through 2017, en banc review seems to have developed some sort of immunity from partisan behavior over time. . . . Our data from 2018–2020 show a dramatic and statistically significant surge in behavior consistent with the weaponizing of en banc review."); Keith Carlson et al., *The Problem of Data Bias in the Pool of Published US Appellate Court Opinions*, 17 J. OF EMPIRICAL LEGAL STUD. 224, 224–61 (2020).

[381] Daniel B. Rodriguez, *Whither the Neutral Agency? Rethinking Bias in Regulatory Administration*, 69 BUFF. L. REV. 375 (2021); Jodi L. Short, *The Politics of Regulatory Enforcement and Compliance: Theorizing and Operationalizing Political Influences*, 15 REGUL. & GOVERNANCE 653, 653–85 (2021).

[382] *See, e.g.*, NILS HOLZENBERGER ET AL., A DATASET FOR STATUTORY REASONING IN TAX LAW ENTAILMENT AND QUESTION ANSWERING (2020).

[383] These may be powerful Law-Informed AIs that interact with capable AIs that are not necessarily themselves law-informed. "Ambition must be made to counteract ambition." THE FEDERALIST NO. 51 (James Madison).

[384] *See supra* Section II.B.

Methodologies for making and interpreting law, which advance shared goals in new circumstances, have been theoretically refined over centuries. One of the primary goals of the *Law Informs Code* agenda is to verify that AI can follow the spirit of the law. This entails leveraging humans for the "law-making" and "contract-drafting" part (*using the theory and practice of law to tell agents what to do*) and building AI capabilities for the interpretation part.

Most research on managing the potential unintended consequences of AI development currently falls into two ends of a spectrum related to assumptions of the imminence of AGI. The research operating under a *high probability* estimate of near-term AGI is focused on how to align AGI with a human's intentions to avoid human extinction. Research operating under a *low probability* estimate of near-term AGI is usually focused more on how to reduce discriminatory and privacy harms by present-day AI. As the state-of-the-art for AI advances, we can set higher bars of demonstrated legal understanding capabilities; if a developer claims their AI has advanced capabilities on tasks, they should demonstrate correspondingly advanced legal comprehension of the AI. *Law Informs Code* research bridges these ends of the AI safety spectrum.

The benefits of law-informed AI could be far-reaching. In addition to more locally useful and societally aligned AI, law-informed AI could power the other two pillars of the existing AI and Law intersection: it is easier for law to govern AI if AI understands the law (all else equal, i.e., holding goal directedness, dishonesty and power-seeking equal), and AI can improve legal services more if it understands the law better.

The practices of making, interpreting, and enforcing law have been battle tested through millions of contracts and legal and regulatory actions that have been memorialized in digital format, providing large data sets of examples and explanations, and millions of well-trained active lawyers from which to elicit machine learning model feedback to embed an evolving comprehension of human goals. However, much more work needs to be done.

For instance, from the theory side, public *law informs code* more through negative than positive directives and therefore it's unclear the extent to which policy – outside of the human-AI "contract and standards" type of alignment we discuss – can inform which goals AI should proactively pursue to improve the world on society's behalf.[385] Legal and ethical theorizing on these questions could help guide research. We should also conduct research on how to systematically differentially weight empirical legal data based on its estimated expressive power (defined broadly to account for historical injustices and how they reduce the extent to which certain areas of law update fast enough to express current human views).

---

[385] This concern is similar to the reinforcement learning research on reward functions that seek to balance a tradeoff between an AI agent doing nothing and causing too much impact in the world. *See, e.g.*, Victoria Krakovna et al., *Avoiding Side Effects by Considering Future Tasks*, 33 ADVANCES IN NEURAL INFO. PROCESSING SYS. 19064 (2020); Alex Turner et al., *Avoiding Side Effects in Complex Environments*, 33 ADVANCES IN NEURAL INFORMATION PROCESSING SYSTEMS 21406 (2020);

This Article developed ways in which U.S. *law informs code*. We need to extend this to scale the approach globally.[386] The evolutionary psychology of law could be useful in determining cross-cultural universals in legal systems that exemplify common ground for human values.[387]

Issues abound. It is unclear how much we need to improve our understanding of the mental states of AI to advance AI legal understanding,[388] in particular the level of intention of an AI.[389] Governments could become increasingly polarized along partisan lines, and the resulting laws thus increasingly a poor representation of an aggregation of citizens

---

values and preferences.[390] In many cases, Foundation Models are not truthful.[391] AI could find legal loopholes and aggressively exploit them.[392]

Fortunately, mainstream AI capabilities research – while unlocking better AI performance on existing tasks and enabling more powerful and general agents – is likely to have positive externalities on legal informatics alignment performance by enabling greater AI legal reasoning capabilities and legal interpretation skills. However, an area of research with less focus relative to the high value it would deliver to legal informatics capabilities is natural language processing (NLP) of long documents.[393] The current focus of most NLP work is on documents shorter than the average legal text.[394] To deploy Foundation Models more successfully on legal text data, we need models to be able to process and comprehend longer documents. There is promising work in this direction.[395]

---

[390] However, it seems there has been more bipartisanship substantive legislation advanced in the past two years than many political experts would have predicted – the system is incredibly resilient.

[391] S. Lin et al., *TruthfulQA: Measuring How Models Mimic Human Falsehoods*, 1 PROC. 60TH ANN. MEETING ASS'N FOR COMPUTATIONAL LINGUISTICS 3214 (2021); Dan Hendrycks et al., *Measuring Massive Multitask Language Understanding* (International Conference on Learning Representations, Conference Paper, May 4, 2021), https://openreview.net/pdf?id=d7KBjmI3GmQ [https://perma.cc/32CY-G5TL]; Jared Kaplan et al., *Scaling Laws for Neural Language Models* (International Conference on Learning Representations, Conference Paper, April 25, 2022), https://openreview.net/pdf?id=hR_SMu8cxCV [https://perma.cc/Q4LX-5VQ2]; Owain Evans et al., *Truthful AI: Developing & Governing AI That Does Not Lie*, ARXIV (Oct. 13, 2021), https://arxiv.org/pdf/2110.06674.pdf [https://perma.cc/HF2J-HLAY]. However, scaling, alone, is unlikely to make the models "fully truthful." See Evans et al., *supra*, at 59 ("The data [the models are trained on] contains many instances of humans being non-truthful and so [the model] will likely be non-truthful in the same contexts. In summary, we have a speculative argument that language modelling (without tweaks or modifications) is unlikely to produce truthful AI systems.").

[392] However, this is a well-known problem, and the legal system organically adopts quick solutions to it.

[393] *See, e.g.*, Cem Anil et al., *Exploring Length Generalization in Large Language Models* (Neural Information Processing Systems, Nov. 28, 2022), https://arxiv.org/abs/2207.04901 ("[N]aively finetuning transformers on length generalization tasks shows significant generalization deficiencies independent of model scale.").

[394] *See generally* CONGRESS.GOV, https://www.congress.gov/bill-texts-received-today [https://perma.cc/82QL-PDPT] (showing that congressional bills are routinely hundreds of pages long); Nay, *Predicting and Understanding Law-Making with Word Vectors and an Ensemble Model*, PLOS ONE 1 (May 10, 2017), https://journals.plos.org/plosone/article?id=10.1371/journal.pone.0176999 [https://perma.cc/L6JZ-DPNZ].

[395] *See e.g.*, Yunyang Xiong et al., *Nyströmformer: A Nyström-Based Algorithm for Approximating Self-Attention*, 35 AAAI CONF. ON A.I. 14138 (2021); Yann Dubois et al., *Location Attention for Extrapolation to Longer Sequences*, 2020 PROC. ANN. MEETING ASS'N FOR COMPUTATIONAL LINGUISTICS 403; Cem Anil et al., *Exploring Length Generalization in Large Language Models* (NeurIPS 2022), https://arxiv.org/pdf/2207.04901.pdf [https://perma.cc/E7YK-5SWS]; Iz Beltagy et al., *Longformer: The Long-Document Transformer*, ARXIV (2020), https://arxiv.org/pdf/2004.05150.pdf [https://perma.cc/X3F8-QMYB]; Nikita Kitaev et al., *Reformer: The Efficient Transformer* (International Conference on Learning Representations, Conference Paper, April 30, 2020), https://arxiv.org/pdf/2001.04451.pdf [https://perma.cc/LN3M-PKSY]; Jason Phang et al., *Investigating*

The legal informatics approach to aligning AI suggest three high-level courses of action. *First*, we should advance legal informatics' unique role in theoretically framing, and technically implementing, improvements to AI by fully embedding a deep understanding of law, the language of alignment, into AI. *Second*, we should increase the probability that law expresses the democratically deliberated views of citizens with fidelity and integrity by reducing regulatory capture, illegal lobbying, the politicization of judicial and agency independence, and impact of AI on law-making (defined broadly to include proposing and enacting legislation, promulgating regulatory agency rules, publishing judicial opinions, enforcing law systematically, and more).[396] *Third*, we should cultivate the spread of democracy globally because then there will be more democratically produced law for AI to learn representative societal values from.[397]

In conclusion, the integration of law into AI is becoming increasingly important as AI technology advances and becomes more widely deployed. While there have been suggestions to embed ethics into AI to increase alignment with humans, it is unclear how to determine these ethics and who gets a say in the process. Instead, we propose that the target of AI alignment should be democratically endorsed law, which provides a legitimate grounding for AI behavior and can serve as a set of methodologies for conveying and interpreting directives and a knowledge base of societal values. By using law as theoretical scaffolding and data, AI can be made safer by design through the Law Informs Code research agenda. The benefits of law-informed AI would be far-reaching and could help power the other two pillars of AI and law: law governing AI, and AI improving legal services.[398]

---

[396]  Reducing the first three have obvious positive externalities beyond increasing AI safety through the Law-informed AI channel.

[397]  If, in general, democracy is the form of government most likely to lead to positive outcomes for the governed, this would have positive externalities beyond increasing AI safety through the Law-informed AI channel.

[398]  This final paragraph was written by a large language model (OpenAI's GPT-3.5) that used much of this paper as context and was prompted to provide a final paragraph for the Article. *See generally* https://platform.openai.com/docs/models/gpt-3 [https://perma.cc/9Q7C-WPY8].